\documentclass[useAMS,usenatbib,usegraphicx]{mn2e}
\usepackage{aas_macros}
\usepackage{times}

\title[An Optical Survey for X-ray Dark Clusters]
{Exploring the Selection of Galaxy Clusters and Groups:\\
  An Optical Survey for X-ray Dark Clusters} 
\author[D. G. Gilbank et al.]  {David
  G.~Gilbank$^1$\thanks{Email: D.G.Gilbank@durham.ac.uk}
  \thanks{Visiting astronomer of the German--Spanish Astronomical
    Center, Calar Alto, operated by the Max--Planck--Institut f\"ur
    Astronomie, Heidelberg, jointly with the Spanish National
    Commission for Astronomy.}, Richard~G.~Bower$^1$\footnotemark[3],
  F.~J.~Castander$^2$ and
  B.~L.~Ziegler$^3$\footnotemark[3]\\
  $^1$Department of Physics, University of Durham, South Road, Durham,
  DH1 3LE, UK\\
  $^2$Institut d'Estudis Espacials de Catalunya/CSIC Gran Capit\`a 2-4,
  08034, Barcelona, Spain.  \\
  $^3$Sternwarte, Geismarlandstr. 11, 37083 G\"ottingen, Germany }
\date{\today}

\def\gsim{\mathrel{\raise0.35ex\hbox{$\scriptstyle >$}\kern-0.6em
\lower0.40ex\hbox{{$\scriptstyle \sim$}}}}
\def\lsim{\mathrel{\raise0.35ex\hbox{$\scriptstyle <$}\kern-0.6em
\lower0.40ex\hbox{{$\scriptstyle \sim$}}}}
\def\sex {{\sc SExtractor}}

\def\hsf { {$h_{64}$}}
\def\hmpc {{h$^{-1}$Mpc}}
\def\h3mpc {{h$^3$Mpc$^{-3}$}}
\def\tx {{T$_X$}}
\def\lx {{L$_X$}}
\def\lell {{L$_E$}~~}
\def\kms { {kms$^{-1}$} }
\def\n05 { {$N_{0.5}$} }
\def\bgc { {$B_{gc}$} }

\def\mstar {{M$^\star$}}

\pagerange{\pageref{firstpage}--\pageref{lastpage}} \pubyear{2002}

\def\LaTeX{L\kern-.36em\raise.3ex\hbox{a}\kern-.15em
    T\kern-.1667em\lower.7ex\hbox{E}\kern-.125emX}

\begin{document}

\label{firstpage}

\maketitle

\begin{abstract}
  
  Data from a new, wide field, coincident optical and X-ray survey, the
  X-ray Dark Cluster Survey (XDCS) are presented. This survey comprises
  simultaneous and independent searches for clusters of
  galaxies in the optical and X-ray passbands. Optical cluster
  detection algorithms implemented on the data are detailed. Two
  distinct optically selected catalogues are constructed, one based on
  I-band overdensity, the other on overdensities of colour-selected
  galaxies.  The superior accuracy of the colour-selection technique
  over that of the single passband method is demonstrated, via internal
  consistency checks and comparison with external spectroscopic
  redshift information.  This is compared with an X-ray selected
  cluster catalogue. In terms of gross numbers, the survey yields
  185 I-band selected, 290 colour selected and 15 X-ray selected
  systems, residing in $\sim$11deg$^2$ of optical $+$ X-ray imaging.
  
  The relationship between optical richness/ luminosity and X-ray
  luminosity is examined, by measuring X-ray luminosities at the
  positions of our 290 colour-selected systems.  Power law correlations
  between the optical richness/ luminosity versus X-ray luminosity are
  fitted, both exhibiting approximately 0.2 dex of intrinsic
  scatter. Interesting outliers in these correlations are discussed in
  greater detail. Spectroscopic follow up of a subsample of X-ray
  underluminous systems confirms their reality.

\end{abstract}


\section{Introduction}
\label{sec:introduction}

Clusters of galaxies are extremely important astrophysical tools. They
are the most massive virialised objects in the Universe.  Since
clusters form from extremely high peaks in the initial density field on
scales of around 10~\hmpc, they are sensitive to the amplitude of the
power spectrum on these scales.  Thus, observations of the cluster
mass function out to large redshifts can place tight constraints on
cosmological parameters \citep[e.g.  $\Omega_m$, $\sigma_8$, $\Lambda$;
][]{ecf96}. They are also powerful laboratories for studying galaxy
formation and evolution. Several different techniques exist for finding
clusters, each relying on different properties of clusters in order to
locate them, and it is important to try to understand how the selection
method may bias the sample and affect the scientific results.

The first attempt at a large, homogeneous survey for galaxy clusters
was conducted by \citet{abell}. This was a phenomenal effort by one
individual to identify overdensities of galaxies by visual inspection
of Palomar Observatory Sky Survey (POSS) photographic plates, yielding
nearly 1700 clusters in his ``homogeneous statistical sample'' and over
2700 in his full sample.  Similar catalogues were constructed by Zwicky
and collaborators \citep{zwicky}.  \citet{abell}'s Northern catalogue
was extended to the Southern hemisphere by \citet{aco}, applying his 
same statistical criteria.

With the advent of space-based X-ray telescopes, such as UHURU, a new
way to discover galaxy clusters was found.  Spatially extended, thermal
X-ray emission was detected and shown to be due to the hot intracluster
medium (ICM) - the plasma trapped in a cluster's potential well
\citep{mitchel76,serlemitsos77}.  This provided a way to show that the
cluster was a genuine physically bound system. Furthermore, the
background signal (produced by X-ray point sources) is lower in the
X-ray sky than the background in the optical, produced by a
much greater surface density of foreground and background galaxies.
Optical selection techniques lost favour: their main disadvantage being
that there was no way, at the selection stage, to distinguish between
genuine clusters and chance projections of less massive galaxy groups
along the lines of sight.  Extensive discussions of this contamination
have been published \citep[e.g.][]{katgert96,vfw}. To reject such
spurious systems, observationally expensive spectroscopy is required to
confirm the overdensities in 3D. Despite the revolutionary new X-ray
techniques, four large optical photographic cluster surveys with follow
up spectroscopy were undertaken in the late 1980s
\citep{gho86,cemm,apmclus1,edcc}.  The first two used visual inspection
of photographic plates, and the second two utilised machines which
automatically measured parameters of objects from photographic plates.
The catalogues derived from plate scanning could be passed to
computerised overdensity detection algorithms, and for the first time
cluster detection advanced beyond subjective visual inspection.

Prior to the construction of large X-ray selected samples of clusters,
it was natural to target the optically selected clusters described
above with X-ray telescopes in an attempt to measure the X-ray
luminosity function (XLF) at high redshift.  \citet{castander94} used
ROSAT to observe cluster candidates in the redshift range 0.7-0.9 from
a 3.5 square degree subsample of \citet{gho86}'s optical cluster
catalogue and also found surprisingly weak X-ray emission
($\approx$10$^{43}\,$erg s$^{-1}$). \citet{bow94} undertook ROSAT X-ray
observations of optically selected clusters from the \citet{cemm}
catalogue (\citealt{cemm} visually selected clusters based on the
density enhancement of galaxies above the mean background, but tested
their method exhaustively against simulated data).  From this 46
deg$^2$ catalogue, \citet{bow94} took clusters with reliable
spectroscopic follow up and X-ray data in the redshift range 0.15 to
0.66, assuming this to be a random subsample of the full catalogue.
The total sky coverage of this survey was 26.8 deg$^2$ and
contained 14 clusters. The X-ray luminosities of all but two of the
clusters was found to be surprisingly weak - less than
5$\times$10$^{43}\,$erg s$^{-1}$. This decrease with respect to the
locally measured value was attributed to evolution in the XLF between
z=0 and $\approx$0.4. The alternative is that if the XLF does not
evolve between these redshifts, then the missing X-ray luminous
clusters must be made up of optically poorer systems, missing from this
sample. This raises the question {\it do optical and X-ray surveys
  sample the same clusters?}

With the advent of high quantum-efficiency, large format
charged-coupled devices (CCDs) in the early 1990s, optical cluster
studies are again becoming attractive.  The first serious attempt at an
automated optical CCD survey with a quantifiable selection function was
carried out with the Palomar Distant Cluster Survey \citep[PDCS,
][]{pdcs}. Their pioneering work involved assuming a model for the
spatial and luminosity distribution of galaxies in a cluster and in the
field, and filtering the data using these models as templates.  Using a
likelihood analysis of the data, with cluster richness and redshift as
free parameters, the most likely cluster candidates could be extracted,
and their redshifts estimated as a by-product of the process.  The
technique is known as the matched-filter (MF) and is discussed in more
detail in \S\ref{sec:optic-clust-detect}. This method reduced spurious
clusters due to projection effects compared with the more traditional
techniques described above, but many still remained (discussed further
below).  The MF need only be used on photometric data from a single
passband, but with an additional filter other techniques are possible.
Algorithms using colour selection have been proposed \citep[e.g.
][]{gal,gy00}. \citet{gal} used mild colour cuts to reduce
contamination due to field-like galaxies. Since elliptical galaxies are
predominantly found in dense environments \citep[the morphology-density
relation, ][]{dressler80}, and exhibit only a narrow range of colours
at a given redshift, in any environment, data can be filtered in colour
to remove galaxies with colours incompatible with ellipticals.
\citet{gy00} took the colour selection a stage further, placing
very strict colour cuts in two-colour data, to only search for
overdensities of galaxies with colours consistent with elliptical
galaxies at a given redshift (see \S\ref{sec:optic-clust-detect}).
This works because in all known clusters for which multi-band
photometry exists (regardless of how the cluster was selected), a tight
relation exists between the colour and magnitude of its early-type
galaxies \citep[e.g. ][]{visv,bow92}.  This relation is clearly visible
over small spatial scales ($\sim$ the size of the cluster core), as
early-type galaxies are predominantly found in the central regions of a
cluster \citep{dressler80}.

With an abundance of new wide-field optical and X-ray
\citep{jones2,vmf98,mason00,romer00} imaging data, it is timely to
directly compare clusters found using these different methods. To
recap, X-ray selected clusters are required to have a hot, dense
intracluster plasma; whereas optical selection just requires an
overdensity of galaxies. To this end, we have undertaken optical and
X-ray imaging surveys in exactly the same regions of sky. We refer
to this as the X-ray Dark Cluster Survey (XDCS), as the project is
specifically aimed at searching for the optically rich but X-ray
underluminous clusters described in \citealt{bow94}, \citealt{bow97}.

A plan for the outline of this paper is as follows. The X-ray field
selection, optical observations and data reduction are presented in
\S\ref{sec:x-ray-dark}.  Two optical cluster detection algorithms are
presented in \S\ref{sec:optic-clust-detect}, the first uses only single
band optical photometry, and is a variant of the now widely-used
Matched Filter algorithm \citep{pdcs}; the second utilises colour
information, and is our implementation of the algorithm of
\citealt{gy00}. \S\ref{sec:richness} discusses measures of optical
richness, and the construction of the optical catalogues.
\S\ref{sec:x-ray-selection} deals with the X-ray selection of clusters.
These samples are cross-compared with the X-ray sample in
\S\ref{sec:comparison-optical-x}.  Spectroscopic follow up for a
subsample of X-ray underluminous clusters is presented in
\S\ref{sec:spectr-observ-x}, and finally our conclusions in \S8.

\section{The X-ray Dark Cluster Survey}
\label{sec:x-ray-dark}

\subsection{Sample Selection and Observations}

X-ray imaging is observationally expensive. Thus we chose to base our
survey on archival X-ray data, which is relatively inexpensive to
follow-up with wide-field optical imaging.  Essentially a random sample
of deep extragalactic X-ray fields was required.  The ROSAT
International X-ray/ Optical Survey \citep[RIXOS, ][]{mason00} provided
an ideal list of such fields.  Their sample was constructed from ROSAT
Position Sensitive Proportional Counter (PSPC) fields which had
exposure times of at least 8ks.  This ensures that sources at the
intended survey flux limit (for a point source) of $3\times10^{-14}$
erg cm$^{-2}$ s$^{-1}$ (0.5-2.0 keV) lie significantly above the
sensitivity threshold of every field.  They also limited the choice of
fields to those which have Galactic latitudes greater than 28$^{\circ}$
in either hemisphere, since RIXOS is primarily intended for
extragalactic source studies.

\begin{figure}
\centering
\includegraphics[width=50mm,clip=t]{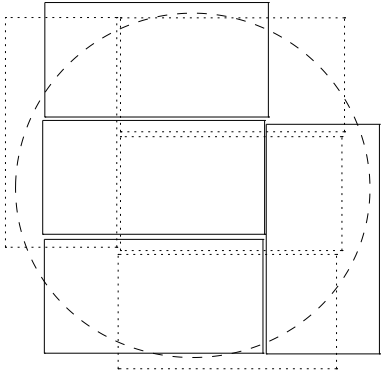} 
\caption{INT WFC Tiling Strategy.  The circle
  represents the inner 19 arcmin radius of the PSPC field.  Rectangles
  show the four CCDs of the WFC.  Solid lines indicate the camera
  configuration in one orientation and broken lines show the camera
  rotated through 180$^\circ$.  The diagram illustrates how the field
  can be efficiently imaged in two pointings.  }
\label{fig:tiling}
\end{figure}

The archival ROSAT fields listed in Table~\ref{table:rosatfields} were
observed in the optical using the Wide Field Camera (WFC) on the Isaac
Newton Telescope (INT), La Palma.  The observations were carried out in
two runs, in June 1998 and January 1999.  The median seeing for the two
runs was around 1.0\arcsec\ and 1.6\arcsec\ respectively. Conditions
for both runs were photometric. The inner $\gsim$19 arc minutes of the
PSPC fields were imaged to depths of V$\sim$24 and I$\sim$23 (50\%
completeness).  This is the region of the PSPC used for X-ray source
identification in RIXOS, to ensure the best X-ray image quality and the
most accurate source positions \citep{mason00}.  For each band, two
exposures were taken, rotating the camera through 180 degrees, and
offsetting the centre of the pointing in order to ensure optimum
coverage of the PSPC, as shown in Fig.~\ref{fig:tiling}.  Hereafter,
images taken with the camera rotator angle set to 0 degrees will be
referred to as {\it A} images; and those with a rotator angle of 180
degrees {\it B} images.  The two independent observations of a large
fraction of the survey area will provide important internal checks of
galaxy photometry and of cluster catalogues generated from independent
data covering the same area of sky. We shall use the term ``mosaic'' to
refer to a single pointing (either A or B) comprising the four WFC
chips.

The WFC comprises four thinned EEV (2048 $\times$ 4100) Charge-Coupled
Device (CCD) chips, at the prime focus of the 2.5 metre INT. The
science devices have 13.5 micron pixels (0.333"/pixel). Each covers an
area of 22.8 arcmin $\times$ 11.4 arcmin on sky. The total sky coverage per
exposure is 0.29 square degrees.  A single exposure covers 76\% of the
ROSAT PSPC area.  By using two exposures, virtually the entire inner 38
arc minute diameter was covered: see Fig.~\ref{fig:tiling}.

\begin{table*}
\caption{List of ROSAT Fields in XDCS
Columns give: RIXOS ID of field; name of the target of the original
ROSAT pointing; RA, Dec; exposure time of field in RIXOS survey;
overlapping X-ray cluster survey (V - VMF, S - SHARC); exposure time
of the VMF or SHARC pointing (to give an indication of the depth to
which they could have searched for X-ray emission).}
\label{table:rosatfields}

\begin{tabular}{ccccccc}
\hline
RIXOS & Target & $\alpha$ (J2000) & $\delta$ (J2000) & T$_{exp}$ & Overlapping &
T$_{exp}$(overlap)\\
ID & & [hh:mm:ss.s] & [dd:mm:ss] & (ks) & Survey & (ks) \\
\hline
R110 & LHS 2924 & 14:28:43.17 & +33:10:45.47 & 18.3 & V  &  28.5 \\
R116 & NOWER2 & 12:03:60.00 & +56:10:11.99 & 30.1 & ---  & ---  \\
R122 & Meaty & 16:29:24.00 & +78:04:48.01 & 38.5 & S  & 34.4  \\
R123 & 1116+215 & 11:19:4.80 & +21:18:36.00 & 25.0 & V  & 32.2  \\
R126 & ON 231 & 12:21:33.60 & +28:13:48.00 & 10.4 &  V & 12.5  \\
R133 & CY UMA & 10:56:55.20 & +49:42:0.00 & 9.4 & V  & 7.9  \\
R205 & P100578 & 23:12:21.60 & +10:46:48.00 & 10.3 & S  & 9.8  \\
R211 & S5 0716+7 & 07:21:52.70 & +71:20:23.99 & 21.0 & S  & 17.3  \\
R213 & IRAS 0759 & 08:04:31.20 & +65:00:0.00 & 8.4 & V  & 6.4  \\
R216 & S4 0917+6 & 09:21:36.00 & +62:15:35.99 & 19.5 & V  & 15.9  \\
R217 & 1411+442 & 14:13:48.00 & +44:00:0.00 & 25.3 & V  & 22.5  \\
R220 & RX J1726. & 17:26:12.00 & +74:31:11.99 & 10.6 & V  & 8.1  \\
R221 & E0845+378 & 08:48:19.20 & +37:40:11.99 & 12.4 & V  & 10.0  \\
R223 & CM DRA & 16:34:24.00 & +57:09:0.01 & 47.5 & V  & 37.3  \\
R224 & HZ43 & 13:16:24.00 & +29:06:0.00 & 34.9 & S  & 18.3  \\
R227 & GD140 & 11:36:33.51 & +29:47:60.00 & 33.9 & V  & 26.6  \\
R228 & GBS0839+3 & 08:38:47.90 & +36:31:12.00 & 11.0 & V  & 9.2  \\
R231 & Survey Fi & 10:10:16.70 & +54:45:0.00 & 16.8 & V  & 14.4  \\
R236 & Q1700+515 & 17:01:23.90 & +51:49:12.00 & 8.2 & V  & 6.5  \\
R245 & H0323+022 & 03:28:25.82 & +02:47:57.84 & 25.7 & S  & 24.5  \\
R248 & 3C216 & 09:09:33.50 & +42:54:0.01 & 23.6 & V  & 19.9  \\
R254 & MRK 273 & 13:44:43.10 & +55:53:24.00 & 17.1 & V  & 28.1  \\
R255 & 0755+37 & 07:58:28.70 & +37:47:24.00 & 16.0 & S  & 15.5  \\
R257 & B2 0902+3 & 09:05:31.10 & +34:07:48.00 & 14.5 & V  & 26.5  \\
R258 & 1115+080 & 11:18:16.70 & +07:46:12.00 & 14.4 & V  & 13.2  \\
R262 & 520 & 01:24:33.50 & +03:47:60.00 & 13.9 & V  & 12.0  \\
R265 & B2 1308+3 & 13:10:28.70 & +32:20:59.99 & 13.0 & V & 7.6  \\
R268 & MRK 463 & 13:56:2.30 & +18:22:12.00 & 11.6 & V  & 18.3  \\
R272 & 3C 371 & 18:06:50.40 & +69:49:12.00 & 10.5 & S  & 8.0  \\
R273 & 1040+123 & 10:42:45.51 & +12:03:36.00 & 10.2 & V  & 8.4  \\
R274 & 1404+226 & 14:06:21.60 & +22:23:60.00 & 10.1 &  V & 6.7  \\
R278 & MKN 789 & 13:32:24.00 & +11:06:36.00 & 9.6 & V  & 9.1  \\
R281 & III ZW2 & 00:10:28.70 & +10:58:12.00 & 9.1 & V  & 16.8  \\
R283 & 1H 0414+0 & 04:16:52.70 & +01:05:24.00 & 9.0 & ---  & ---  \\
R285 & PSR 0940+ & 09:43:43.20 & +16:31:12.00 & 9.0 & V  & 8.1  \\
R287 & MKN 40 & 11:25:36.00 & +54:22:48.00 & 8.8 & V  & 7.7  \\
R292 & GLIESE 70 & 01:43:21.50 & +04:19:48.00 & 8.7 & V  & 5.4  \\
R293 & GD 90 & 08:19:47.90 & +37:31:12.00 & 9.0 & V  & 7.3  \\
R294 & KUV 2316+123 & 23:18:45.0 & +12:36:00.00 & 9.5 & ---  & ---  \\
\hline
\end{tabular}
\end{table*}

\subsection{Optical Data Reduction}
\label{sec:optic-data-reduct}
The data reduction was carried out using mostly standard {\sc
  iraf}\footnote{{\sc IRAF} is distributed by the National Optical Astronomy
  Observatory which is operated by AURA Inc. under contract with the
  NSF.} routines, and is detailed below.

\subsubsection{Debiasing and Linearity Correction}

Bias frames were visually inspected for quality.
Master bias frames were constructed for each night by taking the mean
of all the good data using {\sc zerocombine}.  Next it was necessary to
correct for the non-linear response (due to problems with the ADC) of
the WFC. This was achieved by applying polynomial corrections taken
from the Cambridge Wide Field Survey (WFS) webpage ({\tt
  http://www.ast.cam.ac.uk/$^\sim$wfcsur/}).  The temporally nearest
fits were used. The August 1998 polynomials were used for the XDCS data
taken in June 1998 and the October 1998 values for January 1999 data
(the non-linearity was found to remain stable between October 1998 and
August 1999). Following this correction, the camera residual
non-linearity is estimated by the WFS team to be less than 0.5\% of the
sky level.

\subsubsection{Flatfielding and Defringing} 
Flatfielding was carried out using master flats for each night. These
were constructed from a median combination of all the science data in
the V-band and from twilight sky flats in the I-band. The I-band
science data were not used to construct master flatfields as the
thinned CCDs suffer from fringing. 
The fringing patterns are additive, but their broad structure is largely
stable with time.  Therefore, all the I-band data for
each chip were averaged together to make master fringe frames for each
night.  Although the shape of the lines is stable, the amplitude can
vary considerably (although always at the level of a few percent of
sky).  Thus, a method was needed to scale the amplitude of the fringe
mask to the amplitude of the fringes in each data frame, before
subtraction. This was done with software kindly provided by Mike
Irwin.  
The scaled fringe masks were then subtracted from the I-band science
images. After this procedure the level of fringing is reduced to
$\sim$0.5 percent of sky, which is of the order of the residual
non-linearity.  

\subsubsection{Object Detection and Photometry}

The \sex\ package \citep{sex} v2.2.1 was used to locate and classify
objects in the optical data.  Firstly the seeing was measured for each
frame. This was accomplished by fitting Gaussian PSFs to all bright
($>10\sigma$) detections, classified by \sex as stars.

\sex\ was run in double image mode on each pair of aligned V/I images
using the V-band image as the detection image, and the I-band as the
measurement image. This was done in order to obtain \sex\ MAG\_BEST
magnitudes to measure the total magnitude for I-band objects.  Ideally
one would like to use the redder passband for object detection as the
number counts are shallower for redder bands (i.e. a lower background
for cluster detection), but it was found that the residual fringing in
the I-band posed a problem for object detection in a few frames,
dramatically increasing the numbers of false detections in the most
badly-fringed frames. Although the level of the remaining fringes is
too small to affect the photometry (the photometric error is still
dominated by the linearity correction at the bright end, and Poisson
noise at the faint end), the background estimation method used by \sex\ 
cannot model the fringes.  High residuals pass above the \sex\ 
detection threshold and are classified as extended objects.  As this
method could not be reliably used on all the data, its use was decided
against. The depth of the data is such that, by limiting the object
catalogues to $I = 22.5$, few objects are missed which would have been
detected in the I-band image (see magnitude limits in
Fig.~\ref{fig:slices}) and the use of the V-band to perform detection
is entirely one of operational ease.  The survey is therefore I-band
limited.

After generating a list of object positions from \sex\, the x and y
coordinates of objects in the V frame were logged and used to position
the aperture for photometry. For each V/I frame pair, the frame with
the better seeing was convolved to that of the worst, using Gaussian
convolution. The {\tt IRAF} task {\tt phot} was used to perform
aperture photometry to measure colours, using an aperture of diameter
2.6 $\times$ the seeing \citep[e.g.][]{lcg91}, on the aligned,
convolved V and I frames.

Objects with a \sex\ CLASS\_STAR index of $\leq$0.90 were taken to be
galaxies.  
Detections with {\tt
  FLAGS}$\geq$4 were rejected.  This means blended objects and those with
near neighbours are kept, but those with saturated pixels, or corrupted
data (e.g. due to boundary effects) are rejected.

\subsubsection{Photometric Calibration}

The photometric data were converted to the standard Cousins system
using observations of several \citet{landolt} standard star fields each
night. Due to the small size of these fields, only the central chip
sampled the standard stars well.  Therefore an internal calibration of
the other three chips to the reference central chip was performed by
comparing the relative sky levels in the chips for each observation.
The flux difference was then converted into a relative magnitude
offset. Different offsets were measured for each run, due to servicing
of the instrument between the two runs, which changed the gains of the
devices. The uncertainty on the absolute calibration derived from the
scatter between Landolt stars, and repeat observations of the same
fields was estimated to be better than 0.1 mags.

\subsubsection{Astrometric Calibration}

Relative astrometry between the chips was performed by Mark Taylor in
Cambridge using A and B observations of one of our fields. This gives
an internal astrometric solution converting chip coordinates into
global camera coordinates. This solution is available on the WFS
webpage. A precise external astrometric calibration is not essential,
as clusters are very extended objects, so specifying a cluster centre
to within several arcseconds is sufficient.  Therefore, the pointing
position of the telescope was used as the centre of the instrument and
the internal solution described was used to transform chip coordinates
to sky coordinates.  This gives an external accuracy of around 3
arcseconds or better, but the internal accuracy is better than 0.5
arcseconds. For the purpose of cluster detection it is the relative
positions between galaxies (i.e. the internal solution) that is
important.

\begin{figure}
\includegraphics[width=80mm]{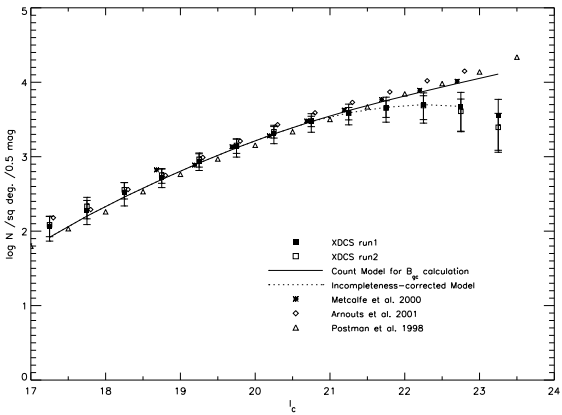}
\caption[I-band galaxy counts]{I-band galaxy counts for both XDCS runs and
  model counts used in derivation of $B_{gc}$.  Error bars are standard
  deviation from field to field. Both runs are found to be in good
  agreement, suggesting that the photometry for XDCS is homogeneous.
  Overplotted is the completeness model \citep[relative
  to][]{metcalfe00} used to allow data to $I=22.5$ to be used.
  Magnitudes are \sex\ MAG\_BEST magnitudes.  I-band counts from the
  literature are overplotted, and found to be in good agreement.  The
  completeness is modelled by a single-sided Gaussian of width 1.90
  mags, centred on $I=20.86$.  }
\label{fig:counts}
\end{figure}

\section{Optical Cluster Detection Algorithms}
\label{sec:optic-clust-detect}

Two different cluster detection methods were applied to the optical
data.  Both use positional information to search for overdensities in
the galaxy catalogues, but they do this in different ways. The first
uses only the I-band photometry, and looks for overdensities of
galaxies which appear to follow the luminosity function of a galaxy
cluster; the second includes the V-band data and uses the V-I colour to
search for the colour-magnitude relation of early-type cluster
galaxies.

\subsection{The Matched Filter Algorithm}
The ``matched-filter'' (MF) was pioneered by \citealt{pdcs}, and
modified by several groups including \citet{kawasaki,kep,lobo}.
Its principle is to assume that galaxy clusters follow some
well-defined model, in both their spatial and luminosity distributions.
i.e.  some universal radial profile is assumed for the distribution of
galaxies in a cluster, which can be projected into 2D.  In the same
way, some universal luminosity function can be assumed for its member
galaxies.  High resolution N-body simulations suggest that 
virialised objects follow a universal density profile \citep[NFW][]{nfw97};
observations show that cluster profiles are compatible with such a
profile, but also with simpler analytic fits \citep{carlberg97,lubin}.
The luminosity function of galaxies in clusters and the field is
well-fitted by a Schechter function with mild luminosity evolution
\citep{ylc99}, or mild luminosity and density evolution \citep{lin99}.
This model can be scaled to any redshift.  The observed radial profile
scales with redshift according to the angular diameter distance, $D$,
to the object, given by
\begin{equation}
\label{equ:zang}
D=(c/H_0)\{q_0z + (q_0-1)[(2q_0z+1)^{1/2}-1]/q_0^2(1+z)^2
\end{equation}
where $c$ is the velocity of light, H$_0$ is the Hubble constant, q$_0$
is the deceleration parameter, and $z$ is the
redshift\footnote{Throughout we assume H=64km~s$^{-1}$Mpc$^{-1}$,
  $q_0=0.1$ unless otherwise stated.  This is to provide consistency
  with the stellar population synthesis models made available to us.
  Adopting the currently favoured cosmological parameters of:
  $\Omega_m=0.3$, $\Omega_\Lambda=0.7$ and $H_0=70$ \kms would affect
  the derived luminosities by $\lsim$4\% and inferred linear sizes by
  $\lsim$2\% over the redshifts considered.}.  The observed luminosity
distribution scales according to the distance modulus formula $m-M =
5\log(d_{Mpc}) + 25 + k + e$ with corrections for bandpass shifting due
to the redshift of the source (k-correction, $k$) and evolutionary
corrections of the stellar populations ($e$).  $m$ and $M$ are the
apparent and absolute magnitudes of the galaxy, and the luminosity
distance, $d$, in Mpc is a factor of (1+$z$)$^2$ times the angular
diameter distance, $D$. The only other free parameter the model needs
is the {\it richness} of the cluster (i.e. some parameterisation of the
number of galaxies it contains).  Thus, a model for the observed
properties of a galaxy cluster of arbitrary richness and redshift is
obtained.  The final aspect to be taken into account is the
distribution of field galaxies.  This model can be derived from the
data itself.  It is assumed they are randomly distributed in position -
explicitly ignoring the correlation between the positions of pairs of
galaxies (see \S\ref{sec:sims}).  The contribution of cluster galaxies
to the total number of galaxies in any dataset will be small, unless
the survey consists of small fields targeted at clusters.  Hence, by
studying the number density and luminosity distribution of the whole
sample, a model for the field galaxies can be deduced.  The luminosity
function in the PDCS method had to be modelled by a power-law due to
assumptions made in the derivation (see paper for details) which is
generally a good fit to the data, depending on the magnitude range
observed (e.g. Metcalfe et al.  2001, Smail et al. 1995, see also
Fig.~\ref{fig:counts}).

Astronomical data comprising positions and photometry can be
searched for regions where the likelihood of the data fitting this
cluster+field model is high.  Since the cluster model is a function of
richness and redshift as a by-product of the detection process, a
most likely richness and redshift for each cluster candidate is
obtained as a by-product.

The \citealt{pdcs} algorithm made several approximations (detailed
in their paper) which have been removed and treated more fully by later
workers.  For example, their main approximation was to assume that the
data (galaxies) could be binned in both position and magnitude in such
a way that each bin had sufficient datapoints that their distribution
was Gaussian.  This was replaced by a more general treatment which
assumed Poisson distribution of the data by \citet{kawasaki}. The
other key assumption which has been followed by all subsequent works
until \citet{lobo} is that the models predict a unique combination
of spatial and luminosity distributions at a given redshift. The
drawback of this approach is that if a cluster is slightly larger or
smaller in angular extent than predicted for its luminosity
distribution, its signal is reduced and the probability of detection
lessened. \citet{lobo} circumvented this problem by choosing the
combination of spatial and luminosity profiles which independently
maximised the signal.

\subsubsection{A New Matched-Filter Algorithm}
The algorithm presented here is closest in spirit to the technique of
\citet{lobo}, in that the assumptions for the distributions in
radial profile and luminosity have been decoupled.  This method offers
many advantages for this project, the main one being that it is
unnecessary to assume a characteristic physical size for the model
cluster, {\it a priori}, which will obviously help if unrelaxed
systems have larger angular extent than virialised systems of the same
richness and redshift (as may intuitively be expected).  This also
offers some computational rewards which will be explained below.

The assumptions for this model are:

$\bullet$ Field galaxies are distributed randomly over the sky (the two
point galaxy-galaxy correlation function is explicitly ignored), and
that their magnitude distribution has the same shape throughout the
survey but changes slightly in normalisation, from field to field (see
Fig.~\ref{fig:counts}).

All magnitudes in this section refer to the I-band.  In principle any
photometric passband can be used, but as red a band as possible is
desired.  This is due to the fact that the field galaxy counts steepen
toward shorter wavelengths, so the contrast between the cluster and the
field is greater at longer wavelengths
\citep[e.g. power-law slopes of 0.28 and 0.40 were
measured in the I- and V-bands, respectively, by ][]{smail95}.

$\bullet$ Galaxy clusters appear as overdensities in this background
distribution, and their visibility can be enhanced by filtering the galaxy
catalogue with a Gaussian filter, the size of which is given by the typical
sizes of galaxy clusters from the literature.

$\bullet$ Galaxy clusters follow a Schechter \citep{schechter}
Luminosity Function (LF), with fixed faint-end slope, and the
normalisation is given by the amplitude of the overdensity (i.e. the
cluster's richness).  The typical apparent magnitude m$^\star$ of the LF is a
function of redshift.

The maximum-likelihood estimator, $C$, of \citet{cash79} is then
applied to the data.

\subsubsection{Implementation}
\label{sec:imp}
The algorithm was run on each mosaic separately.  First, filtering of the
spatial information was performed.  The x and y positions of galaxies
brighter than magnitude 22.5 were read in, and the mosaic filtered
with five Gaussian filters of different widths.  The standard deviation
of each
filter was taken from \citet{lobo}.  The widths shall be referred to
as $W_n$ for the $n^{th}$ filter (to avoid confusion with standard
deviation $\sigma$s, later), but are equivalent to the $\sigma_{ang}$
in their paper.  The widths, $W_1$,...,$W_5$, ranged from $\sim$0.35 to
$\sim$1.42 arcmins in steps of $\sqrt{2}$, these represent the typical
core-widths of clusters in the redshift range $\approx$ 0.2 to 1.0.  A
cut-off radius of 3 $W_n$ is used.  Unlike \citet{lobo}, instead
of using a regular grid, the positions of the galaxies themselves were
used as the grid to centre each of the Gaussian filters. This
adaptive-grid method was also adopted by \citet{kep} and has the
advantage that it ensures adequate resolution in the core of a cluster,
and saves computational expense by performing few calculations where
the galaxy density is low.  For each spatial filter the mean and
standard deviation of the filter amplitude was calculated (the amplitude
follows a Gaussian distribution), and all five filters normalised
onto the same system (by subtracting the mean and dividing by the
standard deviation).

Peaks were then found in each filter, by sorting the list of signal amplitudes,
retaining the highest value, and then searching down the list, removing
detections which fell within a radius $W_n$ of the peak, and
retaining the next highest value which did not.  Values lower than a minimum
threshold of 2.5 $\sigma$ were immediately rejected.

The peaks from the five filters were then sorted and cross-correlated.
If a peak was detected in more than one filter, the highest amplitude was
retained and the duplicate detections removed.  Two peaks were
considered to be the same object if the distance between them was less
than the mean of their scales (i.e. $(W_n+W_n^\prime)/2$) \citep{lobo}.
This resulted in a single list of peaks, each with an associated scale
(the filter width, $W_n$, in which the highest signal was detected).

A richness estimate of the candidate was then required, for use in the
maximum-likelihood estimation.  As a first pass estimate, the number of
galaxies within a fixed angular radius was taken, for all candidates,
regardless of its associated $W_n$.  (The decision to use a fixed
angular search cell is explained below.) This number then had the
number of background galaxies, scaled to the same area, subtracted from
it.  The background galaxy density was found by counting the number of
galaxies in an annulus of radius $3 \times W_n$ to $15 \times W_n$.
The $W_n$ value was used to ensure the annulus is sufficiently far from
the cluster core to avoid contamination with cluster members.

The importance of using a local estimate of the background can be seen
by looking at the field to field variations in our data in the number
counts in Fig.~\ref{fig:counts}.  These variations are due to a
combination of residual offsets in the photometric calibration and
intrinsic cosmic variance.  The local background number density
was also used to re-normalise the expected number counts locally, for
use in the maximum-likelihood calculation.  The cumulative counts were
used at I=20.5, two magnitudes brighter than the limiting magnitude, to
ensure both a high number of objects and high completeness.

In estimating galaxy number densities, the geometry of the mosaic field
needed to be taken into account.  To compensate for border effects,
where the detect cell starts to fall off the edge of the field, a
weighting function was constructed, taking account of the fraction of
the detect cell area lost.  This requires caution, as upweighting the
signal from a few galaxies is likely to result in increased spurious
detections, due to the uncertainty from using fewer galaxies.  Thus a
cut was made, rejecting candidates where the fraction of the area lost
to borders is $>$ 0.20.

The Cash $C$ statistic (below) was then applied to the data within a
radius of $2.5 \times W_1$.  Most other MF algorithms use a search
radius fixed in physical units at the estimated redshift of the
cluster, and \citet{lobo} use the radius which maximises the
signal.  Early experimentation with simulated clusters showed that just
using data within a fixed radius (of the smallest filter) was adequate
and this is done for the sake of simplicity and computational speed.
Since the same galaxies always enter the maximum-likelihood
calculation, this makes the calculation much more stable. It also meant
that a fair estimate of the likelihood could be found by a simple
application of the Cash statistic, without recourse to bootstrap
resampling the detections to determine their significance, as needed by
the \citet{lobo} method.

The results were sorted in order of increasing $C$ (decreasing
likelihood), and overlapping detections removed using a 2D
friends-of-friends groupfinding algorithm \citep{fof}.  The groupfinder
started with the most significant point and searches within a fixed
radius (the linking length) for another point.  If a point was
found, then the search was repeated within the same radius around this
new point.  The search continued, linking all points within the linking
length of a neighbour, until no more points could be linked.  Thus only
the most significant candidate was retained and all linked neighbours
were removed. This method was found to work better than just removing
candidates within a fixed radius of each other, as this latter approach
tended to either remove too many (unassociated) candidates (if the
rejection radius was too large) or leave multiple detections of the
same candidate around the periphery of a rich candidate.  The
friends-of-friends algorithm is a more natural method for associating
related points.  Through experimentation on fields with known clusters,
a linking length of 500 pixels ($\sim$2.8 arcmin) appeared optimal.
Finally, the distance between the highest likelihood point (the
candidate centre) and the most distant point from it joined to the
group was recorded.  This distance was then used as a characteristic
radius to estimate the extent of the group.  This will be important
later for matching up overlapping candidates.

\subsubsection{Maximum-Likelihood Estimation}
\label{sec:cash}
\citet{cash79} originally developed the maximum-likelihood method for
application to general Poissonian problems (see original paper for full
details).  The Cash statistic, $C$, is given by equation~\ref{equ:cash5}.

\begin{equation}
\label{equ:cash5}
C = 2\left(E - \sum_{i=1}^n \ln I_i\right)
\end{equation}

For the application here, $E$ is the expected number of galaxies per
unit area per unit luminosity.  The number of galaxies can be broken up
into cluster and field.  Now, observing over a given area and
luminosity range on the sky, $E=E(\theta,m;z_c)$ where $\theta$ is the
angular radius of the search area, $m$ is the galaxy magnitude, and
$z_c$ is the redshift of the cluster. Thus the number expected within
these spatial and luminosity ranges is:

\begin{equation}
\label{equ:cash3}
E = \int^{m_{lim}}_0\left[ \Lambda \phi_{cl}(m)+b(m)\right] d\Omega dm 
\end{equation}

and 

\begin{equation}
\label{equ:cashi}
I_i =  [\Lambda \phi_{cl}(m_n^{i})+b(m_n^{i})]d\Omega dm 
\end{equation}

where each $i$ is a data point.

$d\Omega$ and $dm$ are elements of solid angle and magnitude,
respectively.  In most MF algorithms this is modelled as a power-law,
as is necessary in \citealt{pdcs} original implementation.  However,
there is no reason to assume this model in our maximum-likelihood
approach, and so the background number counts were taken from the data,
with a model for incompleteness, relative to literature counts, and a
local normalisation (explained above).  $b(m)$ is the model for the
field galaxy counts, as a function of magnitude; $\phi_{cl}(m)$ is a
model for the cluster contribution.  The lower limit of the integral
was replaced with I=16 in practice, as saturation sets in around this
point, and there are so few galaxies this bright within the whole
survey that number counts could not reliably be computed. $\Lambda$ is
a parameterisation for the richness of the cluster, such that the
number of cluster galaxies $n_c = \Lambda \phi_{cl}$, with $\Lambda$
normalised by

\begin{equation}
\label{equ:kappanorm}
\int_0^{m_{lim}} \Lambda \phi_{cl}(m) dm = 1
\end{equation}

The luminosity distribution, $\phi_{cl}$, can be modelled by the Schechter
function \citep{schechter}:

\begin{equation}
\label{equ:schLF}
\phi_{cl}(m)=0.92\phi^\star\exp\{-0.92(\alpha+1)(m-m^\star) -
\exp[-0.92(m-m^\star)]\}
\end{equation}

In practice,
one can determine $\Lambda$ by measuring the excess number of galaxies
in the search cell (see below), and so the most likely value of
$m^\star$ can be sought.  Matched-filters are usually used to provide
redshift estimates of clusters, and each trial $m^\star$ value can be
thought of as a matched-filter redshift estimate $z_{MF}$, using some
relation between $M^\star$ and $m^\star$
\citep[e.g.][]{colless89}. The predicted magnitude of a
passively-evolving $M^\star$ elliptical, from the models of
\citet{ka97} was used here, for consistency with the CMR method,
explained in \S\ref{sec:cmr}.  Most implementations of the MF assume
some radial profile for the model cluster fixed in metric coordinates
\citep[e.g.][]{pdcs,kep}.  This then means that
$d\Omega=d\Omega(z_{MF})$, where $z_{MF}$ is the Matched-Filter
redshift.  \citet{lobo} adopted the novel approach of decoupling
the assumed radial profile from $z_{MF}$, and just used Gaussian
profiles of several different widths.  Here, this is taken a stage
further and a fixed angular size of detect cell $d\Omega$ was used.  One
simply changes the parameterisation of radial shape and cluster
richness to be contained within the $\Lambda$ coefficient, which
becomes the number of cluster galaxies within a fixed angular area.
We estimated this from the number of excess galaxies over the local
background value.  Fixing $d\Omega$ made the $C$ statistic
operationally easier and more stable.

\begin{figure}
\centering
\includegraphics[width=70mm]{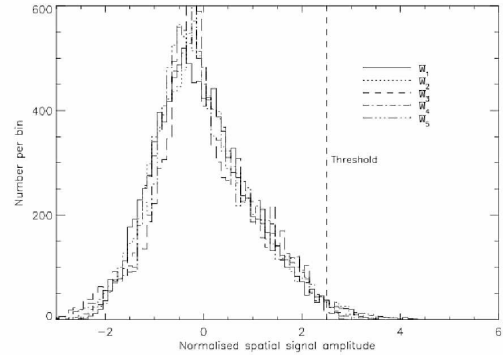}
\caption[Distribution of amplitudes from spatial filter]{Distribution of
normalised amplitudes from spatial filter.  The distribution follows a Gaussian,
with a high-end tail which contains cluster candidates.}
\label{fig:sighists}
\end{figure}

\subsubsection{Simulations}
\label{sec:sims}

In order to test the accuracy, completeness, and spurious detection
rate of the cluster-finder, an extensive set of simulations was run.
First, a cluster model was required.  The fiducial cluster required a
model for the luminosity and spatial distribution of galaxies, since
these are used by the detection algorithm. The spatial profile is given
by the density profile of \citet{nfw97}, projected into 2D using the
prescription of \citet{bartelmann}.

The luminosity profile was taken from the same Schechter function used by
the detection algorithm.  This employed the $M^\star_I$=-20.68 given by
the stellar population synthesis models used (Kodama et al. 1998); we
adopted alpha=-1.1 (as used by Postman et al. 1996.  These authors noted
that varying alpha between -0.6 and 1.6 only alters the FWHM of the
luminosity filter by $\lsim$15\% relative to the nominal alpha=-1.10,
and has a minimal effect on the detectability of cluster candidates).
The normalisation of the LF is varied, and the number of galaxies
brighter than $m_3 +2$ within an Abell radius counted to give the Abell
Richness Class \citep[ARC,][]{abell}.  Clusters from ARC 0 to 3 were simulated.
The cluster models were generated in physical coordinates and then
transformed to different redshifts. For the distribution of field
galaxies, points were put down randomly over the field of the WFC
mosaic.  Each point then had a magnitude assigned to it, extracted from
the observed number counts.  The number of galaxies in each realisation
was allowed to vary according to the range of surface densities seen in
the data.

\begin{figure}
\centering
\includegraphics[width=80mm]{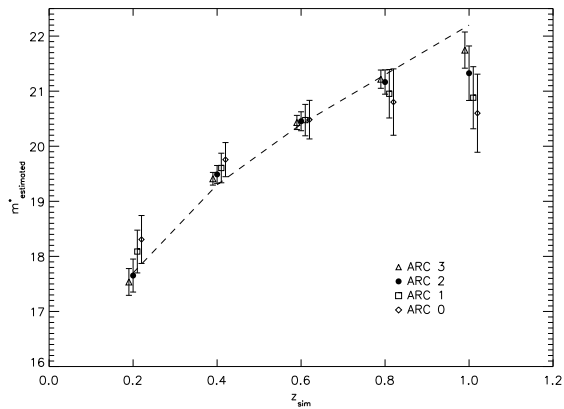}
\caption[Matched-Filter Accuracy]{Matched-Filter Accuracy.  Recovered estimate of
$m^{\star}$ against simulated redshift.  The dashed line shows the model used for
the $m^{\star}$-z relation.  Each point is the mean of 100 simulations.  The points
for the different richness classes are offset slightly in $z_{sim}$, in the plot,
for clarity.  Error bars are 1 $\sigma$ standard deviation between all simulations.}
\label{fig:mf_accuracy}
\end{figure}

For each richness class of cluster, 100 realisations of cluster and
field were produced at each redshift interval from z=0.2 - 1.0 in 0.2
steps.  These mock fields were then passed to the detection algorithm.
The results are plotted in Fig.~\ref{fig:mf_accuracy}.

The completeness was assessed from the same simulations.  For every simulated
cluster, the cluster was considered correctly recovered if a candidate centre
lay within 2.5$W_1$ of the simulated cluster centre, and its Cash $C$ value lay below
the threshold cutoff (explained below).

To assess the number of spurious candidates detected, the algorithm was
run on simulated blank fields.  Other authors
\citep[e.g.][]{kep,pdcs,lobo} have used random realisations of their
data to represent blank (i.e. cluster-free) fields.  However, this is
likely to underestimate the false-positive rate, since the positions of
galaxies on the sky are correlated.  In order to account for the
clustering, mock fields were generated in such a way that the positions
of points obeyed the observed two-point correlation function,
$\omega(\theta)$ - a measure of the number of galaxies at a given
angular separation $\theta$ - usually modelled as a power-law in
$\theta$ \citep{davispeebles}. Such fields were generated using the
iterative tree technique of \citet{sp78}.  This was implemented using
code kindly supplied by Ian Smail. $\omega(\theta)$ is further
discussed in \S\ref{sec:richness}.

\begin{figure*}
\centerline{
\includegraphics[width=75mm]{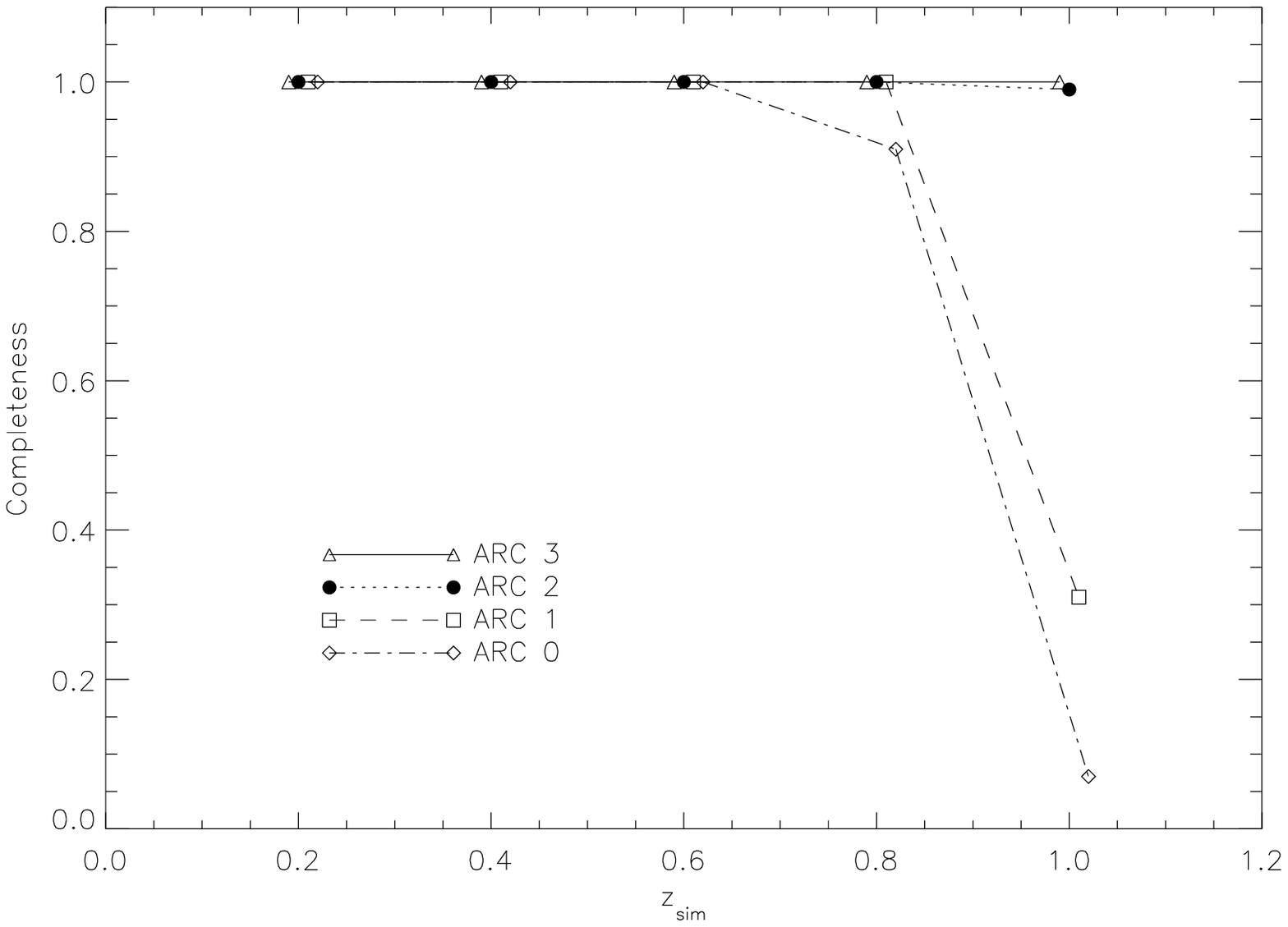}
\hspace{10mm}
\includegraphics[width=75mm]{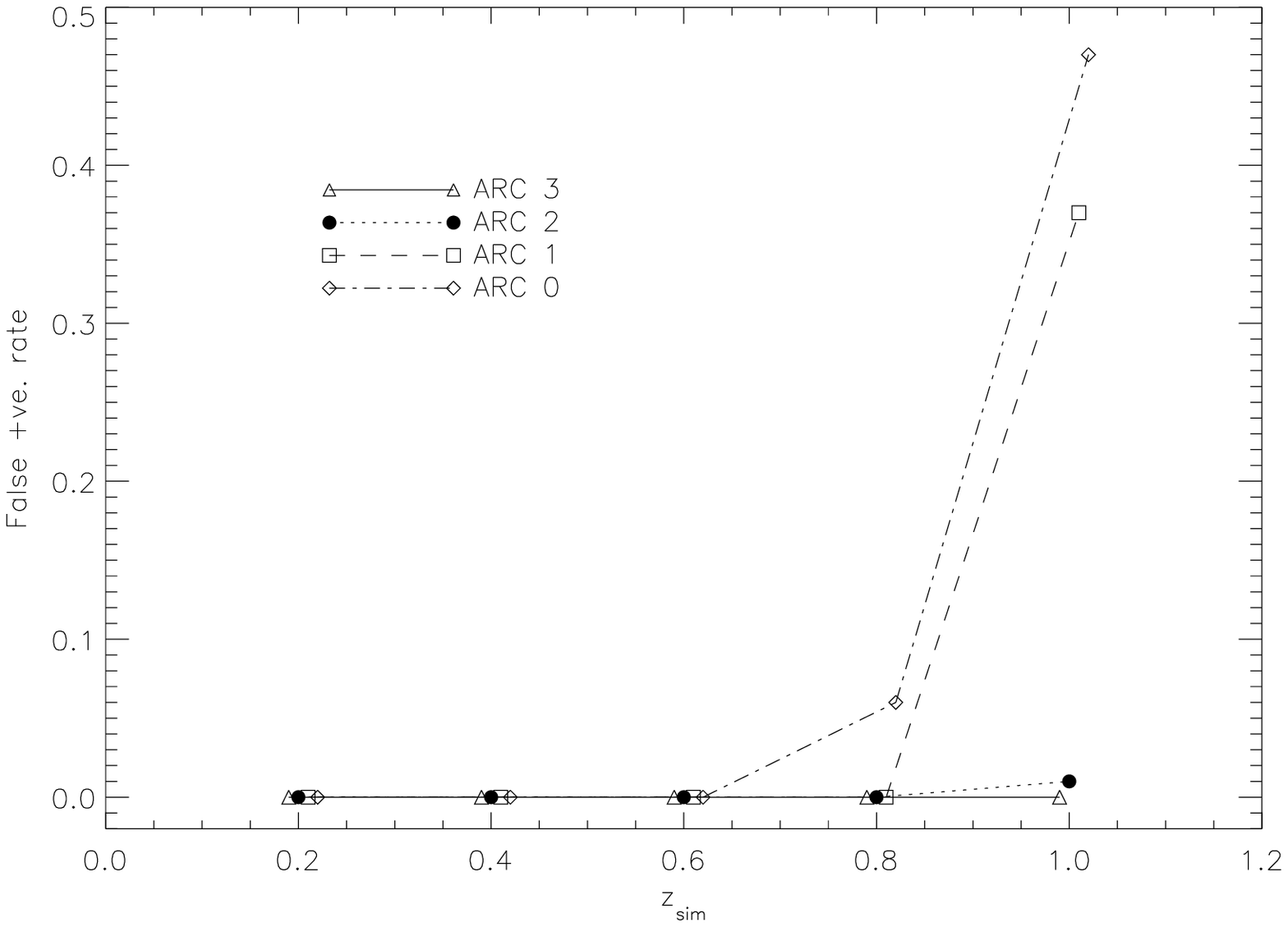}
}
\caption[Matched-Filter Completeness and Spurious Detection Rates]{Matched-Filter
  Completeness (left).  For the same simulations as Fig.~4, the
  fraction of correctly-recovered clusters (see text) was calculated.
  Matched-Filter Spurious Detection Rate (right). As for adjacent plot,
  but now the points represent detections of cluster candidates in a
  blank but correlated field.  See text for details.}
\label{fig:mf_spur}
\end{figure*}

Mock fields were generated to match the geometry of the WFC, and the MF
algorithm applied. The threshold for the Cash $C$ statistic was found by
experimentation until a reasonable compromise was found between
completeness and false-positive detection rate.  The rates for the
final threshold are plotted in Fig.~\ref{fig:mf_spur}. A value of
$C_{thresh}=-155$ in the units set out in the previous section was
chosen.

To summarise, the matched-filter uses only the I-band galaxy positions
and magnitudes to attempt to find systems with luminosity functions
resembling those of galaxy clusters.  This particular algorithm is
designed to generously join points likely to belong to a given
overdensity together to allow for the possibility of irregular,
extended clusters, perhaps as yet unrelaxed.  The performance of this
algorithm is compared with simulated data in
Figs.~\ref{fig:mf_accuracy} and \ref{fig:mf_spur}. Simulated clusters
of various richness classes were generated. The accuracy of the
estimated redshifts is worst for the poorest clusters, but even for
these it should be better than $\Delta$z$=$0.1 for redshifts less than
about 0.7.  Hereafter the accuracy decreases, as a large fraction of
the galaxies drop below the completeness limit of the survey.  Again,
for redshifts less than $\sim$0.7, the MF algorithm should find all
clusters of Abell richness classes (Fig.~\ref{fig:mf_spur}).  The
fraction of false detections is essentially zero below this redshift
and rises (most rapidly for poorer clusters) hereafter.  Thus, this
algorithm should essentially recover all ARC$>$0 clusters with negligible
contamination below z$\sim$0.7.

\subsection{The Colour-Magnitude Relation Algorithm}
\label{sec:cmr}

The CMR finder used was based on the Cluster Red Sequence algorithm of
\citet{gy00}.  Their method is directly applicable to the XDCS data
set, as they tested the algorithm on V- and I$_c$-band data of the
CNOC2 field redshift survey \citep{cnoc2}.  The CNOC2 survey comprises
four fields of similar area and depth to each of the XDCS fields,
although the total area of XDCS is an order of magnitude larger.  The
algorithm works by first filtering the data, leaving only those which
are compatible with galaxies belonging to a model colour slice in
colour-magnitude space.  Then the method proceeds in a similar manner
to the previous methods - convolving the data points with a kernel and
performing density estimation.  However, there is now the added
complication that the overdensity finding has to be done in 3D.

\subsubsection{Model CMRs}
\begin{figure}
\includegraphics[width=80mm]{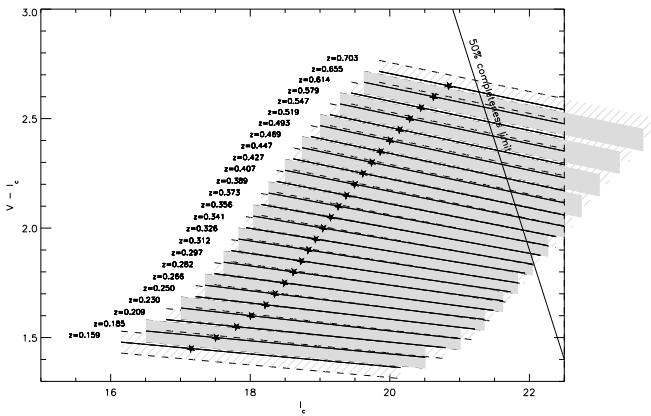}
\caption[Model colour slices used in the CMR finder]
{Model colour slices used in the CMR finder. Thick lines show the CMR at the
redshift given to the left; stars illustrate the position of \mstar; and dashed
lines show the 1$\sigma$ scatter in the CMR, bounding the slice.
Colour slices run from \mstar-1 to \mstar+3 (to a limiting magnitude of $I_c=22.5$).
Alternate slices are shaded differently for clarity.  The 50\% completeness limit
of the photometry is also shown.
}
\label{fig:slices}
\end{figure}

The passive-evolution models of \citet{ka97}, with the cosmology
H$_0=64$\kms Mpc$^{-1}$, and $q_0$=0.1, and a formation redshift of
$z_f=4.4$, were used.  These models reproduce the evolution of the CMR
for clusters to z$\gsim$1 \citep{kaba98}. A redshift is selected and
the model colours as a function of magnitude for this redshift
extracted.  A colour slice of width compatible with the scatter in the
CMR is taken around this line. The slices are selected in colour space
and constructed in such a way that each overlaps by half the width of
the next slice, in order to ensure that cluster CMRs are not lost
between adjacent slices.  This leads to irregular redshift spacing
(e.g. Fig~\ref{fig:slices}).  Slices were chosen between $V - I_c = 1.4$ and $V - I_c
= 2.7$.  Bluer than this limit and the 4000\AA~~break passes below the
limit of the V-band filter, and redder than this limit and the colour
errors become unreasonably large.  The model slices used are shown in
Fig~\ref{fig:slices}. 

\begin{figure}
\centering
\includegraphics[width=60mm]{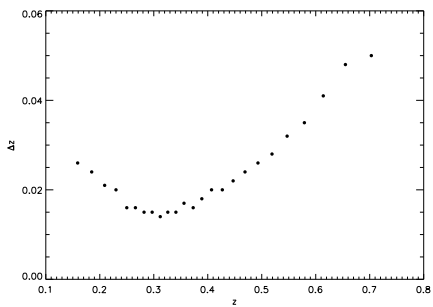}
\caption[The redshift resolution of the CMR-finder]
{The redshift resolution of the CMR-finder.  The binwidth in redshift
of the model slices (spaced constantly in colour), is shown as a
function of redshift.  This illustrates that the method offers greatest
sensitivity at $z\sim0.3$, and the binwidths increase rapidly above
$z\sim0.5$.  Also, increasing colour errors and incompleteness lead to
increasing uncertainty in redshift estimation at the high end.  }
\label{fig:dz}
\end{figure}

Each of the 24 slices illustrated is confronted with the
x,y-position, colour, colour-error and total I$_c$ magnitude data in
turn.  Each galaxy is then given a weight which is the likelihood that
for the given V - I$_c$, $\Delta(V-I_c)$, MAG\_BEST($I_c$), the galaxy
belongs to the model CMR slice (errors in the I$_c$ magnitude are
ignored as the CMRs are virtually horizontal).  This weight shall be
termed the {\it colour weight}.  In practice, many galaxies are so far
from the colour slice that their colour weights can safely be set to
zero, thus galaxies with colour weights of less than 0.1 (i.e.  10 {\it
per cent} probability of belonging to the CMR slice) are ignored.  This
is done for computational efficiency.

As mentioned above, the aim is to run kernel density estimation on the
data, using colour-based weights to amplify the signal from cluster
galaxies.  As can be seen from the number counts, the numbers of
galaxies at faint magnitudes grows rapidly.  Thus, if just the colour
weights were used, spurious detections would be caused due to some
fields containing large numbers of faint objects (many of which would
have the same colour as the CMR slice).  Put simply, brighter (and
rarer) galaxies are more powerful diagnostics of cluster members.
Hence, it is necessary to also apply {\it magnitude weights} to weight
brighter objects more heavily, within each colour slice.  The form of
this weighting function was determined by \citet{gy00} for the CNOC2
data.  This function is the probability that a galaxy of a given
magnitude is a cluster galaxy.  It could be derived from theoretical
models, but requires the cluster galaxy LF; the space density of
clusters and its evolution; and the field galaxy counts: it is simpler
to deduce internally from the data. \citet{gy00} show that whether they
assume 2\% or 20\% of all their galaxies lie in clusters, the fit to
the weighting function only differs in linear slope by a factor of 1.5
(although there is considerable scatter about the relation). Bearing
this in mind, the function chosen here is a fit to the result in their
Fig.~5.

\begin{equation}
 P(M) = \cases{
 0.55                   & $(M^\star-1 < M < M^\star)$ \cr
 -0.08 (M - M*) + 0.55  & $(M^\star < M < M^\star+3)$ \cr
 }
\end{equation}

Once colour weights have been assigned, each galaxy is given a {\it
total weight} for each colour slice which is just the product of the
colour weight and the magnitude weight. The next step is to smooth the
data with a kernel and estimate the density of the weighted points.  In
a change from the previous method, a regular grid is chosen.  This
makes several later stages computationally easier.  A grid fixed in
physical size (for the above cosmology) is constructed with the pixels
spaced in intervals of 0.125 h$^{-1}$Mpc.  The kernel chosen also
differs from the Gaussian kernel used in the above methods.
\citet{gy00} chose to use an exponential kernel of the form
$k(r)=Ae^{(-1.965r)}$ where $A$ is a normalising constant
(although this is unnecessary, as a further
normalisation step is carried out later in the algorithm, and so $A$ is
ignored here) and $r$ is the physical distance between galaxies at the
redshift of the colour slice, in units of 0.33 h$^{-1}$Mpc.  They chose
this kernel as it approaches the shape of the NFW profile at
intermediate radii (for the value of -1.965 chosen) and is constant
provided that $r$ is given in units of the NFW scale radius (a value of
0.33 h$^{-1}$Mpc is suggested by the CNOC1 survey \citep{cnoc96}).

Thus, running the above algorithm results in a series of grid points
distributed over the field of view, each with an associated signal
resulting from the convolution of the exponential kernel with the total
weights.  These signal amplitudes will be referred to as $\delta_{ij}$s
in the notation of \citet{gy00}.  Each colour (redshift) slice
contains a different number of grid points (as the angular size of the
field is fixed, but the physical scale at the redshift of the slice
decreases with increasing redshift), and a different distribution of
$\delta_{ij}$s.  The distribution changes as the fixed physical size
kernel changes apparent size and the mean density of objects differs
between redshift slices.  Thus, the $\delta_{ij}$s need transforming
into some standard measure of significance, correctly normalised
between the colour slices.

\begin{figure}
\centering
\includegraphics[width=80mm]{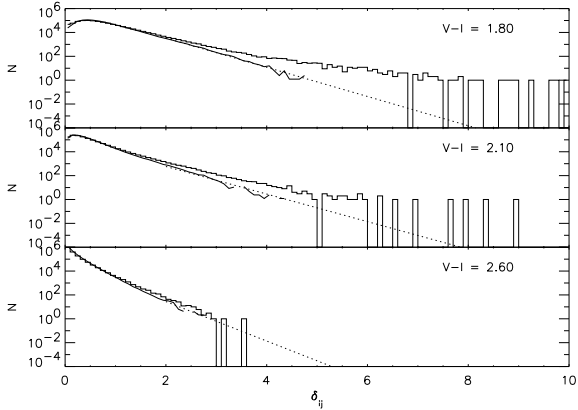}
\caption[Histograms of $\delta_{ij}$ values for CMR-finder]{Histograms
of $\delta_{ij}$ values for CMR-finder.  Three different colour slices
are illustrated.  The solid histogram shows the binned values for the
whole survey.  The thick, solid line shows the values for the
bootstrapped-thresholded data, and the dotted line is the power-law
extrapolation to the bootstrap.  See text.  }
\label{fig:del_ij}
\end{figure}

Several cluster detection algorithms \citep{gal,lobo,gy00} have used
bootstrap resampling techniques to assess the significance of
detections, and this was also done here. As noted by \citet{gy00}, a
direct application of the bootstrap is likely to be incorrect (as the
data contains clusters and is therefore not independently distributed).
So, exclusion cuts of 10\% of the data at the high-$\delta_{ij}$ and
low-$\delta_{ij}$ (to preserve symmetry) ends were performed; and each
data point (comprising an x,y-position, colour, colour-error, and
magnitude) was sampled, with replacement, until the original number of
datapoints had been extracted.  The bootstrapped datapoints were then
run through the algorithm, resulting in a new distribution of
bootstrapped $\delta_{ij}$s.  Each WFC mosaic had one bootstrapped
realisation of its data made, as the process is computationally
expensive.  A power law fit of the high significance tail was
extrapolated to the very rare, highest significance peaks.  This was
found to agree well with tests made for larger number of realisations.
The high-$\delta_{ij}$ tail (where the number of points is low) can be
extrapolated well with a simple linear fit, rather than performing many
more bootstrap resamplings.  See the distributions illustrated in
Fig.~\ref{fig:del_ij}, and cf.  fig.~7 of \citet{gy00}.  The
probability that a given $\delta_{ij}$ occurs at random can be found by
comparing the number of $\delta_{ij}$'s in a given range with the
number in the same range in the clipped-bootstrap sample,
$P(\delta_{ij}) = \frac{N(\delta_{bootstrap} \geq \delta_{ij})}
{N(\delta_{bootstrap})}$.  These can then be stated as the equivalent
Gaussian sigma (denoted $\sigma_{ij}$) for convenience.

The final step in this process is to extract the significant peaks, and
work out which peaks are associated (i.e. part of the same cluster).
The above $\sigma_{ij}$'s form a datacube in x,y,z space, where x,y are
the physical metric coordinates and z is the redshift of each slice.
\citet{gy00} used the clump-finding algorithm {\tt clfind} of
\citet{clfind} to extract significant associations from the data. A
user-friendly IDL version of this algorithm was downloaded from {\tt
  http://www.astro.ufl.edu/$^\sim$williams/clfind/} and run on the
datacubes using the parameters detailed in \citet{gy00}.  Briefly, the
algorithm is a 3D friends-of-friends group-finder which also contours
the data at fixed intervals and looks for clumps in this 4D space. The
code was originally used with temperature maps in radio data, but the
temperature can be replaced with the signal from the CMR-finder, and
the method is identical. The highest peaks are identified first and
traced down in contour levels, their friends above the minimum level
being linked to them at each interval.  Following this through, all
points become joined into one clump as the lowest contour level
approaches the noise within the data. From extensive simulations,
\citet{clfind} recommend the data be contoured at intervals of twice
the {\it rms} noise in the data. \citet{gy00} calculated this value to
be 1.1$\sigma_{ij}$. Tests were performed varying this value. The
resulting groups found were practically identical, but using a value of
1.4$\sigma_{ij}$ seemed a slightly better choice.  Using a lower value
split off clumps around the periphery of higher significance clumps
(described in more detail below). The peaks were traced down to the
lowest possible contour level (1.1$\sigma_{ij}$). This level resulted
in a total catalogue of $\gsim$1000 candidates.  This number was
reduced by setting a higher threshold later, by examining the
repeatability of cluster detections in the overlapping data.  A
threshold of 4.8$\sigma$ was found to result in a reasonable number
($\sim$200) of repeatable candidates, detected in the two independent
images.  This high value may suggest that the bootstrap estimate used
may be an overestimate of the formal significance of candidates.  It
should be noted that changing the size of the high and low exclusion
cuts used in the bootstrap realisations will change the absolute values
of the significances.  However, since the relative significance is
correct, just selecting a subsample of the most significant systems is
a perfectly valid approach.

Thus a list of cluster candidates was extracted from the datacubes.  One
further stage was necessary to clean the resulting catalogues, as a
number of candidates were found in close proximity to more significant
candidates.

These may be genuine groups infalling into larger clusters, or just
spurious detections from increased noise around other candidates.
Failure to remove these would result in the following problem:  when
measuring properties (such as richness, see below) of the detected
systems, if a cluster and poor group are superposed along the same
line, very close together on the sky, then the effect on the richness
measurement of the richer system would be minimal; but the effect on
the poorer system would be to catalogue another rich system (due to
contamination from the richer cluster).  Hence, a minimum distance in
physical and redshift space was imposed to prevent these duplicate
detections, and only the highest peak within two cutoff radii (i.e. 8
times the NFW scale radius of 0.33~\hmpc) and two redshift slices
retained. An estimate of a characteristic radius for each group was
made (as for the MF algorithm) by taking the maximum distance between
the candidate centre and the 3$\sigma$ contour.

\section{Richness Measures}
\label{sec:richness}

One of the simplest observables for a galaxy cluster, in optical data,
is its richness. The original richness classification of \citet{abell}
has been shown to be subject to many biases
\citep[e.g.][]{1983MNRAS.204...33L,katgert96,vfw}. However, more recent
estimates such as the $B_{gc}$ parameter of \citet{ls79} correlate
well with cluster velocity dispersion \citep[e.g.][]{hl91,ylc99}.

\subsubsection{The B$_{gc}$ Measure}
B$_{gc}$ is explained in detail in \citet{ylc99} and references
therein. It has been used by several workers, primarily in studies of
the environments of radio galaxies \citep[for recent
examples:][]{andersen,miller99}. In outline, it is found from the
amplitude of the 3D two-point correlation function.  The 3D correlation
function is difficult to measure; observationally easier is the angular
correlation function, $\omega(\theta)$. This is simply a measure of the
number of galaxies at a given angular separation.  This can be
approximated as a power law $\omega(\theta)=A_{gg}\theta^{1-\gamma}$
\citep[][for example]{davispeebles}, where $A_{gg}$ is the angular
galaxy-galaxy correlation amplitude.  Now, fixing a reference point as
the assumed centre of the cluster, one can measure the two-point
angular galaxy-cluster correlation function.  Its amplitude, $A_{gc}$,
can be calculated by counting the excess number of galaxies (i.e.
background-subtracted), within some radius, $\theta$, of the cluster
centre ($N_{net}=N_{total}-N_{bgd}$).  Assuming fixed $\gamma$,
$A_{gc}=(N_{net}/N_{bgd})[(3-\gamma)/2]\theta^{\gamma-1}$.

$B_{gc}$, the spatial amplitude, can be estimated via deprojection of
the angular correlation function, assuming spherical symmetry, as given
in \citet{ls79}:

\begin{equation}
\label{equ:bgc}
B_{gc}=N_{bgd} \frac{D^{\gamma-3}A_{gc}}{I_\gamma\Psi[M(m_0,z)]}
\end{equation}

where $N_{bgd}$ is the background galaxy counts to apparent magnitude
$m_0$ and $\Psi[M(m_0,z)]$ is the integrated LF of galaxies up to the
absolute magnitude $M$, given by $m_0$ at the cluster redshift $z$.
$I_\gamma$ is an integration constant arising from the deprojection
($I_\gamma=3.78$ for an assumed $\gamma$ of 1.77).  $D$ is the angular
diameter distance to $z$ (Equation \ref{equ:zang}).

\citet{ylc99} discuss extensively the effects of different assumptions/
measurement limits on $B_{gc}$.  The salient points are summarised
here.  If the assumption of a universal LF is not strictly correct,
then the systematic uncertainty this introduces in $B_{gc}$ is
$\sim10\%$.  Changing the parameters of the LF (slope and
normalisation) results in $\lsim20\%$ differences in $B_{gc}$ if
$M^\star$ is incorrect by as much as 0.5 mags, and if $\alpha$ is
incorrect by as much as $\pm0.3$.  $B_{gc}$ is independent of the
sampling area, provided $\gamma$ has been correctly chosen.  $B_{gc}$
is insensitive to the sampling magnitude limit if $m_{lim}$ lies on the
flat part of the LF (between $M^\star$+1 and $M^\star$+2).  The most
important step is ensuring that the cluster LF and background galaxy
counts are determined in a self-consistent manner.

The model LF is the same as that chosen for the cluster in the MF
algorithm.  Fig.~\ref{fig:counts} illustrates how the model LF
assumed for both the cluster and the field translates into field galaxy
counts.  The LF was integrated over 0.05 redshift bins from $z=0.00$ to
$z=2.00$.

The normalisation, $\phi^{\star}$, was fitted to the XDCS data.
$\phi^\star=0.0035$\hsf$^3$Mpc$^{-3}$ was found.  This is consistent
with the R-band value measured by \citet{ylc99}, after correcting the
number density to their cosmology.  The completeness was modelled as
XDCS counts/ literature counts \citep[from][]{metcalfe00} to
$I_{lim}=22.5$ (where the completeness falls to 70\%).  This factor was
then applied to the expected counts for the $B_{gc}$ calculation (as
well as to the MF algorithm, earlier).

The uncertainty in the $B_{gc}$ parameter was computed using the formula from
\citet{ylc99}:

\begin{equation}
\label{equ:err_bgc}
\frac{\Delta B_{gc}}{B_{gc}}=\frac{(N_{net} + 1.3^2 N_{bgd})^{1/2}}{N_{net}}
\end{equation}
The factor $1.3^2$ accounts for the clustered (and so non-Poissonian)
nature of the background counts \citep{yg87}.

\subsubsection{The \lell Measure}
The luminosity of galaxies on the CMR referred to as \lell (since the
galaxies are primarily early-type), has been shown (for a limited
sample) to correlate well with the X-ray temperature of the
cluster \citep{s98}. For a sample of the 10 most X-ray luminous clusters
in the redshift range $z=0.22-0.28$, \citet{s98} investigated the
homogeneity of the stellar populations of cluster early-type galaxies.
One method they used was to compare mass of baryonic material locked up
in stars in early-types (in the form of the luminosity of galaxies on
the CMR) with the total mass of the cluster (using X-ray temperatures
from the literature).  They found a remarkably small scatter about this
relation ($\approx$ 17\% compared to the $\approx$30\% when \lx is used
instead of \tx).  It should be noted that the sample spans a narrow
range in redshift, and relatively narrow range in blue fraction
(i.e. few galaxies statistically belonging to the cluster are {\it not}
red), and mass.  A large sample to characterise an empirical relation
between \lell and \tx (or mass) over a larger range of parameter space
does not currently exist in the literature. However, the evolution of
cluster mass-to-light ratios for a sample of 4 X-ray selected clusters
over a wide range (0.22$<$z$<$0.83) of redshift has been studied by
\citet{hoekstra}.  Using gravitational weak-shear measurements from
HST images, they determined that the mass-to-light ratios in their
sample evolve in a manner consistent purely with luminosity-evolution
of the cluster early-type galaxies.  Thus, inverting this argument,
measuring the rest-frame luminosity of cluster early-type galaxies
(corrected for passive evolution) could potentially provide an estimate
of the total cluster mass.  Again it should be emphasized that the
datasets on which these correlations were based are small and so the
scatter in the relation is not well known.  Furthermore, all the data
came from X-ray selected samples, so the scatter may be further
increased once optically selected clusters are included.

For each cluster candidate, the colour slice from the CMR-finder in
which the candidate was detected was selected.  The galaxies within this
colour slice, and within a radius of 0.45 Mpc (0.5 Mpc in \citealt{s98}
cosmology) brighter than $M_V = -18.5 + 5\log h$ \citep{s98} were
selected, and their apparent I-band magnitudes converted into
rest-frame V luminosity, again using the stellar population synthesis
models of \citet{ka97}.  This magnitude limit is approximately 1.5
magnitudes fainter than L$^\star$ at z$\sim$0.3. Background subtraction
was carried out by calculating the number of galaxies in a surrounding
annulus, scaled to the area of the cluster region, as above, and
subtracting the corresponding luminosity, assuming these galaxies were
at the same redshift as the cluster.  The limits for the maximum and
minimum luminosity were estimated by using all the galaxies whose
photometric colour errors allowed them into or out of the colour slice,
respectively, and summing their luminosities in the same way.  This
gives error estimates in excellent agreement with assuming the error is
entirely due to the error in the estimated redshift (by taking the
redshifts of the next highest and lowest colour slices and
recalculating the luminosities).

\subsection{Construction of Final Cluster Catalogues}
\label{sec:constr-final-clust}
Both the MF and CMR finders were run on each WFC mosaic individually.
Since each field possesses overlapping `A' and `B' data, the next
step is to combine the candidates from the A- and B-rotations, for each
algorithm.  The MF technique is more straightforward, so this will be
discussed first.

The MF catalogue was divided into two catalogues with different
significance thresholds. The higher significance catalogue will be
referred to simply as the MF catalogue, or the {\it final} MF
catalogue, if it is necessary to emphasize the distinction between this
and the lower-significance catalogue: referred to as the {\it full} MF
catalogue. The full MF is that using the thresholding described in
\S\ref{sec:sims}. The final MF catalogue was produced by imposing a
stricter Cash $C$ cut (a value of -280, this time).  The completeness and
spurious rates with this threshold are comparable to those plotted in
Fig.~\ref{fig:mf_spur} for redshifts less than 0.7.  Using
  higher redshift candidates just increases the number of spurious
  detections, and the X-ray data are unlikely to probe enough volume to
  detect clusters at these redshifts. Also, candidates with a MF
group radius of zero were rejected.  Such objects occur when only a
single galaxy (and none of its near neighbours) lies above the Cash $C$
threshold. This reduces the number of clusters detected to a more
manageable number, $\lsim$200, of higher confidence candidates, whilst
the full MF catalogue allows the list of lower significance candidates
to still be retained.  This may prove useful later, if, for example in
cross-comparisons between catalogues, a candidate is not found with
high significance in the MF catalogue; then the full MF catalogue may
be searched.

Next, I-band WFC thumbnail images were produced for all candidates in
the MF catalogue.  These were inspected to see if a candidate was found
due to spurious objects (e.g. satellite trails, haloes of bright
stars).  Those that were spurious were flagged and rejected from the
catalogue.  Finally, the MF catalogue comprising A- and B-rotation
candidates was reduced to a single catalogue by searching for
candidates which overlapped in the two rotations.  Where this occurred,
only the candidate with the larger group radius was retained.  This was
found to be more stable than selecting the highest peak Cash $C$ value
candidate, as the group radius is given by the extent of galaxies
passing the Cash $C$ cut; but the Cash $C$ value noted for each
candidate just comes from the galaxy with the highest individual value
in the candidate. This can be thought of as favouring a larger ``total
likelihood'' over that of a ``peak likelihood'' for each cluster
candidate.

The procedure for producing the CMR catalogue was slightly more
involved.  This was due to the `3D' aspect of finding clusters with
this technique.  Whereas the MF finder just selects overdensities and
fits the most likely redshift to the clump, the CMR finder can, in
principle, detect projections of groups along the line of sight.  The
same first steps as for the MF were followed: I-band thumbnails were
generated, spurious candidates rejected, and a higher and lower
significance catalogue generated.  The final CMR catalogue had a
threshold of 4.8$\sigma$ imposed, as described in \S\ref{sec:cmr},
whereas the full (lower significance) catalogue allowed candidates to
be traced down to the lowest possible contour level with {\tt
  clumpfind} (i.e. 1.4$\sigma$). Candidates in the {\it same} rotation
which showed more than one candidate overlapping (as defined by their
group radii) were flagged as `projection' possibilities.  This check
was performed on an individual frame basis to avoid the possibility
that a single candidate having a significantly different estimated
redshift in the A- and B-rotations would result in one candidate being
artificially classed as a projected system. The final CMR catalogue was
produced by combining the rotations as for the MF.

\begin{figure*}
\centerline{
\includegraphics[width=75mm]{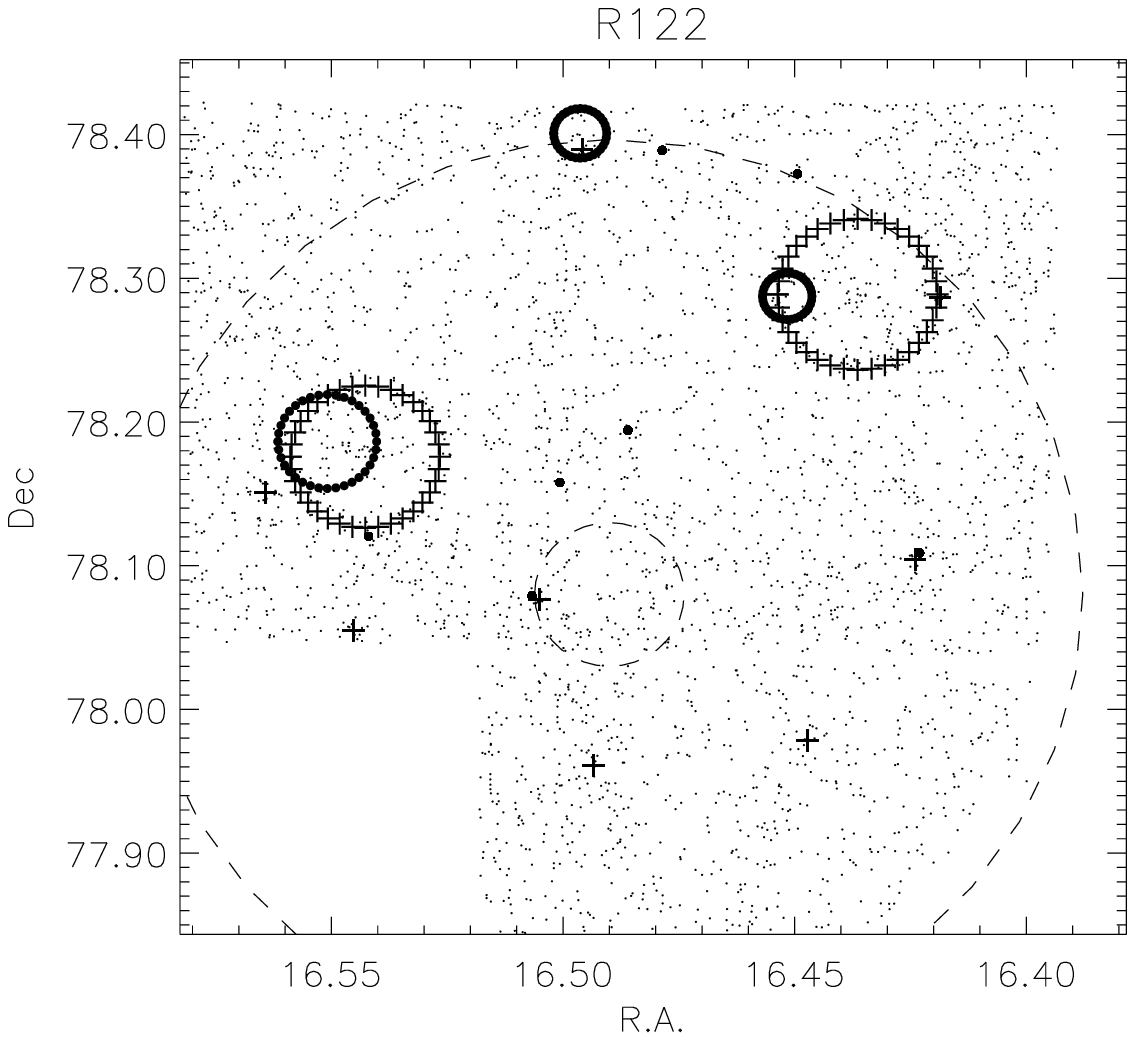}
\hspace{10mm}
\includegraphics[width=75mm]{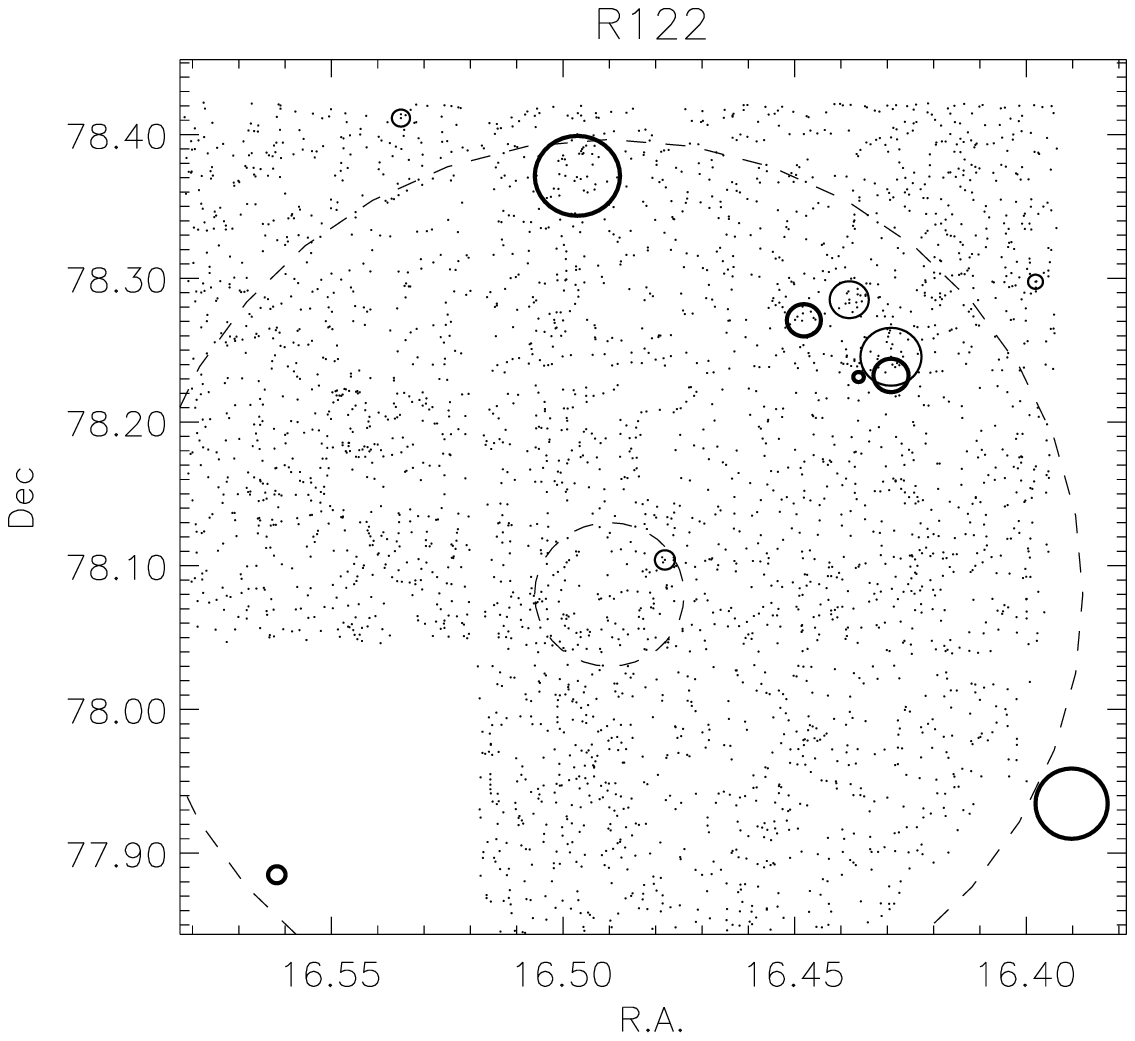}
}
\caption[Full Catalogues in One Field for MF (left) and CMR (right) Algorithms]{Full
catalogues in one field for MF (left) and CMR (right)
algorithms. Points show galaxies with I-band magnitudes brighter than
22.5 (for B-rotation only, for clarity).  Dashed lines denote the
limits of the PSPC field (19 arcmins: outer radius, 3 arcmins: inner
radius). Cluster candidates are outlined by points marking their group
radii (defined in text).  Symbols are: \\ Left panel, MF candidates:
filled circles - A-rotation; crosses - B-rotation.\\ Right panel, CMR
candidates: thick lines - A-rotation; thin lines - B-rotation. \\ Those
candidates which appear as single points have group radii of zero (see
text).}
\end{figure*}

\begin{figure}
\centerline{
\includegraphics[width=75mm]{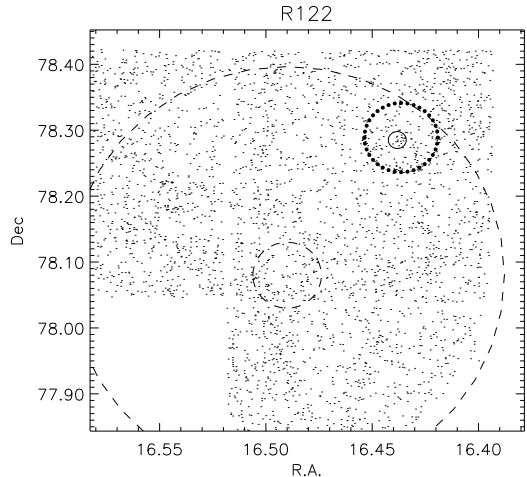}
}
\caption[Final Catalogues for one field]{Final catalogues for this
field. The thresholding described in the text has been applied. Dotted
circle denotes MF candidate, solid line shows CMR candidate. Other
symbols as for previous figure.}
\end{figure}

This provided the two main catalogues for the optical survey: the final
MF catalogue and final CMR catalogue.  These catalogues were then
passed to the richness measuring algorithms described in
\ref{sec:richness}. Finally, both catalogues were cropped to overlap
with the X-ray data which is described below.  To do this, only
candidates within an annulus of 3 to 19 arcmins from the centre of the
X-ray pointing were retained.  At distances greater than this, the
X-ray data are not useful (due to degraded resolution and sensitivity -
see below) and the inner region was excised to avoid objects associated
with the target of the X-ray observation.

\subsubsection{Summary of Catalogues}
In total, the final MF catalogue (constructed and trimmed to the PSPC
field, as described above) contains 185 cluster candidates. The final
CMR catalogue contains 290 candidates. The MF technique fits the most
likely value of M$^\star$ in the range 17.0$\leq$I$_C\leq$21.5. The
bright limit is imposed by there being few galaxies this bright in the
field, and at this redshift (z$=$0.15) the angular diameter of clusters
becomes so large that the contrast of the cluster against the
background is greatly reduced.  The faint limit (corresponding to
z$=$0.9) is set such that the limiting magnitude is one magnitude
fainter than this M$^\star$, thus there are still many galaxies to
which to fit a luminosity function. The CMR method is limited by the
depth of the photometry in {\it both} bands (as illustrated in
Fig.~\ref{fig:slices}), which leads to colour limits
1.45$\leq$V-I$\leq$2.65 corresponding to 0.159$\leq$z$_{est}\leq$0.703.
In practice, the MF catalogue is cropped at the high redshift end to
match the colour limits imposed by our choice of filters used with the
CMR method.  At the low redshift end, only a few clusters lower than
the our CMR colour limit are found by the MF, so these are retained.

\subsection{Internal Check of Redshift Estimates}
\label{sec:intern-check-redsh}

To assess the accuracy with which an estimated redshift can be
assigned, an internal check can be performed comparing the redshifts
estimated for candidates using the A- versus the B-rotation data. Due
to the way in which the object catalogues are generated
(\S\ref{sec:optic-data-reduct}), different data are produced for the
same region of sky using two independent observations. For example, the
main difference between object catalogues for the two rotations was the
star/ galaxy classification. Several objects classified as galaxies in
one rotation were classified as stars in the other, and vice versa. The
neural network classifier of \sex\ uses both the FWHM of a source and
its ellipticity to decide the nature of each source. The effect of FWHM
differences was minimised due to the way in which the data were taken
(observing the same A- and B-rotation fields sequentially) so that,
unless the seeing is changing on very short timescales, the FWHM of
point sources should be the same for the two rotations. Inspection of
objects which changed class between the two frames showed that a
slightly different measure of ellipticity was the primary cause.
Overall, the level of star/ galaxy misclassification should be around
the few percent level. From the spectroscopic observations presented in
\S\ref{sec:spectr-observ-x}, two of the 87 redshifts measured (for
objects brighter than I$\approx$20, classified as galaxies) were due to
stars, or around 2\%.  The contamination is likely to be higher for
fainter objects, where a lower signal makes shape parameters more
difficult to measure. To a lesser extent, the object catalogues between
the two rotations differ due to cosmic rays, diffraction spikes, and
differently deblended objects (as discussed in
\S\ref{sec:optic-data-reduct}).

\begin{figure*}
\centerline{
\includegraphics[width=80mm]{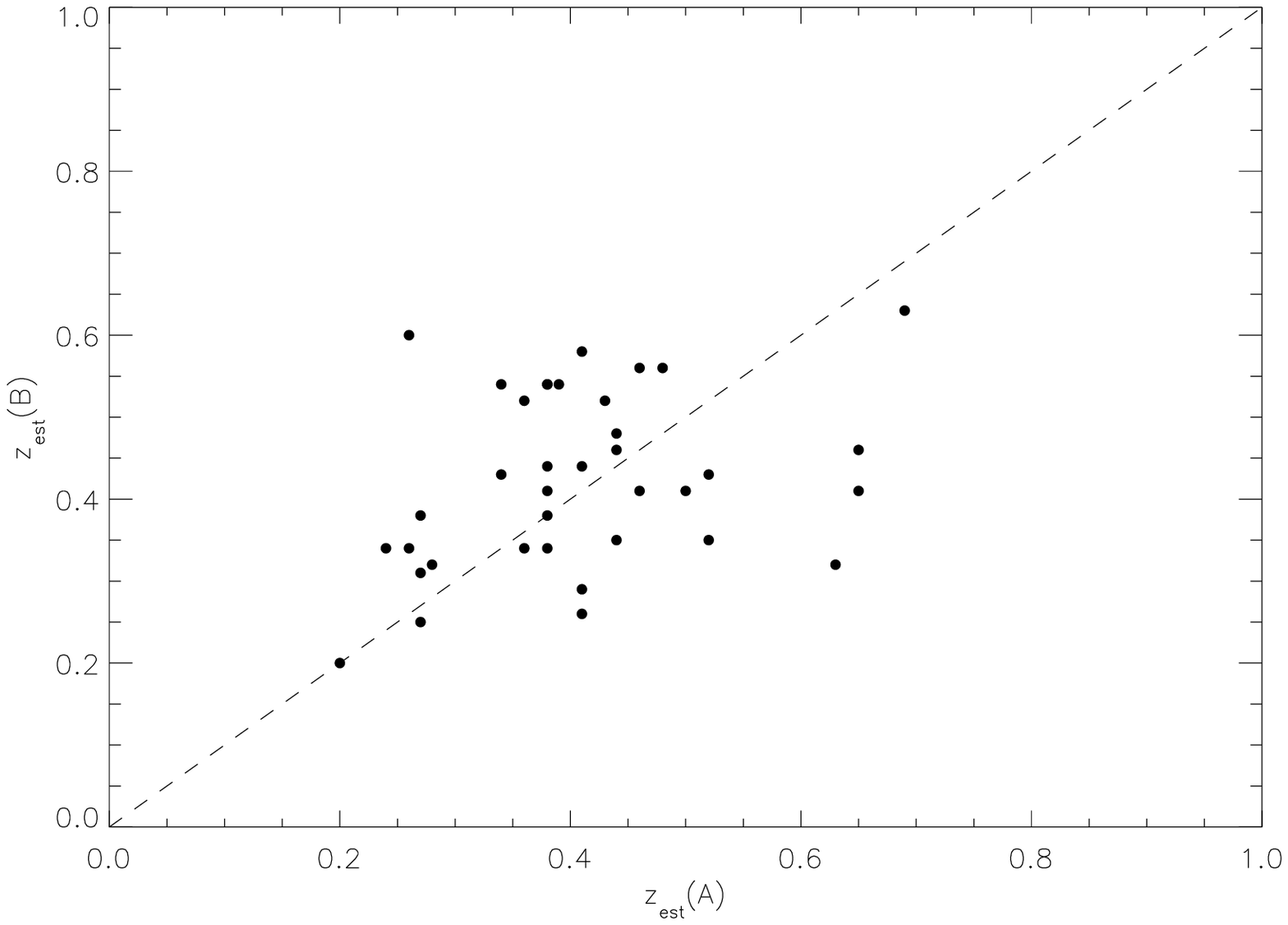}
\includegraphics[width=80mm]{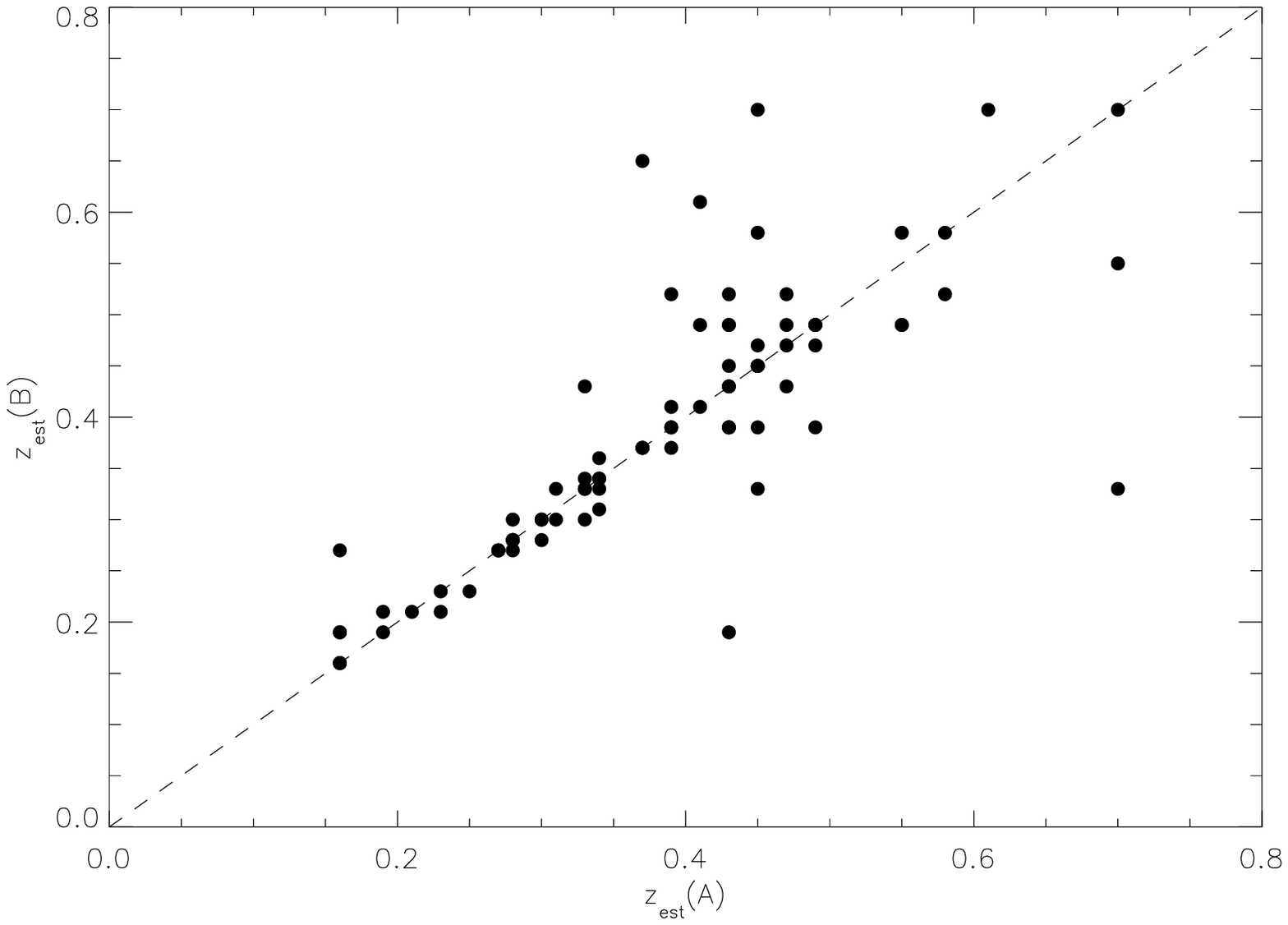}
}
\caption[Comparison of Estimated Redshifts for the MF and CMR
Algorithms from the A versus B Data.]{Comparison of estimated redshifts
for the MF and CMR algorithms from the A versus B data. The left
panel shows the MF estimated redshift in the B field versus the A field
estimate of the same quantity.  
The right panel
shows the A versus B estimated redshift for the CMR algorithm.}
\label{fig:z_ab}
\end{figure*}

A comparison of estimated redshifts between the two independent
observations allows the effect of all these factors to be taken into
account.  This is one of the primary motivations for treating the
repeat observations separately. Candidates for the comparison were
selected in the following manner.  For the MF catalogue, the full
catalogue was compared with the final catalogue.  For each entry in the
final catalogue, if a single candidate from each rotation was present
in the full catalogue, within the final catalogue entry's group radius
(to avoid possible confusion with multiple matches), then the candidate
was selected.  A similar procedure was followed for the CMR catalogue,
with the added condition that the candidate must not be flagged as
comprising projected groups (again to avoid confusion due to multiple
matches). In both cases a limiting radius of 1 arcmin was imposed for
the match, to ensure a high level of confidence that the same candidate
had been selected from the two datasets. The comparison of the
estimated redshifts from each rotation for each cluster detection
algorithm is shown in Fig.~\ref{fig:z_ab}.
Quantifying the bias and scatter in these relationships as:
$\delta z = (z_A - z_B)/(1 + z_A)$ \citep[e.g. ][]{wittman}, the mean
value is -0.004 for the MF and -0.014 for the CMR algorithm; the
standard deviations are 0.097 and 0.081 respectively. This is somewhat
misleading as the majority of the scatter from the CMR comes from a few
outliers, and the majority of points show excellent agreement between
the two independent observations. A large fraction of the outliers were
detected in the final (z$=$0.70) colour slice in one rotation, and thus
could easily be missed and associated with a less significant lower
redshift clump in the corresponding rotation. The few other outliers
can be understood in terms of marginal cases for projected systems. If
each rotation detects two systems, the lower significance candidate
is measured as being more significant in the overlapping rotation, and
the lowest significance system falls below the threshold in both cases,
then a catastrophic failure of the redshift estimate would occur. This
only appears to be the case for seven of the systems in the plot, at
z$_{est}<$0.69. The scatter in the MF estimate is intrinsically large.
Neither estimate shows any significant bias between the two datasets.

\section{X-ray Selection of Clusters}
\label{sec:x-ray-selection}
The X-ray data are archival images taken with the ROSAT Position
Sensitive Proportional Counter (PSPC).  Such observations are divided
into two energy bands: ``hard'' (0.4$-$2.4 keV) and ``soft''
(0.07$-$0.4 keV). The background flux is particularly high at energies
below 0.5 keV and the sensitivity of ROSAT rapidly drops to zero above
2.0 keV.  As a result, most X-ray cluster surveys use the hard band and
cut its range down to 0.5$-$2.0 keV.  All of the fields were taken from
the RIXOS survey \citep{mason00}, which was an international campaign
to follow up in the optical all X-ray sources in a sample of ROSAT PSPC
fields above a point source flux limit of $3\times10^{-14}$ erg
cm$^{-2}$ s$^{-1}$ (0.5-2.0 keV).  Thus, a sample of clusters
discovered in the X-ray in these fields have already been selected and
confirmed.  However, the algorithm with which these were found was
optimised for point source detection, and not for locating extended
emission, as expected for clusters of galaxies.  Thus the RIXOS cluster
catalogue is incomplete.  This resulted in a claim for a deficit of
high redshift clusters in the RIXOS sample \citep{castander95}, when
other investigators found no evolution in the abundance of clusters
\citep[e.g.][]{nichol97}.

The currently favoured technique for the detection of faint, extended
sources in X-ray data is the wavelet method. Sources are detected by
convolving the data with a kernel to enhance the contrast between
objects and the background, in the same way as described in \S
\ref{sec:imp} for the optical algorithms.  The difference with the
wavelet method is that this kernel consists of a positive core and a
negative outer ring (such that its integral over the x,y plane is
zero).  This means that slowly varying backgrounds which can be
approximated by linear functions are completely
subtracted. Furthermore, a wavelet transform of the data reveals
sources bounded by a ring of zero values; the diameter of the
zero-rings gives a measure of the angular extent of the source.  In
practice, a range of kernel values is used (as was done with the
Gaussian filtering of the MF method), and these can be used to infer
the source radius.  An instructive illustration of this technique is
given in fig.~2 of \citet{vmf98}.

Given that several wide-field surveys have also made use of archival
ROSAT data for the serendipitous discovery of clusters
\citep[e.g.][]{jones2,romer00,vmf98}, it is natural to check if any of
these overlap with the fields selected for XDCS.  29 out of the 39
fields were used in the 160 square degree survey of \citet{vmf98}. This
catalogue has the attractive feature that nearly all of the 200$+$
sources detected have been followed up in the optical, many possessing
spectroscopic confirmation.  Furthermore, their spurious detections are
also recorded in their paper, so {\it all} X-ray detected cluster
candidates can be examined, and not just the optically confirmed ones.
19 likely false detections, arising from concentrations of point
sources, were recorded, but none of these occurs in the XDCS fields..
Of the remaining 10 fields, 7 were included in the Bright SHARC Survey
\citep{romer00}. Both of these used wavelet detection algorithms in
their construction.  The SHARC catalogue has had a fairly bright ROSAT
count-rate limit imposed (corresponding to a flux of approximately
10$^{-13}$ erg s$^{-1}$ cm$^{-2}$, or about an order of magnitude
brighter than typical XDCS field limits) in order to reduce the
numbers of clusters found, to make optical follow up achievable in a
reasonable amount of time. Given this limit, 94 sources were found in
460 ROSAT fields.  It is not too surprising, then, that in the seven
fields overlapping with XDCS, no sources are found. The SHARC survey is
not considered hereafter.  The \citet{vmf98} catalogue (VMF), on the
other hand, contains 15 X-ray selected clusters in XDCS fields.  This
is the X-ray selected cluster survey with which the XDCS optical
catalogues will be compared.

The ROSAT fields observed are given in Table \ref{table:rosatfields}.
The VMF clusters in common fields are given in Table \ref{table:vmfclus}.

\begin{table*}
\centering
\caption[VMF clusters in XDCS fields]{VMF clusters in XDCS
fields. Redshifts are given for all but one cluster. The type of
redshift measured by VMF is given in the final column (p -
photometric, s - spectroscopic)}
\label{table:vmfclus}
\begin{tabular}{cccccccc}
\hline
VMF & RIXOS & $\alpha$(J2000) & $\delta$(J2000) & F$_X$ & $\delta$F$_X$
& z & Redshift \\
ID & Field  & [hh:mm:ss.s] & [dd:mm:ss] & 10$^{-14}$ erg s$^{-1}$
cm$^{-2}$ & 10$^{-14}$ erg s$^{-1}$ & & Type \\
\hline
11 & R262 & 01:24:35.1 & +04:00:49 & 7.5   & 2.2 & 0.27  & p \\
62 & R221 & 08:49:11.1 & +37:31:25 & 14.7  & 3.0 & 0.240 & s \\
69 & R248 & 09:10:39.7 & +42:48:41 & 8.3   & 2.0 & ---   & --- \\
73 & R285 & 09:43:32.2 & +16:40:02 & 23.1  & 3.7 & 0.256 &  s \\
74 & R285 & 09:43:44.7 & +16:44:20 & 21.2  & 4.1 & 0.180 &  s \\
84 & R231 & 10:10:14.7 & +54:30:18 & 21.0  & 2.9 & 0.045 &  s \\
86 & R231 & 10:11:26.0 & +54:50:08 & 20.0  & 5.1 & 0.294 &  s \\
94 & R133 & 10:56:12.6 & +49:33:11 & 12.9  & 1.9 & 0.199 &  s \\
97 & R258 & 11:17:26.1 & +07:43:35 & 6.1   & 1.6 & 0.40  &  p \\
99 & R123 & 11:19:43.5 & +21:26:44 & 5.5   & 0.9 & 0.11  &  p \\
131 & R265 & 13:09:55.6 & +32:22:31 & 9.0  & 2.9 & 0.290 &  s \\
132 & R265 & 13:11:12.8 & +32:28:58 & 46.7 & 5.8 & 0.245 &  s \\
150 & R254 & 13:43:29.0 & +55:47:17 & 17.5 & 2.8 & 0.11  &  p \\
181 & R223 & 16:33:40.0 & +57:14:37 & 3.5 & 0.7  & 0.239 &  s \\
194 & R220 & 17:29:01.9 & +74:40:46 & 17.3 & 7.2 & 0.28  &  p \\
\hline
\end{tabular}
\end{table*}

\section{Comparison of Optical and X-ray Selected Clusters}
\label{sec:comparison-optical-x}
The VMF X-ray selected clusters are listed in Table
\ref{table:vmfclus}. 
\citet{vmf98} used several methods
to ``confirm'' their X-ray cluster candidates and it is pertinent to
comment on these here. Aside from the traditional method of requiring
an overdensity of galaxies in the optical, they included another
possible criterion which was that if an elliptical galaxy not included
in the NGC catalogue lay at the peak of the X-ray emission then this
should be considered confirmation.  

This latter point was designed to include ``poor clusters and groups
which fail to produce a significant excess of galaxies over the
background''.  The authors state that it also helps to identify
``fossil groups'' in which galaxies have merged to form a single cD
\citep{fossil,jones}. Such systems appear to be as X-ray luminous as
other bright groups or poor clusters, but with a high percentage of the
optical luminosity arising from the dominant giant elliptical. The
second brightest group member is a factor of 10 fainter than the
brightest (resulting in a gap of 2.5 magnitudes in the LF).  Recent
work estimates that such systems comprise 8$-$20\% of all systems of
the same X-ray luminosity \citep{2003astro.ph..4257J}.

Vikhlinin et al. (priv. comm.) obtained optical follow up either from
second generation Digitized Sky Survey (DSS-II) plates, or R-band (or
sometimes B-, V-, or I-) CCD imaging on 1m class telescopes. Long-slit
spectroscopy was also obtained for some candidates, usually for 2 - 3
galaxies per cluster, and always including the brightest galaxy.

\begin{table*}
\caption{The nearest optically selected candidates to the VMF
clusters. For each X-ray selected cluster, the nearest matching MF and
CMR candidates' details are given.  Candidates in parentheses were not
identified in the final  catalogues. See text for details.
}
\begin{minipage}{12.5cm}
$*$ - candidate matched at a separation greater than its estimated
radius.\\ 
$**$ - candidate matched within its estimated radius, but greater than 2 arcmins.\\
(p) - candidate flagged as exhibiting projection along line-of-sight.
\end{minipage}
\label{table:opt_x_match}
\begin{tabular}{cccccccccc}
\hline
VMF & MF candidate & z$_{VMF}$ & Separation & z$_{MF}$ & $\Delta$z & CMR candidate &
offset & z$_{CMR}$ & $\Delta$z \\
ID &   ID   &           & (arcmin) &    &   & ID  &  (arcmin) &   & \\
\hline
11 & (mfJ012435.6+040107) & 0.270 & (0.328) & (0.422)  & 0.152 & cmJ012437.8+040022 & 0.807 & 0.360 & 0.090 \\
62 & mfJ084914.5+373123 & 0.240 & 0.678 & 0.276 & 0.036 & cmJ084908.8+373036 & 0.939 & 0.190 & 0.050 \\
69 & mfJ091049.4+425002 & ---   & (2.246**) & 0.484 & ---   & cmJ091045.1+424955 & 1.596 & 0.450 & ---\\ 
73 & mfJ094350.5+164034 & 0.256 & (4.421*) & 0.351 & 0.095 & cmJ094329.3+163916 & 1.035 & 0.190 & 0.066 \\
74 & (mfJ094344.0+164500) & 0.180 & (0.691) & (0.293) & (0.113) & --- & --- & --- & --- \\
84 & --- & 0.045 & --- & --- & --- & --- & --- & --- & --- \\
86 & mfJ101137.3+545036 & 0.294 & 1.698 & 0.276 & 0.018 & cmJ101134.1+545014 (p) & 1.168 & 0.330 & 0.036 \\
94 & mfJ105617.5+493237 & 0.199 & 0.972 & 0.157 & 0.042 & --- & --- & --- & --- \\
97 & mfJ111726.2+074316 & 0.400 & 0.306 & 0.232 & 0.168 & cmJ111726.2+074319 & 0.258 & 0.370 & 0.030 \\
99 & --- & 0.110 & --- & --- & --- & --- & --- & --- & --- \\
131 & mfJ131001.9+322110 & 0.290 & (1.889*) & 0.437 & 0.147 & cmJ130954.0+322137 & 0.949 & 0.270 & 0.020 \\
132 & (mfJ131113.2+322843) & 0.245 & (0.259) & (0.422) & (0.177) & cmJ131111.0+322825(p) & 0.664 & 0.210 & 0.035 \\
150 & --- & 0.110 & --- & --- & --- & --- & --- & --- & --- \\
181 & mfJ163334.2+571457 & 0.239 & 0.853 & 0.395 & 0.156 & (cmJ163337.8+571328) & 1.179 & (0.210) & (0.029) \\
194 & mfJ172845.5+743945 & 0.280 & 1.487 & 0.484 & 0.204 & (cmJ172946.3+744238) & 3.474 & (0.190)  & (0.090) \\
\hline
\end{tabular}
\end{table*}

\begin{figure*}
\centering
\includegraphics[width=140mm]{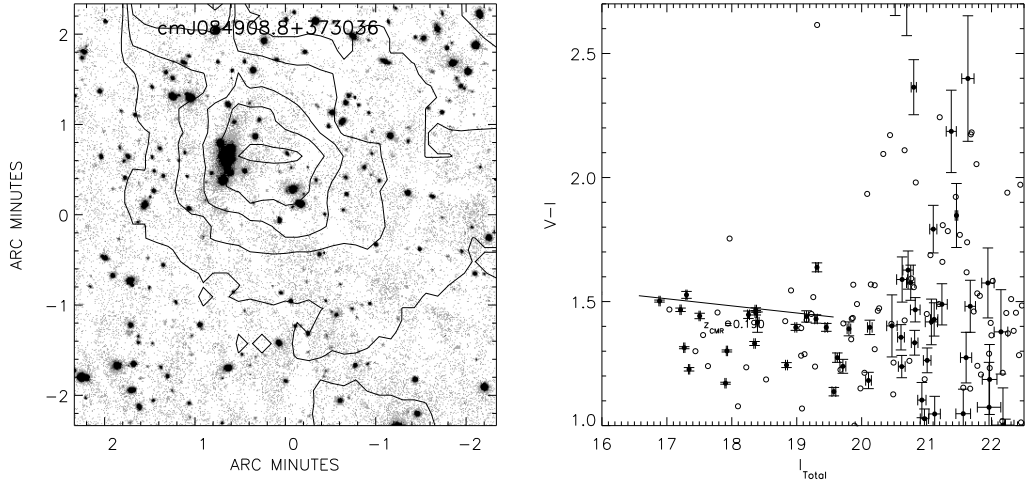}
\caption[X-ray and Optical Images with CMD for VMF62]{VMF62. Left
panel: WFC I-band image with PSPC contours overlaid (images are 5
arcmins on a side and PSPC contours have been smoothed to 30 arcsec -
the approximate PSF. These contours do not correspond to any particular
significance level, and are simply in units of X-ray count rate). Right
panel: CMD centred on VMF position. Filled circles show galaxies drawn
from within 1 arcmin radius of the CMR selected candidate position;
open symbols show galaxies drawn from an equal area, a further arcmin
away.  Note: these fixed angular radii are just chosen as a guide, and
are not the same as the radii chosen by the CMR algorithm (which is
colour dependant).  Photometric error bars are only shown for filled
points for clarity.  Overplotted line shows CMR corresponding to CMR
algorithm's estimated redshift (from M$^\star-1$ to M$^\star+2$).
VMF62 is found in both the final MF and CMR catalogues.  The associated
MF and CMR candidates are listed in Table \ref{table:opt_x_match}.  }
\label{fig:vmf62}
\end{figure*}

\begin{figure*}
\centering
\includegraphics[width=140mm]{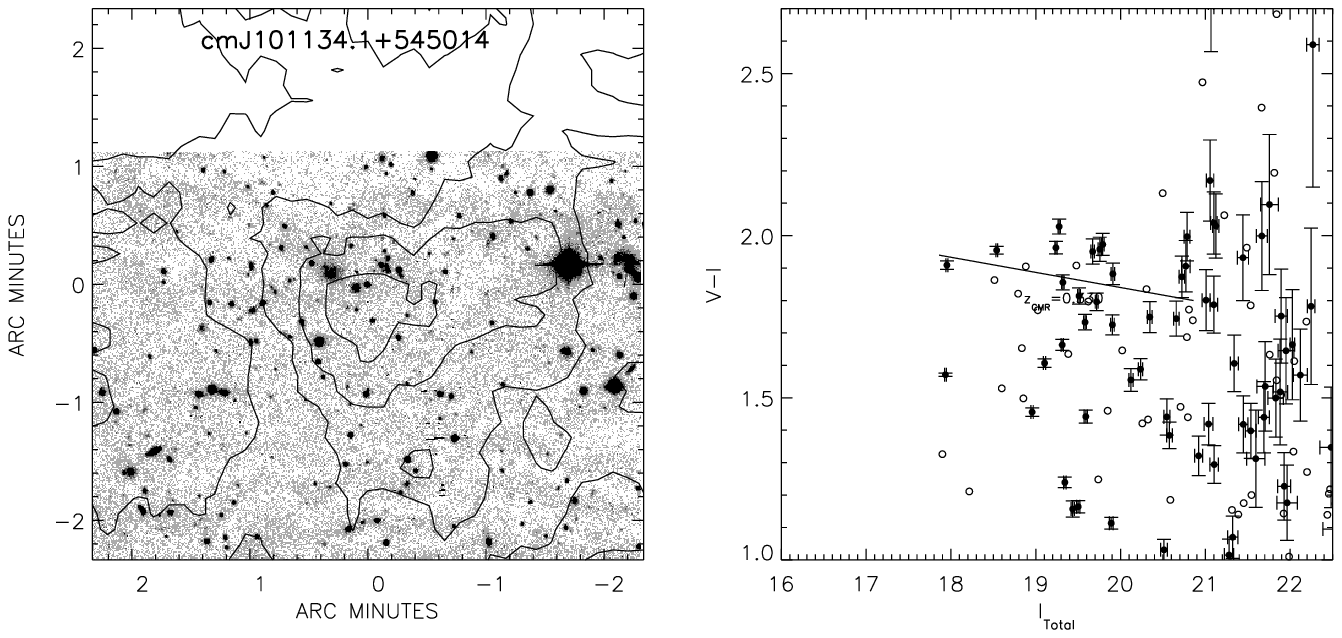}
\caption[X-ray and Optical Images with CMD for VMF86]{VMF86. Panels as
for Fig.~\ref{fig:vmf62}.

VMF86 appears to show a second overdensity even in the DSS image, and
possibly a weak second X-ray source, although this was undetected by
\citet{vmf98} (A. Vikhlinin, priv. comm.).  The second cluster centre
is just visible on the extreme right of the I-band image (at
$\approx$[-2.4,0.0] arcmin offset).  The CMD shows the hint of a second
CMR 0.4 magnitudes blueward of the main candidate, mainly indicated by
open symbols due to its distance from the primary candidate.  Thus,
this appears to be, in fact, two distinct systems at very different
redshifts.  Indeed, this system was flagged by the CMR algorithm as
being a system suffering from projection effects.  }
\label{fig:vmf86}
\end{figure*}

To quantitatively compare the optically selected catalogues with the
X-ray selected clusters, the following method was used. For each of the
final MF and CMR catalogues, cross-correlation with the \citet{vmf98}
catalogue was performed, retaining the nearest match to each X-ray
cluster. If the X-ray cluster lay within the optical candidate's group
radius, it was considered matched (the only caveat is that a minimum
radius of 1 arcmin and a maximum of 2 arcmin was adopted, to ignore
excessively large or small group radii, and also account for the
uncertainty in the X-ray position). These matches are tabulated in
Table \ref{table:opt_x_match}. For the X-ray clusters with no matches
from this process, the full catalogues for each algorithm were checked,
to see if a lower significance candidate is matched. Such matches are
indicated in the table by parentheses. All matches were then inspected
visually and special cases are commented on.  Typical data are shown in
Figs~\ref{fig:vmf62} \& \ref{fig:vmf86}.

\subsection{Summary of Optical Candidates Associated with X-ray Clusters}

In terms of gross numbers with this simplistic matching, the final MF
catalogue contains counterparts to 7 of the 15 X-ray selected clusters,
and the final CMR catalogue contains 8 of the 10 in the redshift range
probed by the algorithm.

In the MF catalogue: VMF11 is not matched in the final catalogue, but
is matched in the full catalogue (just below the Cash $C$ threshold).
VMF69 and 73 were both further than 2 arcmins away from the nearest
candidates, but lay within the candidates' group radii.  Thus, these
were treated cautiously, but visual inspection showed large
overdensities extending this far and thus the association of these
objects seems valid.  VMF74 and VMF132 were detected in the full
catalogue, but did not make the higher significance cut of the final
catalogue.  VMF84 was undetected as its redshift is too low (0.045).
VMF99 and 150 were also undetected. These fields do not show
overdensities of galaxies, and the clusters were possibly `confirmed'
using the single luminous elliptical criterion.  The redshifts given
are also below the expected detection range (both have z=0.11). VMF131
has a match at a distance of 1.9 arcmins: this is outside the
candidate's estimated radius of 0.5 arcmins, but again visual
inspection suggests the association is valid. To summarise, if the
three lowest redshift X-ray clusters (VMF84, 99, and 150) are excluded
(z$\leq$0.11), then the strict automated matching matches 6 of the 12
candidates. Visual matching suggests the recovery rate in the final MF
catalogue should be 9/ 12.  Lowering the significance threshold allows
the remaining X-ray clusters to {\it all} be detected (at the expense
of a higher spurious rate).

In the CMR catalogue: considering all except the 5 X-ray clusters at
z$\leq$0.2 (for the reasons described above), only two are not
immediately matched. VMF181 is matched in the full catalogue and is of
high enough significance ($>$6$\sigma$) to be in the final catalogue,
but was `cleaned' from the catalogue as it had a neighbour of higher
significance. VMF194 shows a counterpart some distance ($\approx$ 3.5
arcmin) from the X-ray cluster, but the X-ray position appears to be at
least an arc minute from a significant overdensity and CMR (the MF
candidate at this position was only matched because it has a large
associated radius). This cluster has been the subject of extensive
follow-up work by Vikhlinin and collaborators (A. Vikhlinin, priv.
comm.) and was difficult to confirm optically. It has very extended
X-ray emission ($\sim$2-3 arcmin) and the galaxy overdensity is
similarly extended.  Spectroscopic follow-up found 4 out of 5
ellipticals in this field at z$=$0.213, and so they consider the
cluster confirmed.  Note that this redshift is in much better agreement
with our CMR-estimated redshift than the initial photometric estimate
of \citet{vmf98}.

The CMR technique, furthermore, allows the possibility of
distinguishing groups projected along the line of sight (the entries in
Table \ref{table:opt_x_match} flagged with a `p'). VMF86 is
identified as two systems (as suspected from the data, illustrated in
Fig.~\ref{fig:vmf86}): the more significant at a redshift of
0.330 and another at 0.230. The quoted spectroscopic redshift of VMF is
0.294; this is within $\Delta$z$=$0.05 of the most significant
candidate. VMF132 also shows two possible further groups, overlapping
with the z$_{CMR}=$0.21 cluster, at higher redshift: z$_{CMR}$=0.37 and
0.65. Thus, to summarise, the CMR matches 8 of the 10 z$>$0.20 X-ray
clusters immediately, and visual matching allows all 10 clusters to be
matched. Another advantage of this method is that it is able to
disentangle projection effects: correctly resolving structure (which is
obvious visually in the CCD images) in one field, and suggesting higher
redshift groups in another.

\subsubsection{Comparison of Estimated Redshifts with VMF Redshifts}
\label{sec:comp-estim-redsh}

\begin{figure*}
\centering
\includegraphics[width=80mm]{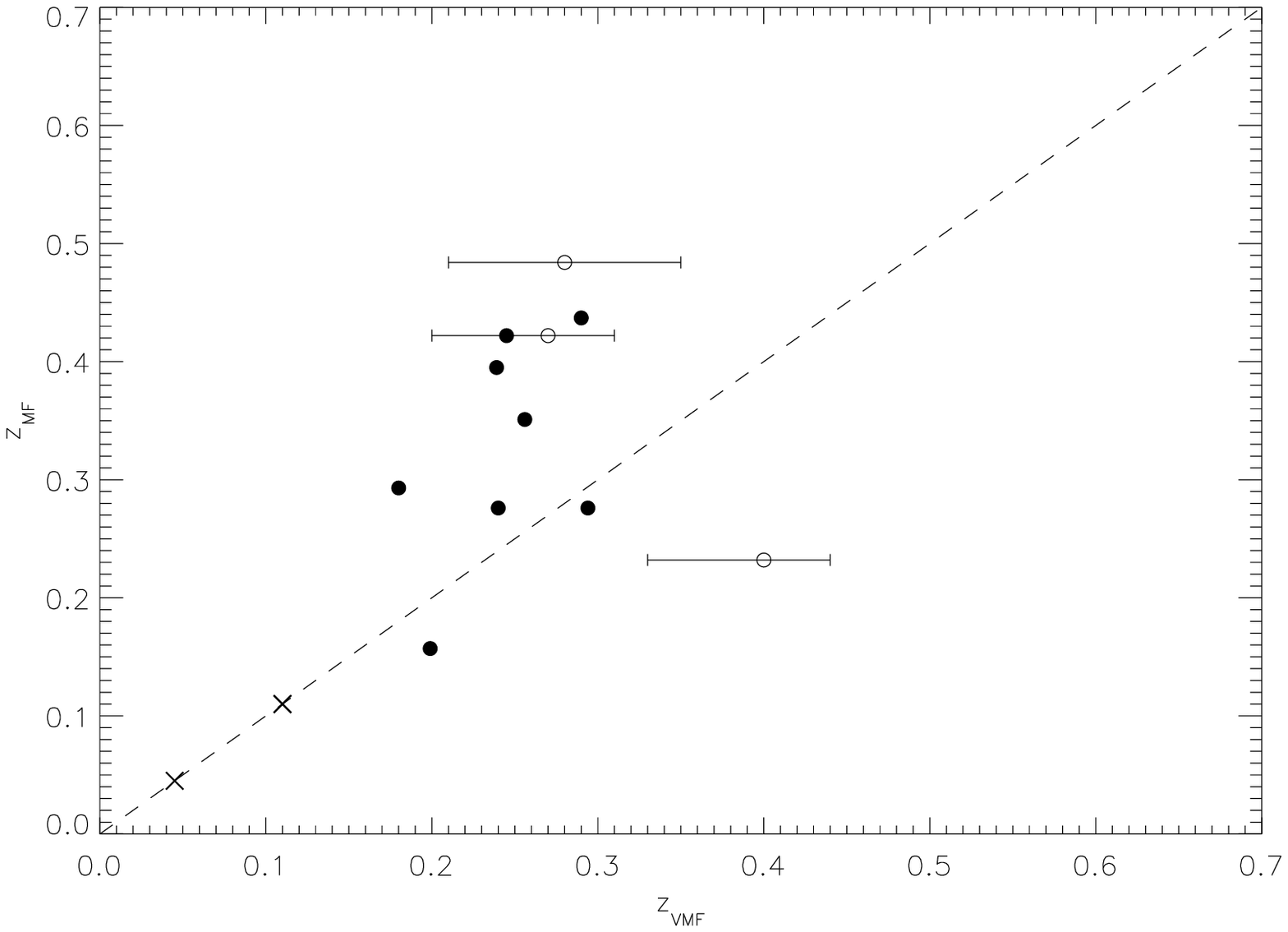}
\includegraphics[width=80mm]{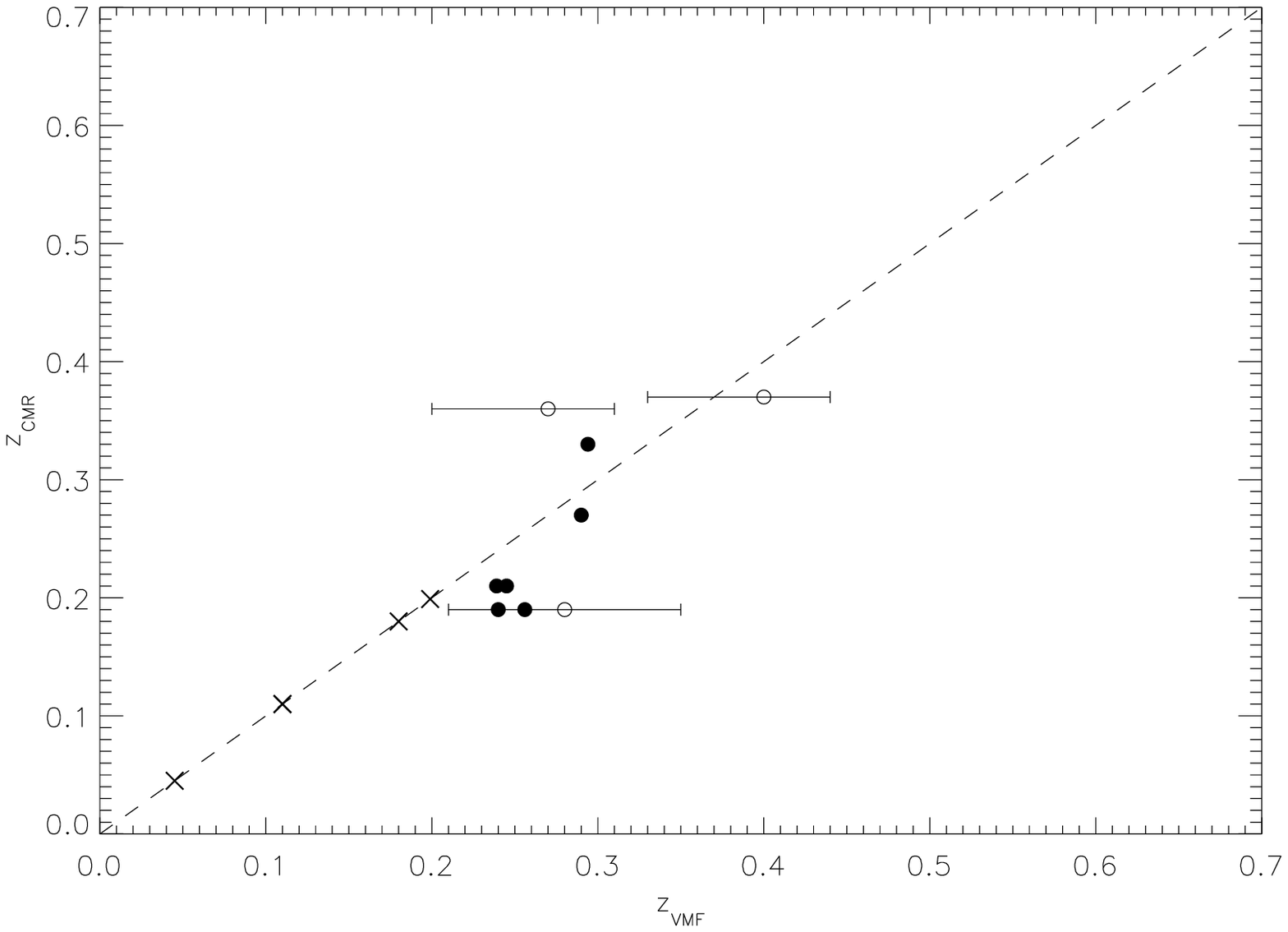}
\caption[Comparison of Redshift Estimates with VMF's
Redshifts]{Comparison of the optical cluster finders' redshift
estimates with Vikhlinin et al.'s redshifts.  Open symbols are for VMF
photometric redshifts with error bars showing their estimated range;
filled points are for spectroscopic redshifts. Dashed line shows the
one to one relation, and crosses on this line indicate undetected
clusters.  These numbers are tabulated in Fig.~\ref{table:opt_x_match}.  The CMR redshift appears to systematically
underestimate the VMF spectroscopic redshift by $\approx$0.03, in all
but one case.  This is for VMF86 which is in fact two systems, as
indicated earlier. The more significant candidate happens to be the
higher redshift one, but if VMF measured a redshift for the lower
redshift system, then this too would be underestimated by a similar
amount.}
\end{figure*}

The average bias in the redshift estimate, defined as (z$_{spec}$ -
z$_{phot}$)/(1 + z$_{spec}$), is 0.067 with a standard deviation of
0.066 for the MF (using 8 spectroscopic redshifts from VMF); the
average bias for the CMR technique is -0.022 with a standard deviation
of 0.028 (from 6 published VMF spectroscopic redshifts). The latter
result compares very favourably with photometric redshifts.
\citet{wittman} find an average bias of -0.027 with standard deviation
of 0.059 for photometric redshifts over a similar range using {\it
  four} photometric passbands. Gladders \& Yee (2000) commented that
redshift determination should include a step to renormalise the stellar
population models to the data with redshifts.  This corrects systematic
offsets such as mismatches between the model and real filters.  Any
remaining 'bias' would be due to photometric calibration errors,
assuming universality of the CMR.

\subsection{Comparison of MF Catalogue with CMR Catalogue}

\begin{figure}
\centering
\includegraphics[width=80mm]{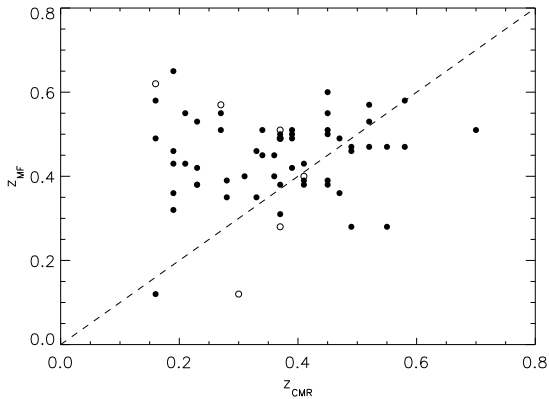}
\caption[Comparison of MF and CMR Estimated Reshifts]{Comparison of MF
and CMR estimated redshifts. Filled points are unique matches; open
points are flagged as line of sight projections in the CMR
catalogue. Dashed line is the one to one relation.
}
\label{fig:zcmrmf}
\end{figure}

Now that both optical techniques have been compared with a
spectroscopic sample, it can be seen that the estimated redshifts from
the CMR technique offer greater precision than the MF estimated
redshifts. Thus, a cross-comparison of the two techniques can be made,
using the CMR catalogue as a reference. The final MF catalogue was
cross-correlated with the final CMR catalogue to determine cluster
candidates in common. To avoid possible confusion from multiple
associations of candidates, only MF candidates with a {\it single} CMR
candidate within the former's radius were considered.  If the
candidates' centres were separated by more than two arc minutes, they
were excluded. Thus, only secure `clean' matches are considered.  Of
the 185 final MF candidates, 62 show unique CMR matches (7 of these are
flagged as line of sight group projections). A comparison between the
estimated redshifts of these techniques is shown in
Fig.~\ref{fig:zcmrmf}. The average bias and scatter (as defined
earlier) in this relation are -0.066 and 0.106 respectively, although
inspection shows that this may equally be due to the MF redshift being
randomly drawn from values between 0.3 and 0.5. This may be due to bias
in the candidates selected for this comparison (those with `clean'
matches between the CMR and MF candidates).  A comparison of the MF
estimated redshifts with those of spectroscopically determined
redshifts (from the X-ray selected clusters in the next section) shows
that the MF estimated redshift is not as bad as this.

A further comparison is to consider the MF candidates {\it not} matched
with CMR candidates. This was done by comparing the full CMR catalogue
with the final MF catalogue and searching for MF candidates with {\it
  no} CMR matches within their radii (or 2 arcmins). 41 of the final MF
candidates show no CMR counterparts at any significance level.  Under
the assumption that {\it all} genuine clusters possess a CMR and that
this technique will find them, this can be used as an approximation to
the number of spurious MF detections. This gives a false detection rate
of around 22\%. This is in general agreement with estimates for other
MF techniques of around 30\% \citep[e.g.][]{cop99}. 22\% is a lower
limit, as some of the CMR matches are of low significance.  Assuming a
spurious CMR rate of around 5$-$10\% \citep[as seems more likely,
e.g.][]{gy00,mythesis} would bring the false positive rate of the MF into
closer agreement with the 30\% value.  Using the fraction of matched
candidates flagged as projections (7 out of 62, above) compares well
with the (spectroscopic) findings of \citet{katgert96} that around 10\%
of Abell clusters comprise two or more significant clusters, projected
along the line of sight.

\subsection{Comparison of Optical Richness Measures with \lx}

Henceforth we consider only the catalogue generated by the CMR
technique.  We have just shown through comparison with the
spectroscopically confirmed sample of \citet{vmf98} and from internal
comparisons within our data the superiority in terms of accuracy and
reliability of the CMR algorithm over the MF approach.

The following procedure was used to measure the X-ray flux for each
optical cluster candidate.

The relationship between each of the richness measures and X-ray
luminosity for cluster candidates in the optically selected catalogues
will now be presented. Since the vast majority of the optically
selected clusters have no X-ray selected counterparts, fluxes/ flux
limit were measured at the positions of the optical candidates. 

\begin{figure}
\includegraphics[width=80mm]{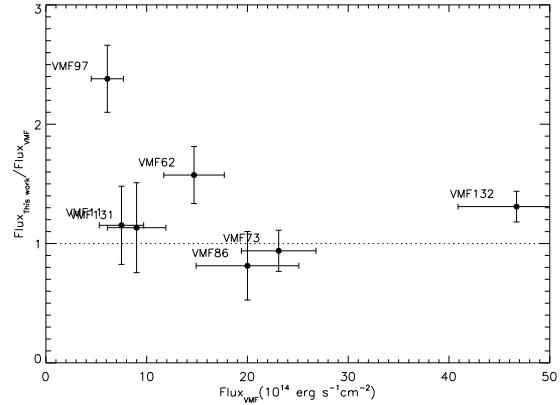}
\caption{Comparison of our aperture X-ray fluxes with the wavelet
  fluxes of VMF98 for clusters in common with their sample.  VMF IDs
  are indicated.}
\label{fig:vmflux}
\end{figure}

Aperture fluxes were measured from the X-ray images. The peak X-ray
flux within a 1.5 arcmin search radius was located, to provide the
centre of the measurement. A 1.5 arcmin radius aperture was used. This
aperture flux was then corrected to a total flux computing the ratio of
total flux to flux within the 1.5 arcmin radius aperture for a cluster
beta model of typical parameters. This method is a compromise between
choosing too large an aperture and increasing the chance of background
contamination from point sources, and choosing too small an aperture
and missing cluster flux. This aperture was found to give the best
agreement with the wavelet X-ray flux measured by \citet{vmf98}
(Fig~\ref{fig:vmflux}).  Also tried was the use of a standard maximum
likelihood technique to fit a cluster beta model to the X-ray surface
brightness profile, but this gave largely poor fits, as the systems we
detect are not well-matched to the standard model assumptions used
(such as spherical symmetry).

The ROSAT count rate in the 0.5$-$2.0 keV band was converted to a
bolometric luminosity correcting for the observed Galactic hydrogen
column density, assuming the CMR estimated redshift and assuming a
cluster temperature of 5 keV.  The luminosity was then iterated once by
assuming a fit to the X-ray luminosity temperature relation given by
$T_x = 2.75 ({L_{bol}}_X/h_{50}^2)^{0.357}$ \citep[a compendium
from][]{1999ApJ...516....1R,1999MNRAS.305..631A,
  1998ApJ...504...27M,1998MNRAS.297L..63A}.  This has little effect on
the resulting luminosity.  The background flux rate was determined for
each object by the maximum likelihood method in which it was a free
fitting parameter. The background flux rate was also measured for each
field by placing 100 apertures randomly around each image and measuring
fluxes in the same way. 3$\sigma$ outliers were rejected from these
estimates to obtain the median and variance. The background values
obtained by both methods agreed well with each other. The significance
of each optical candidate X-ray flux measurement was determined
relative to the variance of the background flux measurements in each
field.  If the measurement was a greater than 3$\sigma$ event, this was
classed as a detection.  For other measurements, a 3$\sigma$ upper
limit was found from the limiting flux for the field.  The typical
3$\sigma$ limiting flux is around 6$\times$10$^{-14}$erg s$^{-1}$
cm$^{-2}$.  X-ray detections at $\geq$3$\sigma$ were visually
inspected, and those showing contamination from an obvious bright point
source were rejected from the analysis.  40 of the 290 apertures
resulted in $\geq$3$\sigma$ X-ray detections.

A plot of each of the richness measures described in
\S\ref{sec:richness} versus the X-ray luminosity is given in
Figs.~\ref{fig:lx_lell} and \ref{fig:lx_bgc}.

Firstly we consider the relation between the total early type galaxy
luminosity, L$_E$ and X-ray luminosity, L$_X$. The two quantities
appear to be correlated, but with large intrinsic scatter.  One concern
about plotting two measures of luminosity against each other is the
fact that the two are correlated through the distance to each object.
In order to show that this is not producing the observed correlation,
Fig.~\ref{fig:lx_lell_z} shows the ratio of the two luminosities as a
function of redshift.  This distance-independent measurement shows that
there is still an intrinsic scatter in the luminosities at each
redshift interval.  We use a Bayesian maximum-likelihood technique to
fit the relations shown in Fig.s~\ref{fig:lx_lell}, \ref{fig:lx_bgc}.
We assume that there is a power-law correlation between $L_e$ and $L_X$
with a large intrinsic scatter ($\sigma$), which we assume to be
Gaussian in $\log\L_X$, and is independent of $L_e$. There are three
model parameters that must be determined: the normalisation and
exponent of the mean relation, and the scatter about the relation. The
statistical problem is an unusual one because the upper limits far
outnumber the detections.  Furthermore, the scatter in the relation is
much larger than the errors in the measurements. To simplify the
fitting procedure, we therefore treat the measurement errors as a
component of the scatter in the model relation.

\begin{figure*}
\includegraphics[width=160mm]{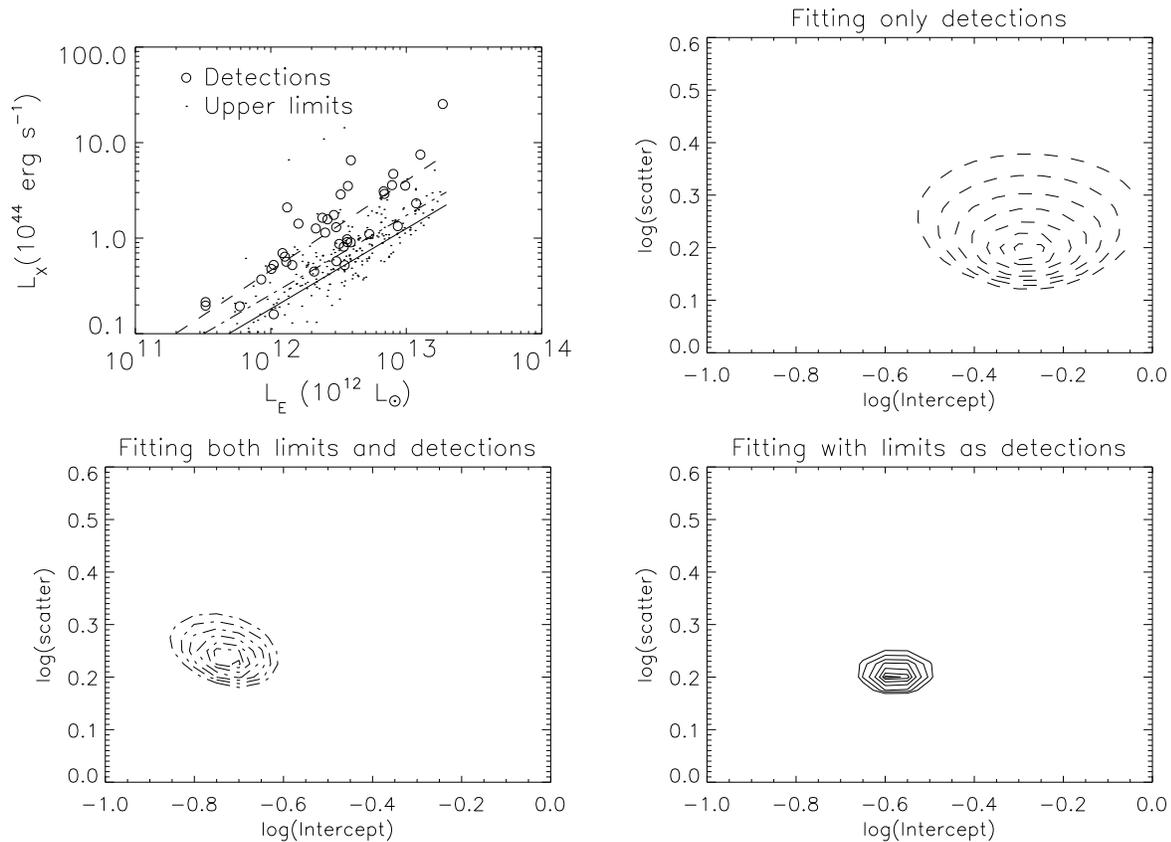}
\caption{Results of different fits to the \lx -- L$_E$ relation.  The data
  are shown in the upper left panel with detections as open circles
  and upper limits as points, for clarity.  Overplotted lines show the
  different fits in the same line style as the contour plots.  Contours
  show confidence limits for the fitted scatter and intercept (at
  L$_E=$10$^{12}L_\odot$) -- upper right panel: fitting only the
  detections; lower left panel: fit to upper limits and detections,
  using Bayesian approach described in text; lower right panel: all
  points fitted assuming upper limits are detections (this is
  essentially the same as assuming detections would lie just below the
  limits).  Confidence levels run from 1-$\sigma$ to 7-$\sigma$ in
  1-$\sigma$ intervals. }
\label{fig:fconts}
\end{figure*}

\begin{figure*}
\centering
\includegraphics[width=160mm]{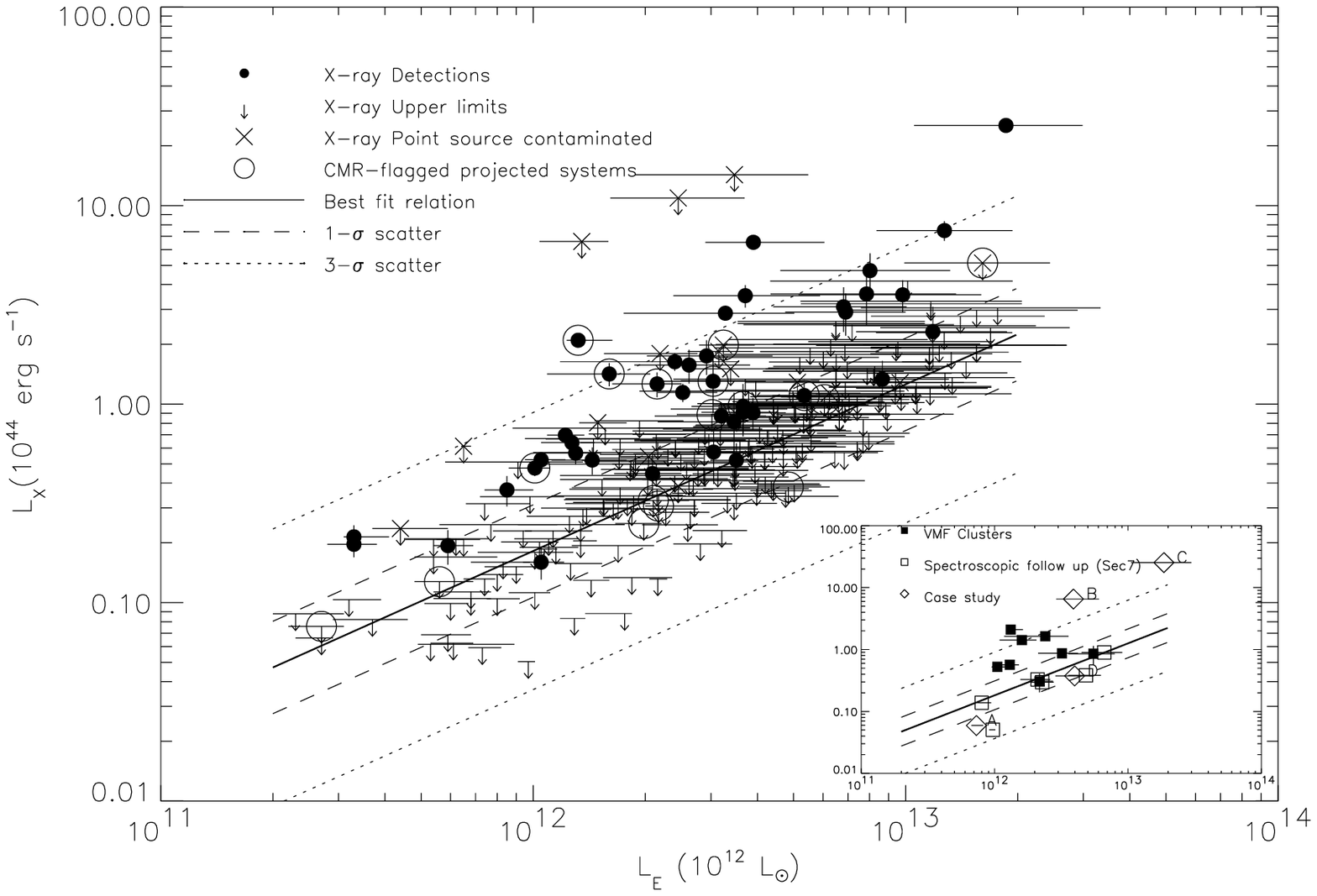}
\caption{X-ray
  luminosity vs the L$_{E}$ richness measure. Filled points are
  $>$3$\sigma$ X-ray detections; downward arrows are 3$\sigma$ upper
  limits. Solid line is the best fit relation of the detections and
  limits using the Bayesian technique described in the text.  Inset
  shows specific systems which are further discussed in text.}
\label{fig:lx_lell}
\end{figure*}

\begin{figure*}
\centering
\includegraphics[width=160mm]{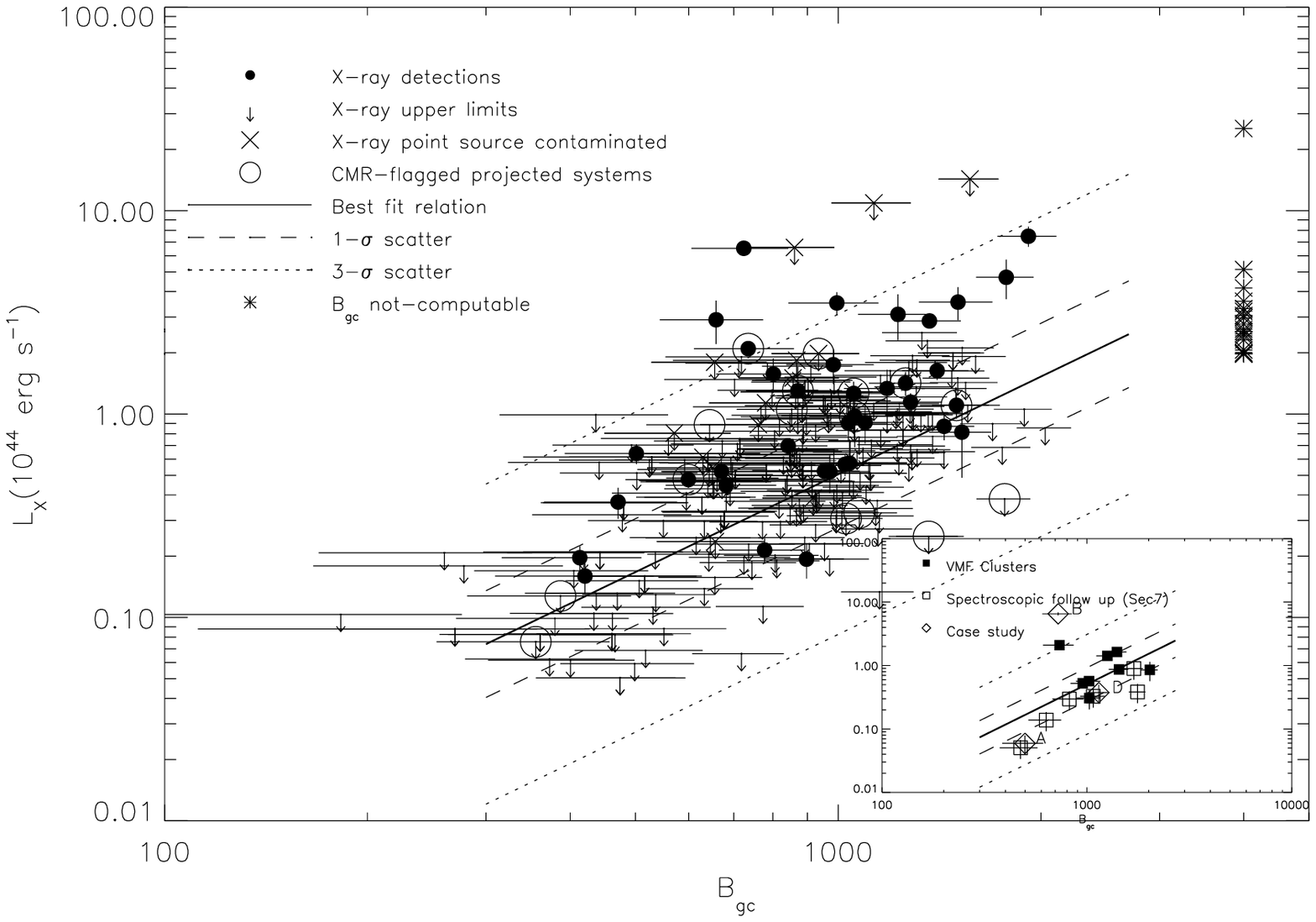}
\caption{X-ray
  luminosity vs the B$_{gc}$ richness measure. Filled points are
  $>$3$\sigma$ X-ray detections; downward arrows are 3$\sigma$ upper
  limits. Solid line is the best fit relation of the detections and
  limits using the Bayesian technique described in the text.  Inset
  shows specific systems which are further discussed in text (note:
  a\bgc value could not be computed for case study C).}
\label{fig:lx_bgc}
\end{figure*}

\begin{figure}
\includegraphics[width=80mm]{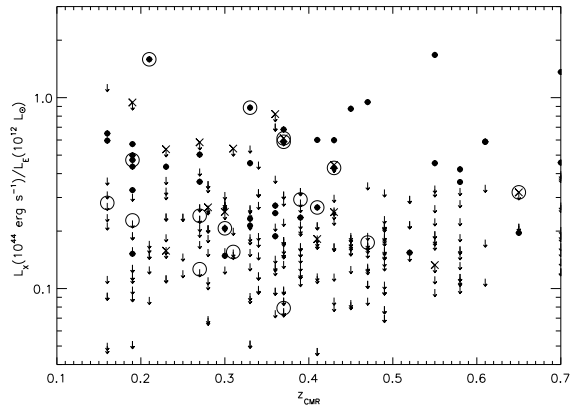}
\caption{X-ray luminosity / L$_{E}$ as a function of CMR-estimated
  redshift. Symbols as for Fig.~\ref{fig:lx_lell}.
} 
\label{fig:lx_lell_z}
\end{figure}

We tried three methods of fitting the data. Firstly, we can use only
the X-ray detections. This gives \lx $=10^{-0.338} L_E~^{0.94}$,
but this model relation fails to take any account of the large number
of upper limits. Secondly, we can incorporate the upper limits by
treating the limit as a data point.  We eliminate the data points
flagged as contaminated by point sources from this fit. This gives a
relation that has lower normalisation and larger scatter. Our ML
estimator gives similar results to conventional regression schemes in
both these cases.  However, while the second approach takes into
account the upper limits, it does not allow for the possibility that
the actual $L_x$ might lie significantly below the upper limit.
Finally, we apply our maximum-likelihood estimator, treating the upper
limits correctly.  We must make two Bayesian assumptions to apply the
method: (1) we assume that the probability distribution for the value
underlying an upper limit is uniform in $L_x$ (rather than $\log L_x$);
(2) we allocate greater prior weight to models with small scatter by
introducing a prior weighting $P_{prior}$($\sigma$) $\propto$
$1\over\sigma$. This is appropriate to values bounded between 0 and
$\infty$ and gives equal probability per decade in $\sigma$. Without this
assumption (and since upper limits greatly outnumber detections), the
likelihood estimator gives too much weight to models with very high
scatter and very low normalisation. This results in the relation \lx
$=10^{-0.740} L_E~^{0.84}$, with a scatter of $10^{0.23}$. The
effects of the different fitting techniques are illustrated in
Fig.~\ref{fig:fconts}.  A similar method for the \bgc parameter
results in the relation \lx $=10^{-5.08} B_{gc}~^{1.59}$, again with a
scatter of $10^{0.23}$, fitting both detections and limits with this
Bayesian approach.

In addition, the richness measure N$_{0.5}$ \citep{bahcall81} similar
to the Abell Richness Class, was studied and found to give an unusably
large scatter.  \citet{ylc99} also found this measure gave unacceptable
scatter.

\begin{figure*}
\centering
\includegraphics[width=140mm]{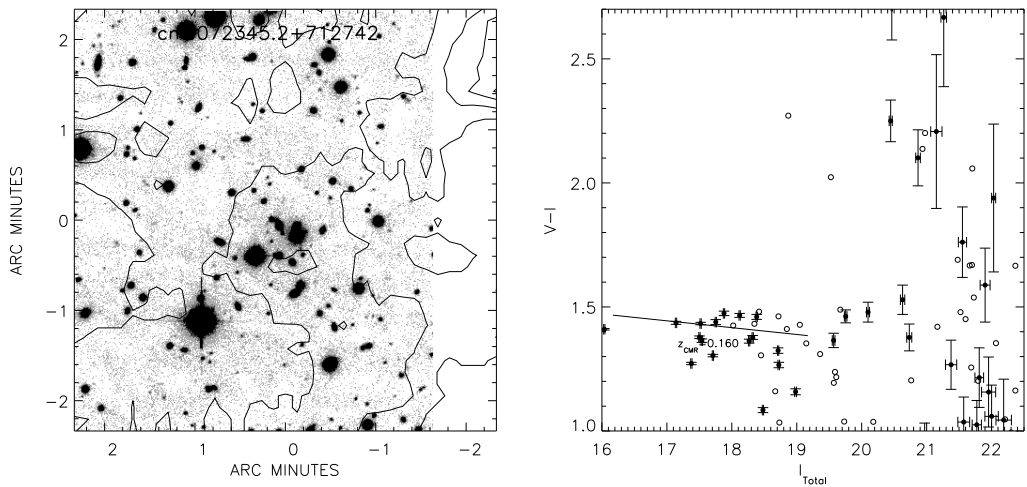}
\caption{Plots as for Fig.~\ref{fig:vmf62} for case study A (candidate
cmJ072345.2+712742). Solid line in CMD indicates model CMR for estimated
redshift.  z$_{CMR}=$0.16. This system is X-ray underluminous, and of
reasonably low optical luminosity.
}
\label{fig:caseA}
\end{figure*}

\begin{figure*}
\centering
\includegraphics[width=140mm]{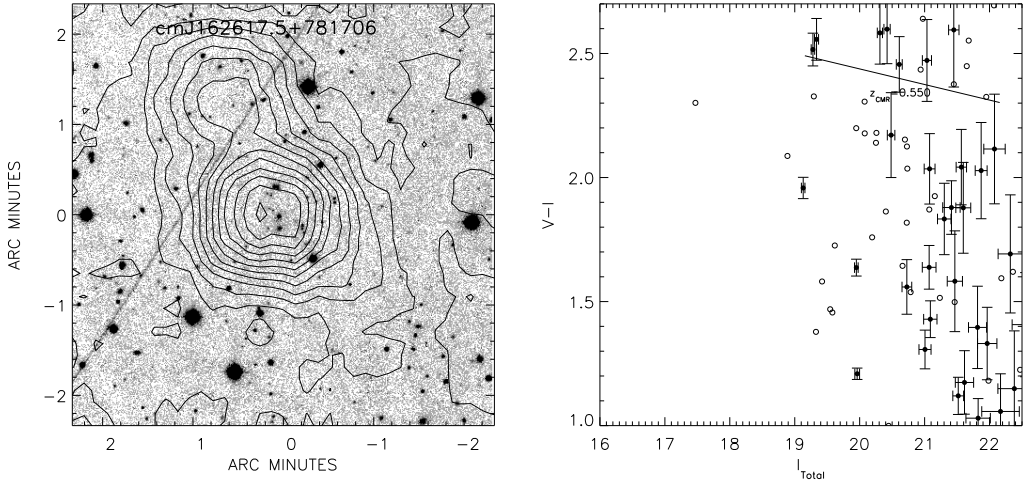}
\caption{Plots as for Fig.~\ref{fig:vmf62} for case study B (candidate
cmJ162617.5+781706). z$_{CMR}=$0.55.  This is a fairly optically
luminous system with a very high X-ray luminosity for its richness.}
\label{fig:caseB}
\end{figure*}

\begin{figure*}
\centering
\includegraphics[width=140mm]{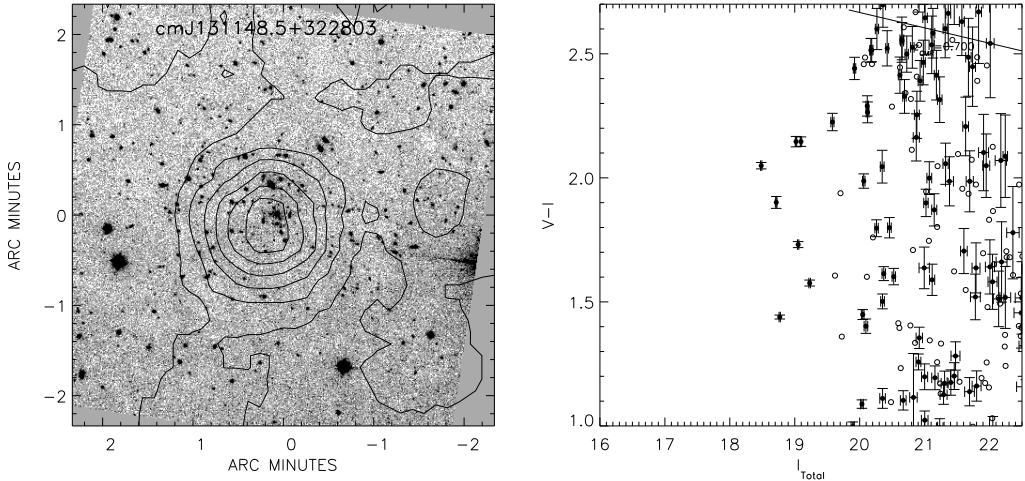}
\caption{Plots as for Fig.~\ref{fig:vmf62} for case study C
  (candidate cmJ162617.5+781706). z$_{CMR}=$0.70.  This is a high
  redshift, optically rich candidate with very high X-ray luminosity. 
}
\label{fig:caseC}
\end{figure*}

\begin{figure*}
\centering
\includegraphics[width=140mm]{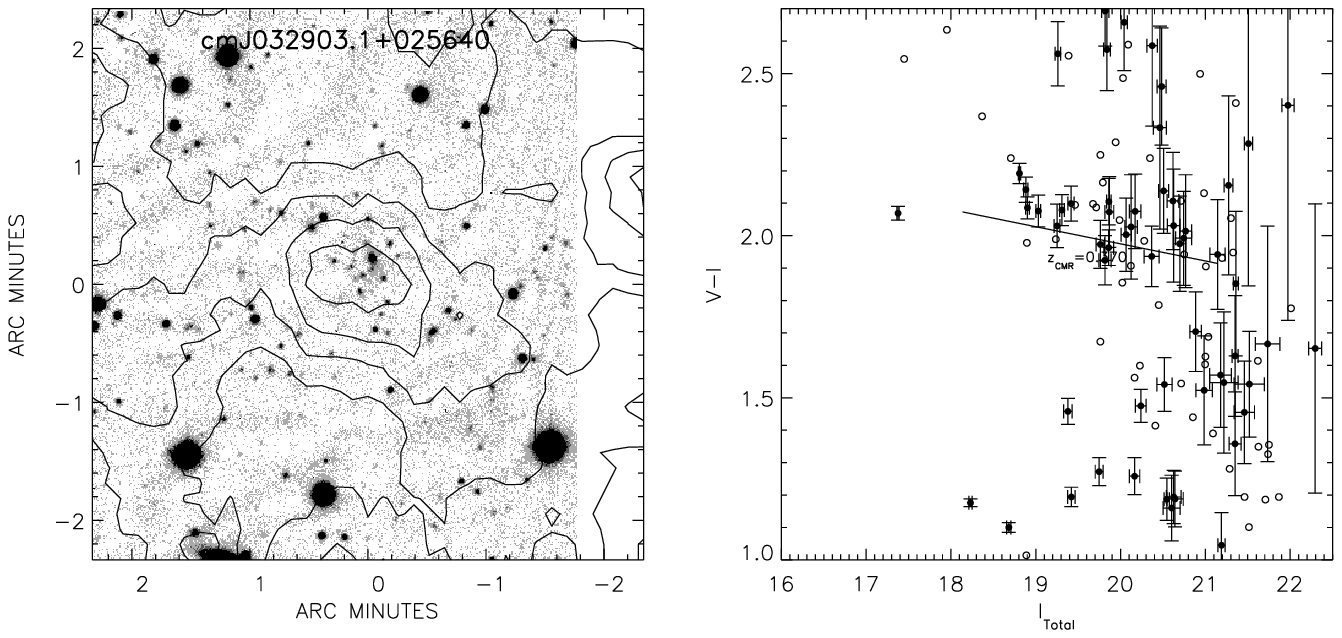}
\caption{Plots as for Fig.~\ref{fig:vmf62} for case study D
  (candidate cmJ032903.1+025640).  z$_{CMR}=$0.37.  This is a candidate
  of similar optical richness to case B (at slightly lower redshift;
  0.37 vs 0.55) but with an order of magnitude lower X-ray luminosity.
}
\label{fig:caseD}
\end{figure*}

Examples of outliers in these relations are now considered, to assess
if clusters with similar richnesses do indeed exhibit very different
X-ray luminosities. The candidates listed in Table~\ref{table:cases}
are chosen as obvious outliers in the \lx - L$_E$ plot.  For brevity,
the candidates are referred to as A, B, C, D. For each of these, an
I-band image with X-ray contours overlaid and a CMD with the fitted CMR
overlaid is shown (Figs~\ref{fig:caseA}-\ref{fig:caseD}), to verify the
properties which locate them in the \lx -- L$_E$ plot. Candidate A is
one of the least X-ray luminous candidates, but is also not very
optically rich. Candidate B is highly X-ray luminous and of moderately
high optical luminosity, laying just above the 3 $\sigma$ upper bound
of the relation.  Candidate C is the most X-ray luminous and optically
luminous candidate; and candidate D has a comparable richness to
candidate B, but is an order of magnitude less X-ray luminous.

The two high \lx~systems (B and C) indicated have very different
optical richnesses. Due to the volume probed and the rarity of such
luminous clusters, these systems are expected to be at the high
redshift end of the survey (the volume between 0.2$<$z$<$0.5 is similar
to that between 0.5$<$z$<$0.7) and, indeed, they are both found to lie
in the range 0.5$<$z$<$0.7. Conversely, in order to be detected, the
faintest systems must lie at low redshift (A). The fairly
optically rich system with a low X-ray upper limit (D) lies
intermediate in redshift to these extremes.

From these figures (\ref{fig:caseA} -- \ref{fig:caseD}), the example
cluster candidates appear to have the properties measured in the
catalogues and shown in Table \ref{table:cases}. For example, the X-ray
detections do not appear contaminated by point sources, and the
redshift estimates seem entirely consistent with the predicted CMRs.

Cases B and D have similar L$_E$ values, but their X-ray
luminosities vary by over an order of magnitude. Examining the
relation using \bgc instead of \lell shows that both these systems 
occupy similar regions of this plot.  

\begin{table}
\centering
\caption{Table of properties for interesting outliers from \lx - \lell
  relation. Letter preceding candidate ID is for brevity when
  discussing these case studies in the text. 
}
\label{table:cases}
\begin{tabular}{lcccc}
\hline
Candidate ID &\lx $^{44}$&L$_E^{12}$   &\bgc   &z$_{est}$ \\
\hline
A -- cmJ072345.2+712742    &$\leq$0.06  & 0.7         & 500   & 0.16\\
B -- cmJ162617.5+781706    &6.5        & 3.9         & 720   & 0.55\\
C -- cmJ131148.5+322803    &25.3       & 18.6        & ---   & 0.70\\
D -- cmJ032903.1+025640    &$\leq$0.4   & 4.0         & 1100  & 0.37\\
\hline
\end{tabular}
\end{table}

The scatter in the relation could be attributable to a number of
factors.  The physical processes involved in the determination of
\lx~have been discussed in \S\ref{sec:introduction}, but will be
reiterated here, along with a discussion of $L_E$.  The X-ray
luminosity is dependent on the temperature and density of the gas.
These in turn depend on the dynamical state of the cluster (which
determines the depth of the gravitational potential, and the densities
that can be reached by the gas). The presence of a cooling flow
increases the luminosity by increasing the gas density. \lx~can also be
increased through unresolved point source contamination.

The optical properties of the cluster candidates obviously depend on
the properties of the member galaxies.  Since these systems were
selected on the presence of a CMR, a population of galaxies which
formed their stars at high (z$>$2) redshift, and terminated star
formation shortly after, is required. 

The interplay between the intracluster medium and the cluster galaxies
is likely to be important and not straightforward to model.  Several
workers \citep[e.g. ][]{pcn,bowerbenson} have recently investigated
such interplay using numerical simulations.  They propose energy
injection at early times from supernovae in cluster galaxies as a
method to `pre-heat' the ICM and produce observed relations such as the
\lx - \tx relation. It is then quite likely that the scatter comes from
variation in the X-ray luminosity rather than the mass to galaxy
luminosity conversion.

\section{Spectroscopic Observations of X-ray Dark Cluster Candidates}
\label{sec:spectr-observ-x}

In order to confirm the reality of candidate cluster systems from the
XDCS, follow-up multi-object spectroscopy was undertaken with the
MOSCA\footnote{ {\tt http://www.mpia-hd.mpg.de/MOSCA/index.html} }
instrument on the Calar Alto 3.5m telescope. MOSCA is a focal reducing
spetrograph, installed at the Richey-Chretien Focus of the 3.5m
telescope on Calar Alto. The reduction ratio of the optical system is
3.7, i.e. the effective focal ratio is f/2.7.  This gives an image
scale of 3 pixels per arcsec and a total FOV of 11x11 arcmin. A thinned
CCD with 2048x4096 15 micron pixels is used as the detector. The
med-green grism was selected.  This gives a wavelength coverage of 4300
- 8200\AA, with a central wavelength of 5500\AA~and a dispersion of
around 2.5\AA/ pixel and resolution of around 10\AA\ FWHM.  This allows
distinctive spectral features to be seen over a wide range of redshifts
from z$\lsim$0.1 to z$\gsim$0.6.

Cluster candidates with no extended X-ray counterpart, detected in the
\citet{vmf98} catalogue fields were chosen.  The only
other criterion applied was that the RA range available for the
observing run meant that the candidates had to come from the subset of
XDCS originally observed during the June 1998 run. This subsample
comprises 16 fields, or approximately 4.5 square degrees. These targets
are listed in Table \ref{table:coords}. For each one, astrometry was
performed using the {\sc starlink} program {\sc astrom} to convert
pixel coordinates into sky coordinates as measured from the APM
catalogue\footnote{ {\tt http://www.ast.cam.ac.uk/$^\sim$mike/apmcat/
  }}, to an {\it rms} accuracy of of $\approx$0.3 arc seconds.
Multi-object slit masks were constructed using a constant slit width of
1.5 arcsec and a slit length of at least 10 arcsec.  In order to be
included as potential slit candidates, galaxies had to be brighter than
I$_C$=20 (often a few galaxies fainter than this were allocated slits
to fill the masks).

\begin{table}
\caption[MOSCA mask centres]{MOSCA mask centres.  Note: the IDs
 are just numbered subfields of the RIXOS fields, and should not be
 confused with the similar nomenclature used for X-ray candidates by
 the RIXOS collaboration.}
\label{table:coords}
\bigskip
\centerline{
\begin{tabular}{lcccc}
\hline
Candidate & $\alpha$ (J2000) & $\delta$ (J2000) \\
\hline
R110\_1 & 14 28 22.0 & +33 07 13 \\
R220\_2 & 17 23 37.9  & +74 43 17 \\
R236\_1 & 17 02 58.9  & +51 53 52 \\
R294\_1 & 23 19 54.5  & +12 32 27 \\
\hline
\end{tabular}
}
\end{table}

\begin{table*}
\caption[XDCS CMR candidates in MOSCA Fields]{XDCS CMR candidates in
MOSCA Fields.  Dist indicates distance from field centre. $^\ddag$ -
(p) in estimated redshift indicates that candidate is flagged as a
line-of-sight projection by the method described in
\S\ref{sec:constr-final-clust}. X-ray fluxes and luminosities (using
estimated redshifts) are 3$\sigma$ upper limits for all candidates,
none of which is detected in X-rays. 
}
\label{table:cmr_mosca}
\bigskip
\centerline{
\begin{tabular}{cccccc}
\hline
Field  & XDCS ID & Dist (arcmin) & $z_{est}$ $^\ddag$ & F$_X$ & L$_X$ \\
       &         &               &                    & (10$^{-14}$erg
       s$^{-1}$ cm$^{-2}$) & (10$^{44}$erg s$^{-1}$) \\
\hline
R1101  & cmJ142812.0+330736  &    2.1   & 0.160    & $\leq$3.0 & $\leq$0.02\\
R2202  & cmJ172333.0+744410  &   1.0  &  0.210    & $\leq$4.5 & $\leq$0.06\\
R2361  & cmJ170244.2+515539 &    2.9 &   0.310 (p) & $\leq$4.6 & $\leq$0.32 \\
R2361  & cmJ170232.6+514922 &     6.1  &  0.300    & $\leq$5.1 & $\leq$0.29 \\
R2361  & cmJ170258.9+514921 &     4.5  &  0.470    & $\leq$3.4 & $\leq$0.90 \\
R2941  & cmJ231951.2+123208 &    0.9 &   0.370 (p)& $\leq$2.0 & $\leq$ 0.08 \\
\hline
\end{tabular}
}
\end{table*}

\subsection{Spectroscopic Observations and Data Reduction}
\label{sec:mosca_data}
The spectra were secured over six nights of observations in July 2000
using MOSCA on the Calar Alto 3.5m. A log of the observations is
presented in Table~\ref{table:mosca_log}.  

\begin{table}
\centering
\caption[Calar Alto Observing Log]{Log of observations of cluster candidates from Calar Alto.}
\label{table:mosca_log}
\begin{tabular}{lccc}
\hline
\small
Night & Field & Mask & Exposure time (s) \\
\hline
$27-28/07/00$ & R110\_1 & 1 & 3 $\times$ 1800 \\
$28-29/07/00$ & R236\_1 & 1 & 3 $\times$ 1800 \\
$29-30/07/00$ & R220\_2 & 2 & 2 $\times$ 1800 \\
$29-30/07/00$ & R236\_1 & 2 & 3 $\times$ 1800 \\
$30-31/07/00$ & R294\_1 & 1 & 3 $\times$ 1800 \\
$31/07-01/08/00$ & R294\_1 & 2 & 3 $\times$ 1800  \\
\hline
\end{tabular}
\end{table}
\vspace*{0.5cm}

Before each night of observation, a series of bias frames (typically
five) was taken.  For the purpose of wavelength calibration (WLC) two
different comparison arcs were observed.  This was done to ensure an
adequate number of emission lines over the full-wavelength range
covered by the MOSCA med-green grism. A 15s exposure of the HgAr/Ne arc
was taken.  This was augmented by a 120s exposure of the Ar lamp using
the 472/78 filter.  The long exposure was used to make weak emission
lines clearly visible, and the (BV-band) filter was used to suppress
lines at the red end of the spectrum, which would otherwise become
saturated.  A combination of the HgAr/Ne spectrum and 100$\times$ the
Ar spectrum was found to provide a good reference spectrum for WLC
(hereafter, WLC refers to this composite arc spectrum image).  Such
calibration frames were taken before twilight, at the start of each
run.  Flatfield frames were taken using MOSCA's internal tungsten lamp.
For the science observations, three exposures were made of each mask,
each of 30 minutes duration.  After each series of science frames,
whilst the telescope was still pointing at the object, another HgAr/Ne
frame was taken (although it was not needed, see later).  This was done
in case flexure in the instrument due to the telescope's different
pointing position affected the arcs taken earlier in the evening (while
the telescope was parked, and therefore pointing at zenith). The Ar arc
was not repeated during the night as its longer exposure time added an
unacceptable overhead.

\subsection{Data Reduction}
Data reduction was carried out using standard {\tt IRAF} routines.
Master biases were created for each night by combining several bias
images.  Full 2D bias removal was necessary as the bias frames
displayed banded structure. Flatfielding was attempted but found to
offer no improvement in the identification of spectral features and so
was omitted. The three science exposures of each image were averaged to
reject cosmic rays. The spectra were then extracted using {\tt apall}.
Wavelength calibration was carried out using the composite arcs. No flux
calibration was performed.

\subsection{Spectroscopic Redshift Determination}
The Fourier cross-correlation technique of \citet{td79} was
applied to the wavelength-calibrated spectra.  This technique
continuum-subtracts and Fourier transforms the galaxy spectrum and a
reference template, applies high- and low-pass filtering, and looks for
peaks in the cross-correlation function of the two.

The template used was a de-redshifted E/S0 (as used by the CNOC
collaboration, courtesy of E. Ellingson), hence no emission lines were
used in the initial redshift determination (emission line objects are
considered below).  Each mask was run through {\tt fxcor}, and the
redshift from the highest cross-correlation peak logged.  Each spectrum
was then de-redshifted using the {\tt fxcor} redshift.  The
de-redshifted spectrum then had the position of prominent absorption
lines (Ca H+K, G-band, H-beta and Mgb) and emission lines OII and
H$\beta$ overplotted.  The spectrum was then visually inspected and a
quality flag assigned to it, either: 2 - the redshift is confident; 1 -
the redshift is less certain but looks compatible with the positions of
the lines; or 0 - no redshift is possible (usually due to too low S/N).

Some spectra were also flagged for re-processing through {\tt fxcor},
if the redshift was clearly wrong, and sufficient signal was present to
get a better redshift estimate.  The main reason for an incorrect
redshift was the presence of large residuals from the subtraction of
bright night sky lines.  Note that in the Fourier cross-correlation,
the direction (i.e. absorption or emission) of lines is not taken into
account; therefore, night sky residuals which approximate the positions
of absorption features in the template can be confused.

Absorption lines in the galaxy spectrum were logged, and if a possible
emission line was present, the 2D spectrum was inspected to check if
the emission was a genuine galaxy feature, or a residual sky line.

For several spectra, for which the redshifts were readily apparent, and
very strong emission was seen, the E/S0 template gave a poor redshift
estimate.  In this case an Sab/Scd template was substituted and was
found to give a much better fit.  Emission line objects are clearly
noted in the tables of results, below.

Furthermore, once groupings in redshift space had been located
(\S\ref{sec:groupings}), all the spectra which failed to yield a
redshift were re-examined to see if they were compatible with the
redshift of any groupings.  This yielded one extra redshift which had
been missed previously.

\begin{table}
\caption{Summary of Spectral Quality. 2 - secure redshift; 1 - less
confident; 0 - rejected.}
\centerline{
\begin{tabular}{l|ccc}
\hline
Field & \multicolumn{3}{c}{Number of Spectra} \\
& Class 0 & Class 1 & Class 2 \\
\hline
R110\_1 & 2 & 5 & 8 \\
R220\_2 & 14 & 7 & 14 \\
R236\_1 & 8 & 6 & 24 \\
R294\_1 & 7 & 8 & 17 \\
\hline
\end{tabular}
}
\end{table}

\begin{figure}
\hspace{0.5cm}
\centering
\includegraphics[width=80mm]{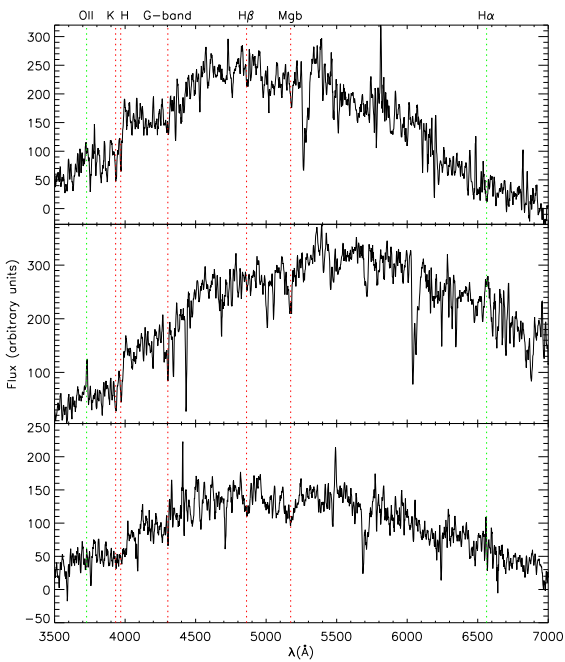}
\caption[Representative Spectra]{Representative de-redshifted spectra.
From top to bottom: spectrum with confident redshift; spectrum with
confident redshift and emission lines; spectrum with less confident
redshift.}  
\label{fig:typ_spec}
\end{figure}

All redshifts were confirmed by visual inspection, by overplotting the
spectral features shown in Fig.~\ref{fig:typ_spec} on the de-redshifted
spectra.  In order to be considered a confident redshift, two or more
spectral lines had to be clearly visible and other features had to have
some good reason for not being seen (e.g. strong sky residuals
concealing a feature which should have been present).

\begin{table*}
\centering
\small
\caption[Galaxy Groupings in Redshift Space]{Galaxy Groupings in Redshift Space}
\label{table:zgroups}
\bigskip
\begin{tabular}{ccccccccc}
\hline
Field & Galaxy& \multicolumn{2}{c}{Centroid$^a$} &N$^b$ &
N$_{TOT}$$^c$& median &$\Delta$z$^d$ & N$_{Absp}^e$\\
& grouping&$\alpha$ (J2000)&$\delta$ (J2000)& & &z & & \\
\hline
R110\_1 & 1a & 14 28 27.9    & 33 05 24 &3 &8(13) & 0.196 & $<$0.001 & 3\\
R220\_2 & 2a & 17 23 25.1 & 74 43 45 &7 &14(21) & 0.260 & 0.003 & 5\\
R236\_1 & 3a & 17 02 59.2 & 51 53 25 &6(7) &24(30) & 0.297(0.297)&
0.007(0.007) & 3(4)\\
R236\_1 & 3b & 17 02 52.4 & 51 54 00 &11 &24(30) & 0.347 & 0.004 &11\\
R294\_1 & 4a & 23 20 00.2 & 12 32 06 &3 &17(25) & 0.268 & 0.001 &1\\
R294\_1 & 4b & 23 19 48.0 & 12 32 58 &3 &17(25) & 0.325 & 0.003 &3\\
R294\_1 & 4c & 23 19 55.6 & 12 32 20 &5 &17(25) & 0.454 & 0.010 &4\\
\hline
\end{tabular}
\begin{minipage}{10.5cm}
\bigskip
$^a$ Centroid of members of grouping, using class 2 redshifts.\\
$^b$ Number of galaxies in grouping -- class 2 spectra (class 1 \& 2 spectra).\\
$^c$ Total number of class 2 (class 1 \& 2) spectra in field.\\
$^d$ Maximum redshift separation between a galaxy in the grouping and the
median redshift of the grouping.\\
$^e$ The number of absorption line only (i.e. no emission) galaxies in
the grouping.\\
\hspace*{0.3cm}
Note: galaxies must be within 1500\kms in the rest-frame, at the median redshift, to
be considered members of the grouping (see text).\\
\end{minipage}
\end{table*}

\subsection{Significance of Clustering in Redshift Space}
\label{sec:groupings}
Groupings in redshift space were extracted by searching for 3 or more
secure redshifts separated by 1500\kms or less.  This is the same
method adopted by \citet{cop99} and corresponds to 3 $\times$ the
typical cluster velocity dispersions they measured.  It should be noted
that a larger value was also tried, but 3$\sigma$ clipping (described
later) removed any extra galaxies added. The results of groupings found
by this technique are illustrated in Fig.~\ref{fig:zhists} and analysed
below.

\begin{figure}
\centering
\includegraphics[width=80mm]{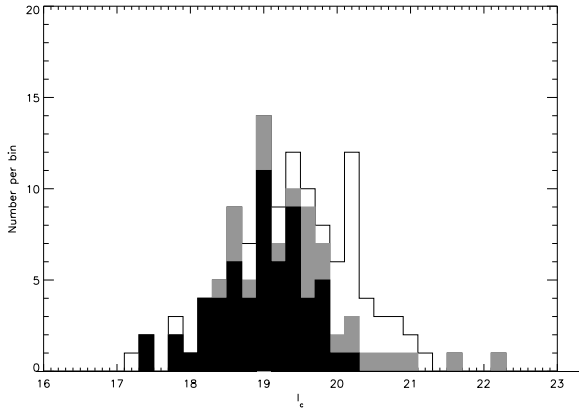}
\caption{The I-band magnitude distribution of galaxies for which
  spectroscopy was attempted.  The empty histogram shows galaxies for
  which no redshift was determined; the black histogram shows galaxies
  with confident redshifts; and the grey histogram shows galaxies with
  less confident redshifts.  }
\label{fig:success}
\end{figure}

\begin{figure}
\vspace{0.5cm}
\centerline{\hbox{
\includegraphics[width=40mm]{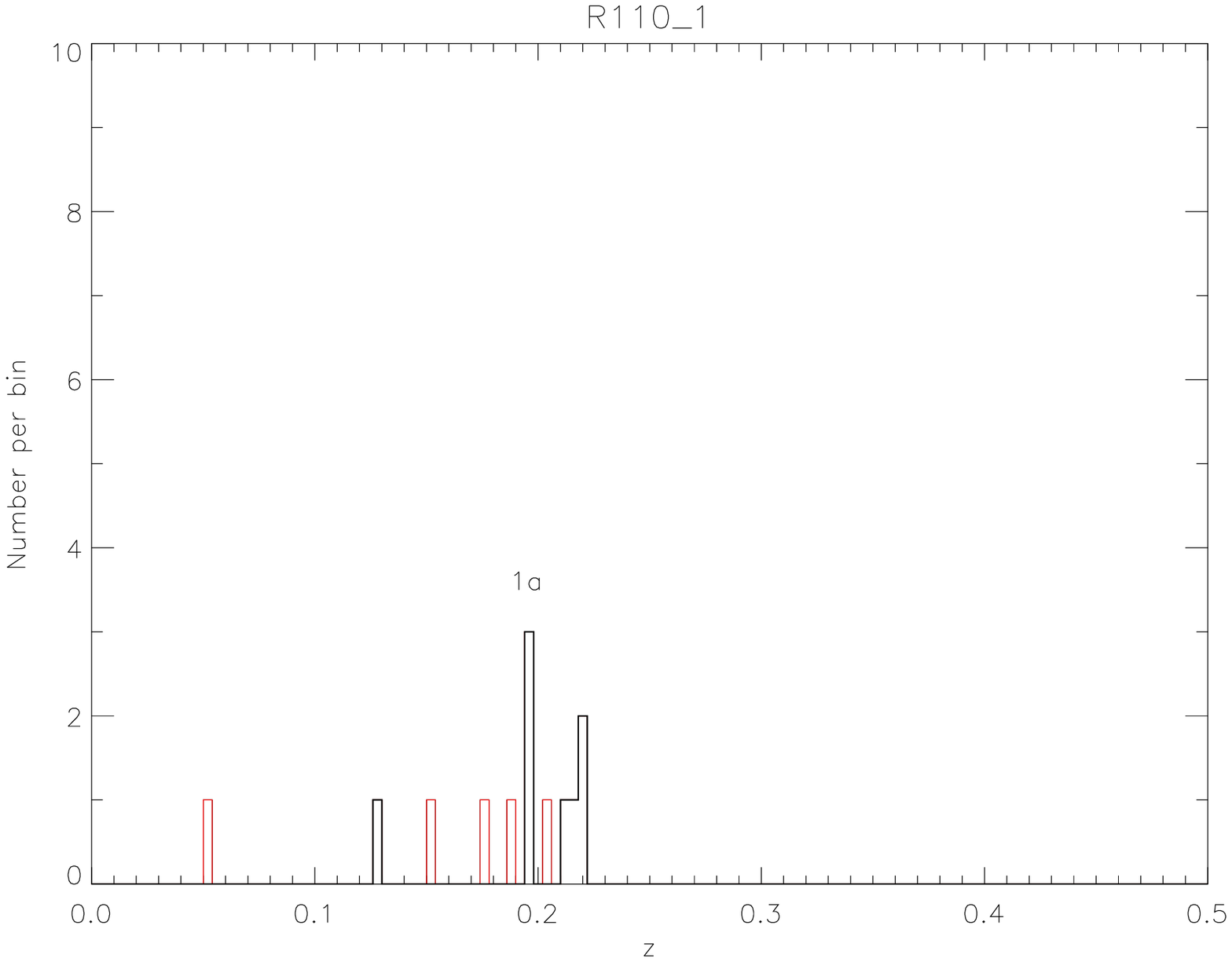}
\vspace*{0.5cm}
\includegraphics[width=40mm]{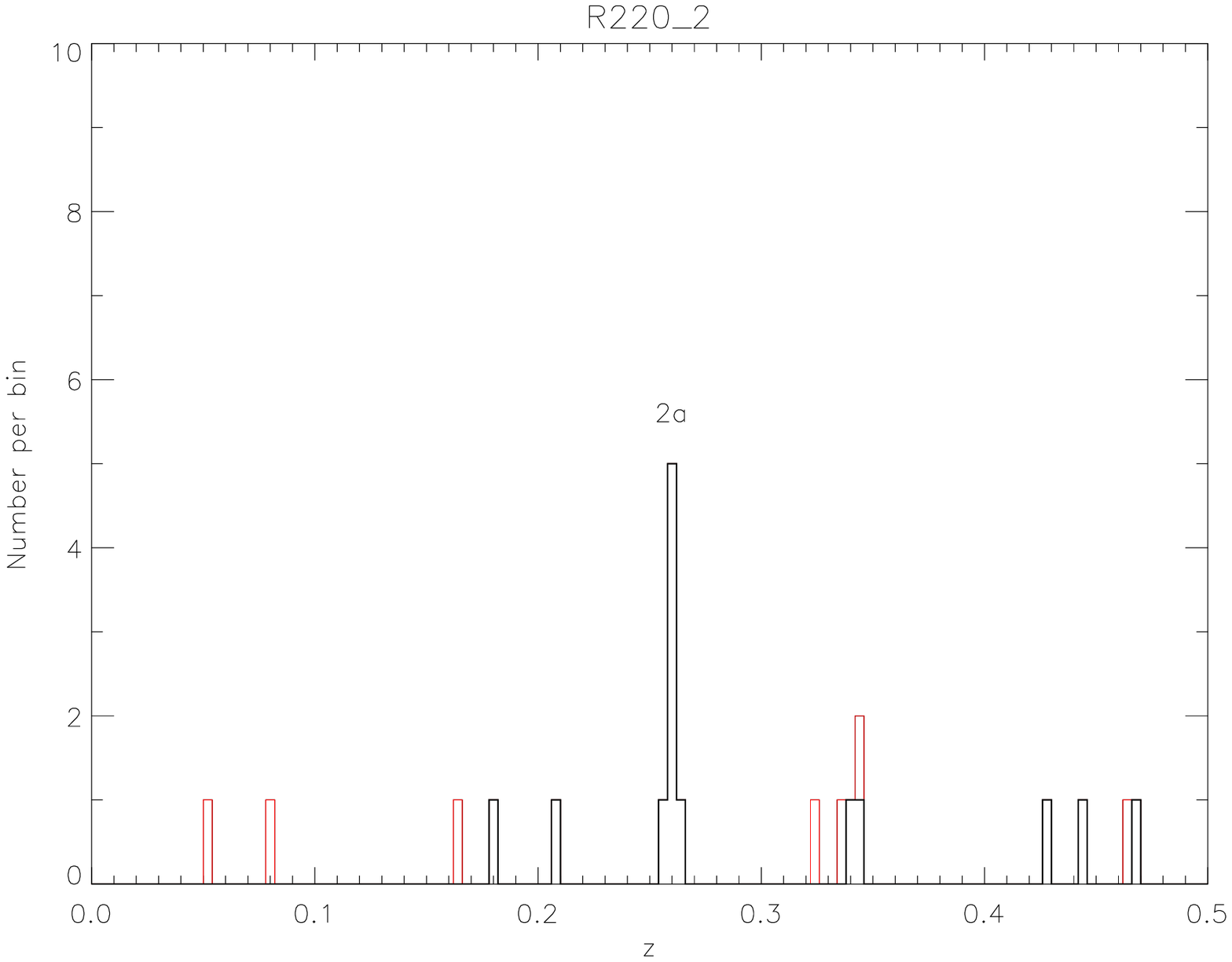}
}
}
\vspace{0.5cm}
\centerline{\hbox{
\includegraphics[width=40mm]{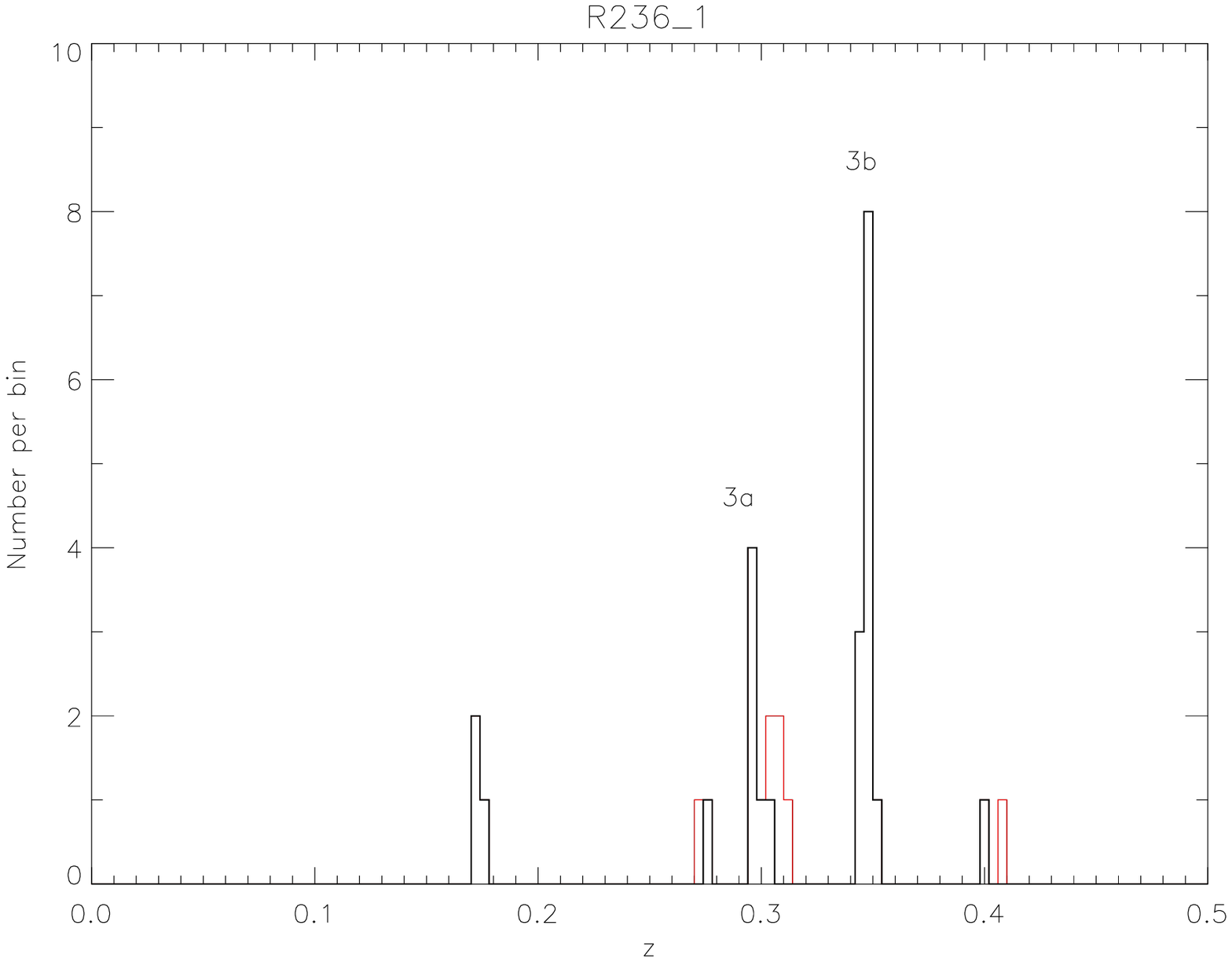}
\vspace*{0.5cm}
\includegraphics[width=40mm]{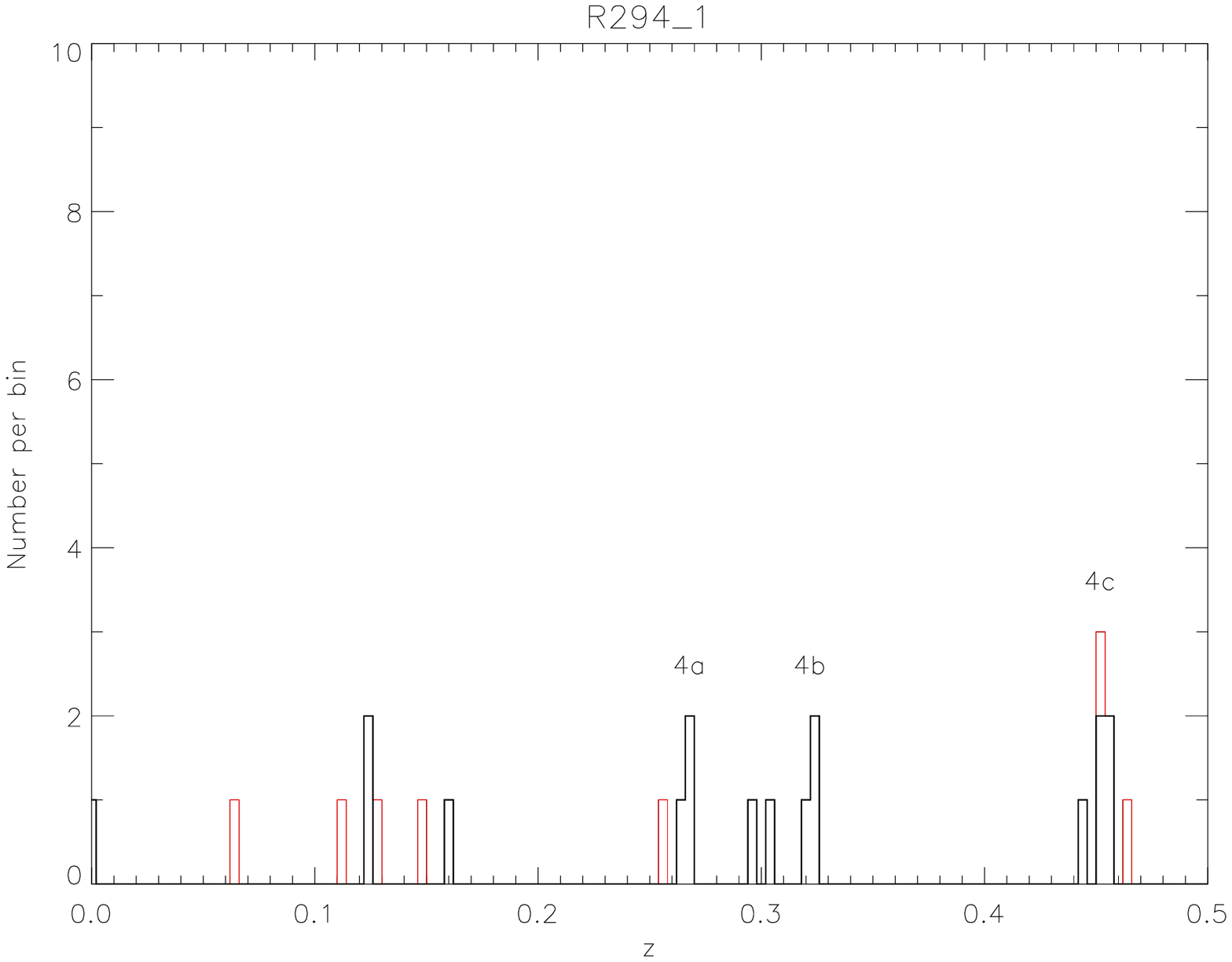}
}
}
\caption[Large Scale Redshift Histograms]{Large scale redshift distributions
 for cluster candidates: R110\_1 (upper left), R220\_2 (upper right),
R236\_1 (lower left), R294\_1 (lower right).  Line thickness indicates
confidence in the redshifts.  Bold lines are confident redshifts.  The
bin size is 0.004, which corresponds to a rest frame velocity of 1200
to 800 kms$^{-1}$ at the left and right sides of the plot,
respectively.  }
\label{fig:zhists}
\end{figure}

\citet{ram00} and \citet{cop99} used similar techniques to assess the
significance of clustering in redshift space.  \citet{ram00}'s method
is followed here.  The selection function was calculated as follows.
Fig.~\ref{fig:success} shows the number of galaxies for which redshift
measurements were possible, as a function of magnitude.  Henceforth,
only secure redshifts will be considered.  Note that all cluster
members in the sample are class 2 (i.e. secure) redshifts, except one
which is class 1. Of the 121 spectroscopic targets, 61 resulted in
secure redshifts, and a further 26 with less secure measurements.  Two
of these objects were stars. The majority of galaxies which fail to
yield a redshift are fainter than $I_c=20.0$.

As with the 2D data in \S\ref{sec:sims}, it is important to model the
clustering of field galaxies when constructing mock galaxy
distributions.  By using the Canada France Redshift Survey
\citep[CFRS,][]{cfrs1} to construct a simulated field redshift
distribution \citep[as was done by][]{cop99}, and bootstrap resampling
sets of galaxies, an estimate can be made of the fraction of spurious
clustering detected in redshift space.  Sets of 15 galaxies were
extracted - the mean number per field (2 masks) for which confident
redshifts were secured.  10000 galaxy sets were generated, applying
bootstrap resampling, and the fraction of sets containing a grouping of
more than 4 galaxies within 1500\kms of their median redshift found.
This occurs by chance $\sim$6\% of the time.  For field R110\_1, only 3
galaxies were found within this velocity difference, but fewer than
average redshifts (8) were obtained (due to only one, rather than two,
masks being used).  The velocity difference between these 3 is less
than 1000 km s$^{-1}$.  This also occurs about 6\% of the time, and is
therefore approximately as significant.  If the velocity difference is
reduced to 1000 km s$^{-1}$, the likelihood of finding 4 or more
galaxies this close together in an observation of a 15 galaxy set is
only 2\%.  These numbers are used as a guideline to the significance of
groupings in redshift space.

\begin{figure}
\centering
\includegraphics[width=80mm]{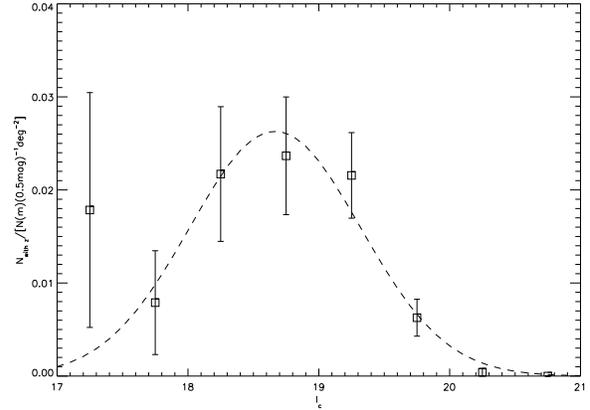}
\caption{The magnitude selection function, $s(m)$, for our
  spectroscopic sample. See text for details.}
\label{fig:s_of_m}
\end{figure}

\citet{ram00} take this technique further by trying to reproduce
more accurately the magnitude selection function.  To do this they take
the histogram of magnitudes for which spectra were obtained and divide
this by the total number of galaxies in the same area in the same
magnitude bin (i.e. the histogram shown in Fig.~\ref{fig:success} is divided
by the field galaxy number counts - Fig.~\ref{fig:counts} - the result is
shown in Fig.~\ref{fig:s_of_m}).

\begin{figure}
\centering
\includegraphics[width=80mm]{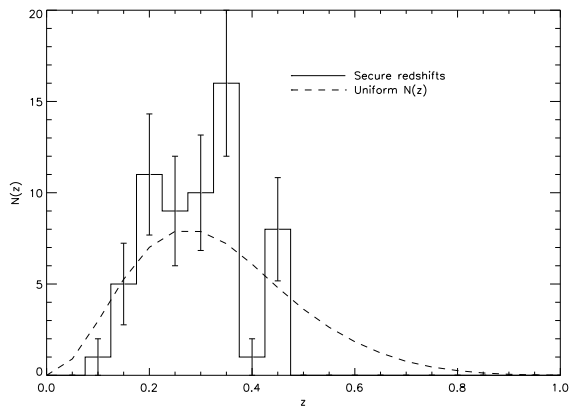}
\caption{The redshift distribution function for our spectroscopic sample.  The dashed line shows the N(z)
  described by Equation \ref{equ:n_of_z}.}
\label{fig:n_of_z}
\end{figure}

If the luminosity function is universal and the local normalisation is
the same everywhere (i.e. no clustering), then the redshift distribution
is given by:

\begin{equation}
\label{equ:n_of_z}
N(z) = \frac{dV(z)}{dz} \int^{m_2}_{m_1} f[L(m,z)]s(m)dm
\end{equation}
\citep{ram00} where dV is the volume element, $m_1$,$m_2$ are the
magnitude limits and $f$ is the Schechter LF given in \S\ref{sec:imp}.
Applying the selection function, $s(m)$ gives the redshift distribution
shown in Fig.~\ref{fig:n_of_z}. \citet{ram00} note that using such a
distribution gives a false impression of the significance of groupings,
as the clustering in redshift space must be accounted for. In order to
do this, the CFRS is again bootstrap resampled, but this time using the
magnitude selection function $s(m)$. In \citet{ram00}'s method, they
compare the N(z) distribution of their data with that of the CFRS using
their selection function, and state that a Kolmogorov-Smirnov test
shows the two datasets have similar distributions.  Using the XDCS
spectroscopic sample, however, a KS test shows the bootstrapped CFRS
N(z) and the N(z) in Fig.~\ref{fig:n_of_z} are significantly different
at the $>$90\% level.  It appears that this is because a larger
fraction of the redshifts in the XDCS sample are cluster members.  The
\citet{ram00} sample targeted higher redshift candidates, and so
recovered a lower fraction of cluster members than the data presented
here (most of the XDCS cluster candidates lie in the range
0.2$<$z$<$0.4).  Therefore \citet{ram00}'s data follow the CFRS N(z) as
most of their data are field galaxies (so it is correct when they state
that magnitude selection is the main process leading to the inclusion
of galaxies in the sample).  In the XDCS sample, however, significant
groupings of 6 or more redshifts (more than the individual groupings in
the \citet{ram00} data) are present, and thus the total (cluster $+$
field) sample is not represented by the field survey of the CFRS.  This
difference in N(z)'s provides reassurance that significant clusters
have been found.

Using Gaussian $s(m)$ leads to different probabilities of false
detections.  For n$_{req}$=4, P(false)=0.19; and for n$_{req}$=5,
P(false)=0.03.  This illustrates that magnitude selection has a big
effect on the significance assigned.

To summarise these tests: the CFRS has been used to simulate the
redshift distribution of field galaxies.  Two different magnitude
selection functions have been used to sample this survey.  Bootstrap
resampling of the data is used to calculate the probability, P(false),
of incorrectly identifying a grouping of $n_{req}$ galaxies in redshift
space - the galaxies being selected in the same way as for the MOSCA
targets. For the simplest selection function (a step function in
magnitude, selecting galaxies brighter than $m_{lim}$),
P(false)$\approx$0.06 for $n_{req}=$4. For the best-fit Gaussian
magnitude selection function, P(false)$\approx$0.19 for $n_{req}=$4,
and P(false)$\approx$0.02 for $n_{req}=$5.

These are the most recent techniques used in the literature for
estimating the significance of redshift groupings found from optical
cluster surveys.  However, there are several problems with them.
Firstly, using the CFRS to model the field does not take any account of
the presence of groups within it.  Thus, for finding the lowest
velocity dispersion systems, which will be numerous in any wide-field
survey (and therefore the CFRS), these methods underestimate the
significance of systems found.  Secondly, not all the available
information is used.  The most basic spectral properties (i.e. whether
or not the cluster members possess emission lines) and the colours can
be used to infer the types of galaxies in the sample.  Absorption line
systems with red colours at the given redshift are highly indicative
that a galaxy is of early-type.  Since these systems dominate the
cores of known clusters, but are much less common in the field, their
presence increases the likelihood of a cluster. This technique,
however, would be biased against systems not containing early-type
galaxies. 

It should be noted that using a mean $N(z)$ from many fields would tend
to overestimate the significance of clustering found in any one pencil
beam survey, especially if galaxies lie in sheets along the line of
sight.

By taking the field galaxy luminosity function from the CNOC2 survey
\citep{lin99}, the expected number of early-type galaxies in a given
volume can be calculated.  \citet{lin99}'s parameters for the
R$_C$-band luminosity function in the q$_0$=0.1 cosmology are
taken. This is a Schechter LF with two additional parameters to model
the evolution in luminosity and density. \citet{lin99} state that
to convert their LFs to another band, a good approximation is to just
apply an appropriate offset in M$^\star$ based on the mean rest-frame
colour for that galaxy type.  Thus, to calculate the LF at z$=$0.3 for
early-type galaxies, a rest-frame colour of R$_C$-I$_C$=0.71 is used
\citep{ka97} to correct M$^\star$ to the I$_C$-band, along with a
correction for the different value of $h$ ($+5\log h$). The value of
$\Phi^\star$ is taken from the R$_c$ value, only correcting for the $h$
difference, and applying the evolutionary parameters $P$ and $Q$
\citep[Table 2,][]{lin99} to correct to a redshift of 0.3. The
difference in $\Phi^\star$ over the range of interest for the extreme
XDCS spectroscopic candidates (i.e. 0.2 $\lsim z \lsim$ 0.5) is less
than a factor of 2, and so, for simplicity, a fiducial redshift of 0.3
is used here.  This results in a space density of early-type galaxies
in the field at redshift 0.3 of 0.04 Mpc$^{-3}$.

The number of galaxies in each redshift grouping showing only
absorption features is given in Table \ref{table:zgroups}. Since these
systems do not show emission lines and have colours consistent with
early-type galaxies at the cluster redshift (Table
\ref{table:zgroups}), these are taken to be early-type galaxies. A
generous estimate of the volume from which each of these redshift
groupings is drawn is to take an angular size of 5 arcmins
(approximately the maximum separation on the sky between galaxies in
the same redshift grouping) at a redshift 0.3, and to assume the volume
is a sphere of this radius (this is about the same size as given by a
line-of-sight velocity difference of a 1000 km s$^{-1}$, again a
generous value for these groupings).  This translates into a volume of
$\approx$ 8 h$^{-1}$Mpc$^3$.  Since the space density of early-type
galaxies in the field at this redshift is of 0.04 h$^{-1}$Mpc$^3$, the
expected number in such a volume is 0.32.  Assuming the field can be
modelled by a Poisson distribution with this expectation value, the
likelihood of finding 11 early-type galaxies (the maximum found -
candidate 3b) is only $\approx$1$\times$10$^{-8}$. The likelihood of
finding 1 (the minimum - candidate 4a) is 25\%; and the likelihood of
finding 3 (the minimum of all the remaining candidates) is 0.3\%.

This argument is an over-simplification as firstly it assumes the field
can be modelled as a Poisson distribution, which is not strictly
correct because of clustering, but not a bad approximation (it just
raises the expectation value slightly). Secondly, the candidates were
selected to be overdense in galaxies of the same colour. This means the
fields selected were not typical regions of space. However, selecting
galaxies on the CMR was not guaranteed to select early-type galaxies at
the same redshift, but this is the result which was found from the
spectroscopy. Therefore, the method is still valid.  Thus, by this
simple argument, it seems reasonable to assume that all the groups
containing at least 3 early-type galaxies are significant. This method
therefore rejects grouping 4a. In summary, all the redshift space
groupings are found to be significant by this technique, except
candidate 4a which contains 3 galaxies, but only 1 of which is
early-type (i.e. red and emission-free). This candidate is found to be
significant by the previous two methods which do not take colour/ type
into account.  Clusters of this nature (i.e. not dominated by
early-type galaxies) have not been observed before, so this system must
be treated cautiously.

\subsection{Comparison of Significant Redshift Groupings with Cluster Candidates}

Now that the groupings in redshift space have been identified and their
significances assessed, the final step is to compare these with the
candidates detected with the cluster-finding algorithm.  Firstly, a
simple comparison will be made by just finding the nearest candidate in
the catalogue (Table~\ref{table:cmr_mosca}) with the centroids of the
groupings found with MOSCA (Table \ref{table:zgroups}). These are
tabulated below for the CMR algorithm (Table \ref{table:cmr_zgps}).

For the CMR-finder, the offset between measured and estimated redshift
for the two isolated groups (R110\_1 and R220\_2) is
$\Delta$z$\leq$0.05 for both. 
For the multiple systems, although only one estimated redshift is given
in the table for each field (for the most significant candidate), these
candidates were flagged as line-of-sight-projections in Table
\ref{table:cmr_mosca}. The estimated redshift of the most significant
CMR candidate is always intermediate between the spectroscopic
redshifts, and always $\lsim$0.1. The full catalogues may be examined
for these projected systems, to see how well these agree with the
spectroscopically determined groups.

\begin{table}
\tiny
\centering
\caption[Nearest CMR candidate to each MOSCA group]{Nearest CMR candidate
to each MOSCA group. 
\begin{minipage}{8.5cm}
\smallskip
$^a$ Separation between the centroid of the spectroscopic grouping and
the nearest MF candidate in arc minutes and physical distance (Mpc) at
z$_{spec}$.\\ 
$^b$ Spectroscopic redshift of the former grouping.\\
$^c$ CMR estimated redshift\\
$^d$ Difference between these two redshifts.
\end{minipage}
}
\label{table:cmr_zgps}
\begin{tabular}{lcccccc}
\hline
Field & ID & CMR Candidate & Separation$^a$ & z$_{spec}^b$ & z$_{est}^c$ &
  $\Delta$z$^d$ \\ & & ID & (arcmin/ Mpc)& & & \\
\hline
R110 & 1a & cmJ142812.0+330736 & 3.995/0.81 & 0.196 & 0.160 & 0.036 \\
R220 & 2a & cmJ172333.0+744410 & 0.670/0.17 & 0.260 & 0.210 & 0.050 \\
R236 & 3a & cmJ170244.2+515539 & 3.228/0.88 & 0.297 & 0.310 & 0.013 \\
R236 & 3b & cmJ170244.2+515539 & 2.092/0.63 & 0.347 & 0.310 & 0.037 \\
R294 & 4a & cmJ231951.2+123208 & 2.187/0.56 & 0.268 & 0.370 & 0.102 \\
R294 & 4b & cmJ231951.2+123208 & 1.147/0.33 & 0.325 & 0.370 & 0.045 \\
R294 & 4c & cmJ231951.2+123208 & 1.082/0.38 & 0.454 & 0.370 & 0.084 \\
\hline
\end{tabular}
\end{table}

The full CMR catalogues in the region of the R236\_1 and R294\_1 fields
are given in Table \ref{table:cmr_mosca_full}. The candidates are split
between A and B rotation results for R236\_1, as here the two rotations
overlap\footnote{R110\_1 also has overlapping A and B rotation,
although the V-band A-rotation data is slightly trailed and so rejected
from this analysis.}.  This is not the case for R294\_1. 

\begin{figure}
\centering
\includegraphics[width=40mm]{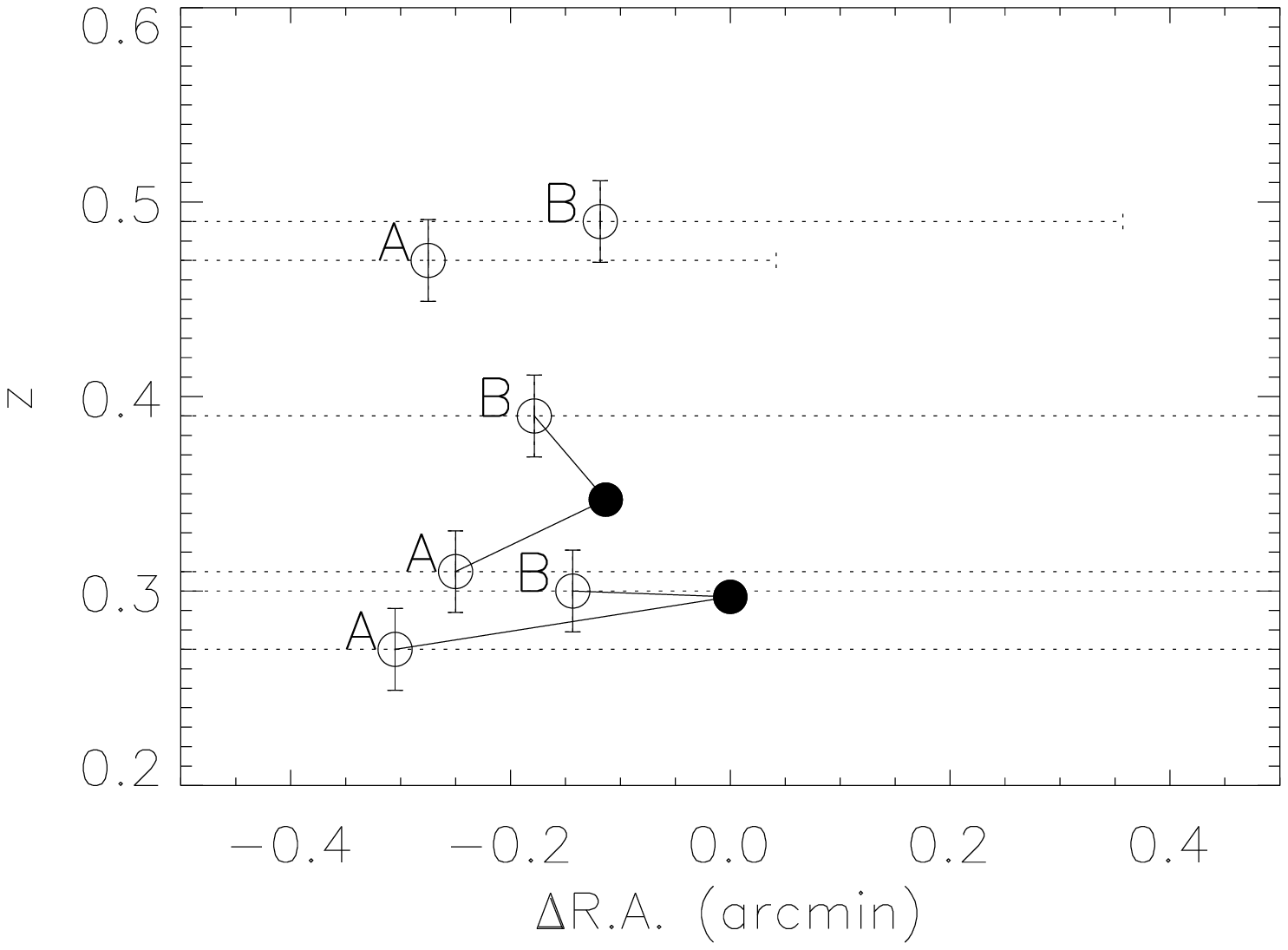}
\includegraphics[width=40mm]{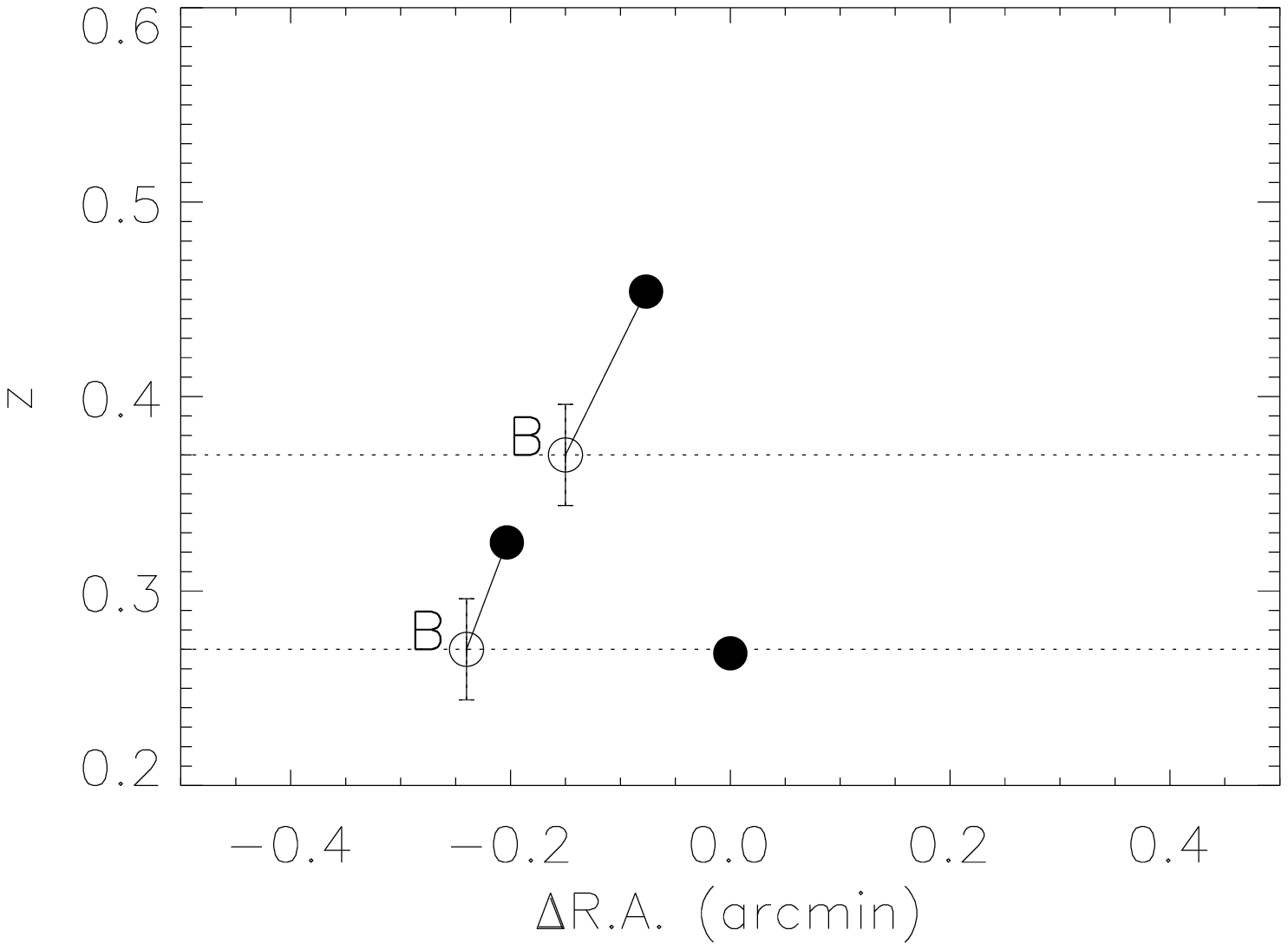}
\caption{R.A. vs z (declination slice) plot for MOSCA
(spectroscopically determined) groups and CMR candidates R236\_1
(left) and R294\_1 (right).  Filled circles are MOSCA groups (note:
the lowest-z point in the R294\_1 is not significant in terms of
early-type galaxies); open circles are CMR candidates from A- and
B-rotation data (as labelled).  Broken horizontal error bars denote
optical radii of candidates (typically $\sim$1 arcmin); solid lines
connect optical candidates to nearest spectroscopically confirmed
group; solid vertical error bars denote width of redshift slice
(i.e. CMR estimated redshift error, c.f. Fig.~7).  For R236\_1, two
candidates in each rotation are seen near the spectroscopically
determined group.  Also, a higher redshift candidate is seen in both
rotations.  The spectroscopy may not probe deep enough to have found
members of this group. In R294\_1 two candidates are found within
$\Delta$z$\lsim$0.08 of the most significant (i.e. the two highest
redshift) spectroscopic groups. }
\label{fig:raz}
\end{figure}

A cross-correlation analysis between the position of each CMR candidate
and each MOSCA candidate (position given by centroid of redshifts), and
their respective redshifts, was performed.  In this way, the closest
match in projected and redshift space was located.  Fig.~\ref{fig:raz}
shows this data projected into 2D along a line of constant declination
(so the R.A. offset gives the approximate sky-plane offset). It can be
seen that, for R236\_1, both the A and B catalogues identify 3
candidates at approximately the same redshifts: two close to the MOSCA
groups and one at slightly higher redshift.  This illustrates that
agreement between the A and B redshift estimates is good, and the
agreement with the spectroscopic redshifts is also good ($\leq$0.05).
The possibility of a higher redshift candidate, not reached by the
depth of the MOSCA spectroscopy, is likely, given that it is identified
independently in both rotations, and because the lower redshift groups
agree so well with the spectroscopy.  For R294\_1, two candidates are
found.  Given that the lowest redshift of the three MOSCA groups in
this field (4a) is not significant from the space density of early-type
galaxies analysis (and will not be found by the CMR algorithm, because
it does not contain significant numbers of early-type galaxies), the
most likely interpretation is that the CMR-finder detects the two
highest redshift groups and underestimates the redshifts of both
(albeit by only $\lsim$0.08). Thus, the candidates from the full
catalogue are naturally associated with the nearest groups in 2D space,
and these are then found to be also the nearest in redshift space.
Therefore, the CMR finder performs excellently, correctly finding and
separating all the systems identified spectroscopically.

\begin{table}
\caption{Groups from the full CMR-catalogue for those systems
flagged as ``projections''.  $\sigma$ is the significance from the CMR algorithm.}
\label{table:cmr_mosca_full}
\centerline{
\begin{tabular}{lccc}
\hline
Field & CMR candidate ID & z$_{est}$ & $\sigma$ \\
\hline
R236A  & cmJ170240.9+515512 &  0.270  & 5.15 \\
R236A  & cmJ170242.7+515222 &  0.470  & 4.35 \\
R236A  & cmJ170244.2+515539 &  0.310  & 5.55 \\
R236B  & cmJ170248.5+515051 &  0.390  & 4.55 \\
R236B  & cmJ170250.6+515506 &  0.300  & 6.85 \\
R236B  & cmJ170252.1+515717 &  0.490  & 4.85 \\
R294B  & cmJ231945.8+123304 &  0.270  & 4.85 \\
R294B  & cmJ231951.2+123208 &  0.370  & 5.05 \\
\hline
\end{tabular}
}
\end{table}

\subsection{Cluster Velocity Dispersion Estimates}

The cluster redshift and velocity dispersion
were calculated following \citet{bfg90}.  They recommend using the
median and standard deviation when dealing with tiny (n$\sim$5)
datasets.  Only the secure redshifts were considered.  Redshifts within
$\sim$2000 km s$^{-1}$ of the peak in the redshift histogram were
extracted and the median value was taken to be the cluster redshift.
The standard deviation was computed, and any value exceeding 3 standard
deviations from the median excluded (this was only the case for the
clusters in field R236\_1: one value was rejected from each), and the
standard deviation then re-computed.  This was then transformed to the
velocity dispersion in the cluster's rest-frame.  The confidence
interval for the velocity dispersion was found by applying the
statistical jack-knife technique to the data \citep[for
example][]{cnoc96}.  This simple resampling technique uses
`pseudo-values' $\delta_i$ of the data, by calculating the difference
between a statistical measure, $f$, calculated for the whole dataset,
and for the dataset with one value removed $\delta_i = f(x_1,...,x_n) -
f(x_1,...,x_i,x_{i+1},....,x_n)$.  The estimate of the variance is
$[n/(n-1)\sum_i{\delta_i^2}]^{1/2}$ \citep{efron81}. For very small N
($\sim3$) this error estimate is likely to be highly biased as only two
data points are being resampled each time and the factor of
$\sqrt{n/(n-1)}$ is likely to be an underestimate. These values must be
treated cautiously for the three groupings with only 3 galaxies.

\begin{table}
\centering
\caption{Cluster velocity dispersion estimates. 
Columns are: ID of redshift grouping; Field ID;
Number of galaxies (N) used in redshift determination; velocity
dispersion ($\sigma_z$); velocity dispersion in cluster rest-frame; and
error on this quantity from jack-knife estimate (see text). This error
estimate is likely to be biased in the presence of very small N
(i.e. $\sim$3), and so such error estimates should be treated
cautiously. Most of the more reliable systems show velocity dispersions
in the range 300 \kms $-$ 700 kms$^{-1}$, typical of massive groups and low -
intermediate mass clusters.} 
\tiny

\bigskip
\begin{tabular}{lcccccc}
\hline
Grouping & Field & N & z & $\sigma_z$ & $\sigma_v^{rest}$ &
        $\Delta\sigma_v^{rest}$ \\ ID & & & & & (kms$^{-1}$) &
        (kms$^{-1}$) \\
\hline
1a & R110\_1  & 3 & 0.196440 & 0.00031 & 78 & 95 \\
2a & R220\_2 & 7 & 0.259740 & 0.00143 & 341  & 346 \\
3a & R236\_1 & 6 & 0.297210 & 0.00315 & 728  & 504 \\
3b & R236\_1 & 11 & 0.347100 & 0.00179 & 398  & 262 \\
4a & R294\_1 & 3 & 0.268450 & 0.00259 & 612  & 750 \\
4b & R294\_1 & 3 & 0.325470 & 0.00224 & 506  & 620 \\
4c & R294\_1 & 5 & 0.453810 & 0.00467 & 962  & 1051 \\
\hline
\end{tabular}
\end{table}

As described in \S\ref{sec:richness}, \citet{s98} plotted the L$_E$
measure against the X-ray temperature of the most X-ray luminous
clusters at z$\sim$0.2, and found a good correlation. This suggests
that L$_E$ may be a good tracer of cluster mass.  Using the relation
between velocity dispersion and X-ray temperature of \citet{wu} to
transform the \citet{s98} data yields a power law fit which is
consistent with our data, although given the large uncertainties on our
velocity dispersions, we cannot constrain this relation.  We simply
note that given our limited data, we cannot tell if the relation
between L$_E$ and $\sigma$ for X-ray luminous clusters still holds for
X-ray underluminous systems.

Spectroscopic observations are underway for several of the XDCS
clusters as part of other projects, and so these should allow much more
accurate estimates of the velocity dispersions, further investigation
of this relation, and the use of L$_E$ as a mass estimator.

\subsection{Summary of Spectroscopic Results}
Spectroscopy has been undertaken for four XDCS subfields, containing
cluster candidates not showing significant X-ray emission.  Candidate
groups in redshift space were identified, and the significance of these
groups evaluated by three different techniques.  The first two involved
bootstrap resampling the Canada-France Redshift Survey using different
selection functions.  Using a simple magnitude limited selection showed
(in general agreement with \citet{cop99}'s method) that 3 concurrent
redshifts was a significant grouping; using a Gaussian magnitude
selection greatly reduced the significance of these groups (showing the
technique is very sensitive to the simulated selection function), but
the drawbacks of both these techniques were discussed.  An argument
based on the space-density of early-type galaxies showed that three
early-type galaxies constituted a robust group.  Using this latter
argument, one group was detected in each of two of the fields, and two
groups were detected in the other two fields.  The colour-magnitude
finder correctly separated line of sight projected systems and also
detected a higher redshift system, not revealed by the spectroscopy
(most likely as galaxies sufficiently faint were not targeted).  The
CMR redshift error is around 0.04, for groups in the redshift range 0.2
- 0.4. Velocity dispersions for most of these systems are around 300 -
700 \kms (corresponding to massive groups and low - intermediate mass
clusters) but these are estimated from tiny numbers of galaxies
($\approx$5), and have jack-knife estimated errors of around 60 - 100\%
(and for the systems with only 3 redshifts, these errors are likely to
be underestimated). Finally, the luminosity in early-type galaxies
versus the velocity dispersion was compared with the relation taken for
the high X-ray luminosity cluster sample of \citet{s98}, and seen to
be consistent, although the errors on the XDCS velocity dispersions are
very large.

\section{Discussion, Conclusions \& Future Work}
\label{sec:discussion}

The original motivation for this work was the studies of
\citet{bow94} and \citet{bow97}, which found that in an optically
selected survey of galaxy clusters at z$\sim$0.4, the X-ray emission
was systematically lower than expected for a non-evolving X-ray
luminosity function, relative to local samples.  Their spectroscopic
analysis indicated that these systems had velocity dispersions
comparable to those of more X-ray luminous systems, which suggested that
if the clusters were virialised then they had dynamical masses similar
to the more X-ray luminous/ massive systems; or that the systems were
in fact unrelaxed and their velocity dispersions were thus inflated
above that of a relaxed system.

This work has constructed similar optically selected samples, albeit
from a smaller area (11 deg$^2$ versus 27 deg$^2$) but with a more
quantifiable selection function and using more efficient selection
techniques.  The relationship between X-ray luminosity and richness (as
measured three different ways) shows considerable scatter.  

During the course of this work, results from a similar study by
\citet{don01} were published. They conducted an optical and X-ray
survey in 23 deep ROSAT fields (4.8 deg$^2$) using
\citealt{pdcs}'s Matched Filter algorithm on I-band data. The
depth of their photometry was about 0.5 magnitudes deeper than that of
the XDCS, although their areal coverage was lower by more than a factor
of two. \citet{don01} detected 57 X-ray candidate clusters and 152
candidates in the optical.  Their MF algorithm detected 74\% (26 out of
35) of the most reliable X-ray candidates. This number is in good
agreement with the 75\% (9 out of 12) found with the MF algorithm used
here.  We have shown that an even higher recovery rate is
possible using CMR techniques (potentially 10 out of 10) and with much
more reliable redshift estimates.

As in \citet{don01}, we find that within their optically selected
sample, optical and X-ray luminosity are correlated, with considerable
scatter.  Their measure of richness is essentially the number of
L$^\star$ galaxies ($\Lambda$ in equation \ref{equ:kappanorm})
contributing to the cluster signal at their MF estimated redshift.  We
show that the MF estimated redshifts are much poorer than those
estimated from the CMR finder.  This will potentially increase the
scatter of the relation. They state that although there is significant
scatter within the relation, there is no need to impose a bimodal
distribution of X-ray luminous and X-ray faint clusters.  This seems to
be borne out by this work, as the distribution of detections in
Fig.~\ref{fig:lx_lell} appears continuous.

We find a scatter of 0.2 dex (a factor of 1.6) in the
relation. Clearly this is important if all the systems (both X-ray dark
and bright) are needed for cosmological models. Our X-ray dark clusters
are certainly convincing and in \S\ref{sec:spectr-observ-x} we show
that they are confirmed by spectroscopy. How we deal with the scatter
in cosmological surveys depends on its origin. We consider this below.

Possible reasons for this scatter include: 

1) Variations in the efficiency of galaxy formation.  If galaxy
formation is more efficient at a given epoch/ environment, then for a
given mass of gas, a higher fraction can be converted to stars,
increasing the light to mass ratio of a cluster. Furthermore, this
leaves less gas available for production of X-ray emission, decreasing
the X-ray luminosity. So, higher galaxy formation efficiency leads to
increased optical luminosity and decreased X-ray luminosity. This,
however, is not seen in semi-analytic galaxy formation models such as
\citet{2000MNRAS.319..168C}.

2) The dynamical state of the cluster.  As mentioned before, if a
cluster is dynamically unrelaxed then the hot intracluster gas will not
be centrally concentrated to densities sufficient for X-ray emission
(\S\ref{sec:introduction}).  If the cluster galaxies are already in
place \citep[as seems to be suggested by][]{2001ApJ...552..504S} then
such a cluster would have an unusually low X-ray luminosity for its
optical luminosity. High resolution observations with Chandra have
shown cluster cores are far from relaxed \citep{2001ApJ...555..205M}
but what we require here is an even more widespread distribution of
X-ray properties.

3) Thermal history of the gas.  The presence of cooling gas in the
cluster raises the ICM density and initially increases X-ray
luminosity.  Conversely, injecting energy into the ICM at early times
[e.g. by AGN or through supernovae/ feedback from galaxy
formation \citep{pcn,1998MNRAS.301L..20W,2002ApJ...576..601V,bowerbenson}]
decreases the ICM density and lowers X-ray luminosity. Both these
effects could contribute to scatter in the optical -- X-ray luminosity
relation. The scatter may also reflect different levels of preheating
from cluster to cluster \citep{1997ApJ...482L..13M}.

4) Projection effects. Groups of galaxies projected along the line of
  sight would appear as higher optical luminosity clusters (since the
  number of galaxies observed is simply additive); whereas the X-ray
  luminosity would appear extremely low for a cluster of such optical
  richness, as the X-ray luminosity scales as the square of the gas
  density. This was shown to probably not be a significant factor in
  \S\ref{sec:comparison-optical-x}, by separately considering optical
  cluster candidates flagged as projections.  Although, again, the
  volume probed by this survey is relatively small, so large scale
  filaments viewed `end-on' may be too rare to be included.
  
  These mechanisms all assume that the fundamental parameter is the
  cluster mass. The best measurement for the cluster mass in this paper
  is the velocity dispersion.  This suggested that within the (large)
  errors, a sample of optically selected, X-ray underluminous clusters
  had optical luminosities consistent with those of the most X-ray
  luminous clusters. Clearly, better mass estimates are required for a
  larger number of clusters. Recently, \citet{ye03} have examined the
  CNOC1 sample of X-ray luminous clusters, and found that $B_{gc}$,
  $T_X$ and $L_X$ can be used to infer the dynamical mass of these
  systems to within 30\%. Other possibile mass estimators include
  gravitational lensing \citep{hoekstra} or total K-band galaxy
  luminosity \citep{2003astro.ph..4033L}.

  \section*{acknowledgements}
  
  Based on observations made with the Isaac Newton Telescope operated
  on the island of La Palma by the Isaac Newton Group in the Spanish
  Observatorio del Roque de los Muchachos of the Instituto de
  Astrofisica de Canarias.

  We thank the referee, M. Gladders, for a careful reading which
  improved the clarity of this manuscript.

  The authors acknowledge useful discussions with Alastair Edge, Mike
  Gladders, Mike Irwin, Laurence Jones, Catarina Lobo, Pasquale
  Mazzotta, Nigel Metcalfe, Ray Sharples, Ian Smail, and Alexey
  Vikhlinin. Thanks for technical help with various stages of this
  project go to Mike Balogh, Eric Bell, Harald Kuntschner, and Peder
  Norberg; and to Taddy Kodama for providing results of his stellar
  population synthesis models.
    
  Particular thanks go to Peter Draper and Mark Taylor for assistance
  with the WFC distortion correction.
    
  We are grateful to the staff of the ING and Calar Alto observatories
  for assistance with the observations presented here.  DGG is
  supported by the Leverhulme Trust and was supported by a PPARC
  studentship for the bulk of this work. BLZ acknowldeges financial
  support from the Volkswagen Foundation (I/76\,520) and the Deutsche
  Forschungsgemeinschaft (ZI\,663/2--1).

\label{lastpage}

\end{document}